%% file: THDMren-rev2.tex
\documentclass[a4paper,11pt]{article}

\usepackage[utf8]{inputenc}
\usepackage[numbers,sort&compress]{natbib}

\usepackage[margin=1.0in]{geometry}

\usepackage{graphicx}
\usepackage{amsmath}
\usepackage{hyperref}
\usepackage{placeins}
\usepackage{slashed}
\usepackage[colorinlistoftodos, shadow]{todonotes}
\usepackage{titlesec}
\usepackage{gensymb}
\usepackage{textcomp}
\usepackage{amssymb}
\usepackage{comment}
\usepackage{subfigure} 
\usepackage{feynarts} 

\allowdisplaybreaks

\def\refeq#1{\mbox{(\ref{#1})}}
\def\reffi#1{\mbox{Fig.~\ref{#1}}}

\def\refta#1{\mbox{Tab.~\ref{#1}}}

\def\refse#1{\mbox{Sec.~\ref{#1}}}

\def\citere#1{\mbox{Ref.~\cite{#1}}}
\def\citeres#1{\mbox{Refs.~\cite{#1}}}

\newcommand{\im}{\mathrm{i}}

\newcommand{\GeV}{\unskip\,\mathrm{GeV}}

\def\mathswitch#1{\relax\ifmmode#1\else$#1$\fi}
\def\mathswitchr#1{\relax\ifmmode{\mathrm{#1}}\else$\mathrm{#1}$\fi}
\def\mathswitchit#1{\relax\ifmmode{#1}\else$#1$\fi}

\newcommand{\order}[1]{\ensuremath{ \mathcal{O}( #1 ) }}

\newcommand{\Ph}{\mathswitchr h}
\newcommand{\PH}{\mathswitchr H}
\newcommand{\PAO}{\mathswitchr {A_0}}
\newcommand{\PHP}{\mathswitchr {H^+}}

\newcommand{\Pu}{\mathswitchr u}
\newcommand{\Pd}{\mathswitchr d}
\newcommand{\Ps}{\mathswitchr s}
\newcommand{\Pc}{\mathswitchr c}
\newcommand{\Pt}{\mathswitchr t}

\newcommand{\Pe}{\mathswitchr e}

\newcommand{\PW}{\mathswitchr W}
\newcommand{\PZ}{\mathswitchr Z}

\newcommand{\PG}{\mathswitchr {G}}
\newcommand{\PGO}{\mathswitchr {G_0}}
\newcommand{\PGP}{\mathswitchr {G^+}}

\newcommand{\PA}{\mathswitchr A}

\newcommand{\MW}{\mathswitch {M_\PW}}

\newcommand{\MZ}{\mathswitch {M_\PZ}}
\newcommand{\MH}{\mathswitch {M_\PH}}
\newcommand{\Mh}{\mathswitch {M_\Ph}}
\newcommand{\MAO}{\mathswitch {M_\PAO}}
\newcommand{\MHP}{\mathswitch {M_\PHP}}
\newcommand{\Me}{\mathswitch {m_\Pe}}

\newcommand{\Md}{\mathswitch {m_\Pd}}
\newcommand{\Mu}{\mathswitch {m_\Pu}}
\newcommand{\Ms}{\mathswitch {m_\Ps}}
\newcommand{\Mc}{\mathswitch {m_\Pc}}
\newcommand{\Mb}{\mathswitch {m_\Pb}}
\newcommand{\Mt}{\mathswitch {m_\Pt}}

\newcommand{\scrs}{\scriptscriptstyle}
\newcommand{\sw}{\mathswitch {s_{\scrs\PW}}}
\newcommand{\cw}{\mathswitch {c_{\scrs\PW}}}
\newcommand{\sa}{\mathswitch {s_\alpha}}
\newcommand{\ca}{\mathswitch {c_\alpha}}
\newcommand{\cb}{\mathswitch {c_\beta}}
\renewcommand{\sb}{\mathswitch {s_\beta}}

\newcommand{\GF}{\mathswitch {G_\mu}}

\newcommand{\ri}{{\mathrm{i}}}

\newcommand{\rS}{{\mathrm{S}}}
\newcommand{\rR}{{\mathrm{R}}}
\newcommand{\rL}{{\mathrm{L}}}
\newcommand{\rT}{{\mathrm{T}}}

\def\Re{\mathop{\mathrm{Re}}\nolimits}

\newcommand{\lsim}
{\;\raisebox{-.3em}{$\stackrel{\displaystyle <}{\sim}$}\;}

\newcommand{\MSbar}{{$\overline{\mr{MS}}$}}

\newcommand{\Pb}{\mathswitchr b}

\newcommand{\mr}{\mathrm}

\def\bfi{\begin{figure}}
\def\efi{\end{figure}}

\def\draftdate{\relax}
\def\mda{\relax}
\def\mua{\relax}
\def\mla{\relax}
\def\draft{
\def\thtystars{******************************}
\def\sixtystars{\thtystars\thtystars}
\typeout{}
\typeout{\sixtystars**}
\typeout{* Draft mode!
         For final version remove \protect\draft\space in source file *}
\typeout{\sixtystars**}
\typeout{}
\def\draftdate{\today}
\def\mua{\marginpar[\boldmath\hfil$\uparrow$]%
                   {\boldmath$\uparrow$\hfil}%
                    \typeout{marginpar: $\uparrow$}\ignorespaces}
\def\mda{\marginpar[\boldmath\hfil$\downarrow$]%
                   {\boldmath$\downarrow$\hfil}%
                    \typeout{marginpar: $\downarrow$}\ignorespaces}
\def\mla{\marginpar[\boldmath\hfil$\rightarrow$]%
                   {\boldmath$\leftarrow $\hfil}%
                    \typeout{marginpar: $\leftrightarrow$}\ignorespaces}
\def\Mua{\marginpar[\boldmath\hfil$\Uparrow$]%
                   {\boldmath$\Uparrow$\hfil}%
                    \typeout{marginpar: $\uparrow$}\ignorespaces}
\def\Mda{\marginpar[\boldmath\hfil$\Downarrow$]%
                   {\boldmath$\Downarrow$\hfil}%
                    \typeout{marginpar: $\downarrow$}\ignorespaces}
\def\Mla{\marginpar[\boldmath\hfil$\Rightarrow$]%
                   {\boldmath$\Leftarrow $\hfil}%
                    \typeout{marginpar: $\leftrightarrow$}\ignorespaces}
\def\muua{\marginpar[\boldmath\hfil$\upuparrows$]%
                   {\boldmath$\upuparrows$\hfil}%
                    \typeout{marginpar: $\upuparrows$}\ignorespaces}
\def\mdda{\marginpar[\boldmath\hfil$\downdownarrows$]%
                   {\boldmath$\downdownarrows$\hfil}%
                    \typeout{marginpar: $\downdownarrows$}\ignorespaces}
\def\mlla{\marginpar[\boldmath\hfil$\leftleftarrows$]%
                   {\boldmath$\leftleftarrows $\hfil}%
                    \typeout{marginpar: $\leftleftarrows$}\ignorespaces}                    
\overfullrule 5pt
\oddsidemargin -15mm
\marginparwidth 29mm
}

\numberwithin{equation}{section}

\begin{document}

\thispagestyle{empty}
\def\thefootnote{\fnsymbol{footnote}}
\setcounter{footnote}{1}
\null
\draftdate\hfill FR-PHENO-2017-003, CP3-Origins-2017-012 DNRF90
\vfill
\begin{center}
  {\Large {\boldmath\bf {Renormalization schemes for the Two-Higgs-Doublet Model
\\[0.5em] 
and applications to $\Ph \to \PW\PW/\PZ\PZ \to 4$ fermions}
\par} \vskip 2.5em
{\large
{\sc Lukas Altenkamp$^{1}$, Stefan Dittmaier$^{1}$, Heidi Rzehak$^{2}$
     }\\[2ex]
{\normalsize \it 
$^1$Albert-Ludwigs-Universit\"at Freiburg, Physikalisches Institut \\
79104 Freiburg, Germany
}\\[2ex]
{\normalsize \it 
$^2$University of Southern Denmark, $\text{CP}^3$-Origins \\
Campusvej 55, DK-5230 Odense M, Denmark}
}}
\par \vskip 1em
\end{center}\par
\vskip .0cm \vfill {\bf Abstract:} 
\par 

We perform the renormalization of different types of Two-Higgs-Doublet Models
for the calculation of observables at next-to-leading order. 
In detail, we suggest four different renormalization schemes based on on-shell renormalization 
conditions as far as possible and on $\overline{\mr{MS}}$ prescriptions for the remaining 
field-mixing parameters where no distinguished on-shell condition exists and
make contact to existing schemes in the literature. 
In particular, we treat the tadpole diagrams in different ways 
and discuss issues of gauge independence and perturbative stability in the considered schemes.
The renormalization group equations for the $\overline{\mr{MS}}$ parameters are solved in each 
scheme, so that a consistent renormalization scale variation can be performed. 
We have implemented all Feynman rules including counterterms and the renormalization conditions 
into a \textsc{FeynArts} model file, so that amplitudes and squared matrix elements can be generated automatically. 
As an application we compute the decay of the light, CP-even Higgs boson of the 
Two-Higgs-Doublet Model 
into four fermions at next-to-leading order. The comparison of different schemes and the investigation of the renormalization scale dependence allows us to test the perturbative consistency in each of the 
renormalization schemes, and to get a better estimate of the theoretical uncertainty that arises due to the truncation of the perturbation series. 

\par
\vskip 1cm
\noindent April 2017
\par
\null
\setcounter{page}{0}
\clearpage
\def\thefootnote{\arabic{footnote}}
\setcounter{footnote}{0}

\section{Introduction}
\label{se:intro}

After the discovery of a Higgs boson at the Large Hadron Collider (LHC) \cite{Aad:2012tfa,Chatrchyan:2012ufa} at CERN, the complete identification of this particle is ongoing. The  properties of the discovered particle, such as its couplings, are determined experimentally in order to fully identify its nature. For the endeavour of the identification of this particle, 
input from the theory side is needed in form of precise predictions for the production and decay processes in the Standard Model (SM) as well as in its extensions that are to be tested. It is also crucial to provide reliable uncertainty estimates of the theoretical predictions. 
Underestimating this uncertainty might lead to wrong conclusions.
In the SM, predictions and error estimates are well advanced, and in SM extensions they are
consolidating as well (see, e.g., the reviews in 
Refs.~\cite{Dittmaier2011,Dittmaier2012,Dittmaier:2012nh,Heinemeyer:2013tqa,deFlorian:2016spz,%
Spira:2016ztx}).

One of the simplest extensions of the SM 
is the Two-Higgs-Doublet Model (THDM) \cite{Lee:1973iz,HHGuide:1990} where a second Higgs doublet is added 
to the SM field content. The underlying gauge group 
$\mr{SU}(3)_{\text{C}} \times \mr{SU}(2)_{\text{W}} \times \mr{U}(1)_{\text{Y}}$ as well as the fermion content of the SM are kept. After spontaneous symmetry breaking, there are five physical Higgs bosons where three of them are neutral and two are charged. In the CP-conserving case, which we consider, one of the neutral Higgs bosons is CP-odd and two are CP-even with one of them being SM-like. 

Even such a simple extension of the SM 
can help solving some questions that are unanswered in the SM. For example, CP-violation in the Higgs sector could provide solutions to the problem of baryogenesis~\cite{Turok:1990zg,Davies:1994id,Cline:1995dg,Fromme:2006cm}, and inert THDMs contain a dark matter candidate \cite{PhysRevD.18.2574,Dolle:2009fn}. An even larger motivation comes from the embedding of the THDM into more complex models, such as axion \cite{Kim:1986ax, Peccei:1977hh} or supersymmetric models \cite{Haber:1984rc}. Some of the latter are promising candidates for a fundamental theory, and supersymmetric Higgs sectors contain a THDM (in which the doublets have opposite hypercharges). Even though the THDM is unlikely to be the fundamental theory of nature, it provides a rich phenomenology, which can be used in the search for a non-minimal Higgs sector without being limited by constraints from a 
more fundamental theory.

In this sense, it is obvious that the THDM should be tested against data,
and phenomenological studies have been performed recently, e.g., 
{in Refs.~\cite{Celis:2013rcs,Celis:2013ixa,Chang:2013ona,Harlander:2013mla,%
Hespel:2014sla,Baglio:2014nea,Broggio:2014mna,%
Bernon:2015qea,Bernon:2015wef,%
Goncalves:2016qhh,Dorsch:2016tab,%
Cacciapaglia:2016tlr,Cacchio:2016qyh,Aggleton:2016tdd,Han:2017pfo}.} 
In order to provide precise predictions within this model, not only leading-order 
(LO), but also next-to-leading-order (NLO) contributions have to be taken into account. For the calculation of NLO contributions, a proper definition of a renormalization scheme is mandatory. There is no unique choice, and applying different renormalization schemes can help to estimate the theoretical uncertainty of the prediction that originates from the truncation of the perturbative series. 
The renormalization of the THDM has already been tackled in several publications:
First, in Ref.~\cite{Santos:1996vt}, the fields and masses were renormalized in the on-shell scheme, 
however, the prescription given there does not cover all parameters.
In Refs.~\cite{Kanemura:2004mg,LopezVal:2009qy}, a minimal field renormalization was applied, and the field mixing conditions were used to fix some of the mixing angles. 
In view of an automation of NLO predictions within the THDM a tool was written by 
Degrande~\cite{Degrande:2014vpa} 
where all finite rational terms and all
divergent terms are computed using on-shell conditions or conditions within the ``modified minimal subtraction scheme'' ($\overline{\mr{MS}}$). 
Though automation is very helpful, often specific problems occur depending on the model, the process, or the renormalization scheme considered, and, it might be necessary to solve these ``manually''. Specifically, spontaneously broken gauge theories with extended scalar sectors pose issues with the renormalization of vacuum expectation values and the related ``tadpoles'', 
jeopardizing gauge independence and perturbative stability in predictions.
Renormalization schemes employing a gauge-independent treatment of the tadpole terms were described recently in Refs.~\cite{Krause:2016oke,Denner:2016etu,Krause:2016xku}. 

In this paper, we perform the renormalization of various types of THDMs 
(Type I, Type II, ``lepton-specific'', and ``flipped''),
describe four different renormalization schemes
(for each type), 
and provide explicit results facilitating their application in NLO calculations. The comparison of results obtained in these renormalization schemes allows for checking their perturbative consistency, i.e.\ whether the expansion point for the perturbation series is chosen well and no unphysically large corrections are introduced. Knowing in which parts of the parameter space a renormalization scheme leads to a stable perturbative behaviour  is important for the applicability of the scheme. 
In addition, we investigate the dependence on the renormalization scale $\mu_\mr{r}$ which is introduced by defining some parameters via \MSbar{} conditions. In order to investigate the $\mu_\mr{r}$ dependence consistently, we solve the renormalization group equations (RGEs) and include the running effects. We also make contact to different renormalization schemes suggested in the literature, including the recent formulations \cite{Krause:2016oke,Denner:2016etu} 
with gauge-independent treatments of tadpoles.%
\footnote{In our work we do not consider the ``tadpole-pinched'' scheme suggested
in \citere{Krause:2016oke}.
Following the arguments of \citeres{Denner:1994nn,Denner:1994xt}
we consider the ``pinch technique'' just as one of many physically
equivalent choices to fix the gauge arbitrariness in off-shell
quantities (related to the 't~Hooft--Feynman gauge of the
quantum fields in the background-field gauge)
rather than singling out ``its gauge-invariant part'' in any sense.}
To facilitate NLO calculations in practice, we have implemented our renormalization schemes
for the THDM into a \textsc{FeynArts} \cite{Hahn2001} model file, so that amplitudes and squared matrix elements can be generated straightforwardly. 
Finally, we apply the proposed renormalization schemes in the NLO calculation of the partial decay width of the 
lighter CP-even Higgs boson decaying into four fermions, $\Ph \to \PW\PW/\PZ\PZ\to 4f$,
a process class that is a cornerstone in the experimental determination of Higgs-boson
couplings, 
{but for which electroweak corrections in the THDM are not yet known in the literature.}
The impact of NLO corrections on Higgs couplings in the THDM was, for instance, investigated more
globally in \citeres{Kanemura:2014dja,Kanemura:2015mxa}.
{However, a full set of electroweak corrections to all Higgs-boson decay processes in the THDM 
does not yet exist in the literature, so that current predictions (see, e.g., \citere{Harlander:2013qxa})
for THDM Higgs analyses globally neglect electroweak higher-order effects.}
{Our calculation, thus, contributes to overcome this shortcoming;
electroweak corrections to some $1\to2$ particle decays of heavy
Higgs bosons were presented in \citere{Krause:2016oke}.}%
\footnote{Since we consider the decays of the light Higgs boson $\Ph$ via
$\PW$- or $\PZ$-boson pairs, where at least one of the gauge bosons
is off its mass shell, we have to consider the full $1\to4$ process
with all off-shell and decay effects, rendering a comparison to
results on $\PH\to\PW\PW/\PZ\PZ$ not meaningful.}

The structure of the paper is as follows. We introduce the four considered types of THDMs
and our conventions in Sect.~\ref{sec:THDM}, and the derivation of the counterterm Lagrangian is performed in Sect.~\ref{sec:CTLagrangian}. Afterwards we fix the renormalization constants with renormalization conditions (Sect.~\ref{sec:renconditions}). The on-shell conditions, where the renormalized parameters correspond to measurable quantities, are  described and applied in Sect.~\ref{sec:onshellrenconditions}. The renormalization constants of parameters that do not directly correspond to physical quantities are fixed in the $\overline{\mr{MS}}$ scheme, so that they contain only the UV divergences and no finite terms. We describe different renormalization schemes based on different definitions of the $\overline{\mr{MS}}$-renormalized parameters  in Sect.~\ref{sec:diffrenschemes}. 
The RGEs of the \MSbar{}-renormalized parameters 
are derived and numerically solved in Sect.~\ref{sec:running}. 
The implementation of the results into an automated matrix element generator is described in Sect.~\ref{sec:FAmodelfile}, and numerical results for 
the partial decay width $\Ph\rightarrow 4 f$ are 
presented in Sect.~\ref{se:numerics}. 
Finally, we conclude in Sect.~\ref{se:conclusion}, 
and further details on the renormalization prescription as well as 
some counterterms are given in the appendix.

\section{The Two-Higgs-Doublet Model}
\label{sec:THDM}

The Lagrangian of the THDM,  $\mathcal{L}_\mathrm{THDM}$, is composed of the following parts,
\begin{align}
\mathcal{L}_\mathrm{THDM}=\mathcal{L}_\mathrm{Gauge}+\mathcal{L}_\mathrm{Fermion}+\mathcal{L}_\mathrm{Higgs}+\mathcal{L}_\mathrm{Yukawa}+\mathcal{L}_\mathrm{Fix}+\mathcal{L}_\mathrm{Ghost}. 
\label{eq:LTHDM}
\end{align}
The gauge, fermionic, gauge-fixing, and ghost parts 
can be obtained in a straightforward way from the SM ones, e.g., 
given in Ref.~\cite{Denner:1991kt}. The Higgs Lagrangian and the Yukawa couplings to the fermions are discussed in the following and are mostly affected by the additional degrees of freedom 
of the THDM. A very elaborate and complete discussion of the THDM Higgs and Yukawa Lagrangians, including general and specific cases, can, e.g., 
be found in Refs.~\cite{Gunion:1989we,Branco:2011iw}.

\subsection{The Higgs Lagrangian}
\label{sec:HiggsL}

The Higgs Lagrangian, $ \mathcal{L}_\mathrm{Higgs}$, contains the kinetic terms and a potential $V$,
\begin{align}
  \mathcal{L}_\mathrm{Higgs}= (D_\mu\Phi_1)^\dagger (D^\mu \Phi_1)+(D_\mu\Phi_2)^\dagger (D^\mu \Phi_2)- V(\Phi_1^\dagger \Phi_1,\Phi_2^\dagger \Phi_2,\Phi_2^\dagger \Phi_1,\Phi_1^\dagger \Phi_2),
\label{eq:Lhiggs}
\end{align}
with the complex scalar doublets $\Phi_{1,2}$ 
of hypercharge $Y_\mr{W}=1$, 
\begin{align}
\Phi_1=\begin{pmatrix} \phi_1^+ \\ \phi_1^0\end{pmatrix}, \qquad \Phi_2=\begin{pmatrix}\phi_2^+ \\ \phi^0_2\end{pmatrix},
\label{eq:doublets}
\end{align}
and the covariant derivative 
\begin{align}
  D_\mu=\partial_\mu \mp \im g_2 I^a_\mr{W} W^a_\mu +\im g_1 \frac{Y_\mr{W}}{2}B_\mu,
\label{eq:EWcovderivative}
\end{align}
where $I^a_\mr{W}$ ($a = 1,2,3$) are the generators of the weak isospin. 
The SU(2) and 
U(1) gauge fields are denoted 
$W^a_\mu$ and $B_\mu$ with the corresponding gauge couplings $g_2$ and $g_1$, respectively.
The sign in the 
$g_2$~term is negative in the conventions of B\"ohm, Hollik and Spiesberger 
(BHS)~\cite{Denner:1991kt,Bohm:1986rj} and positive in the convention of Haber and Kane 
(HK)~\cite{Haber:1984rc}. We implemented both sign conventions, but used the former one as default. 
In general, the potential involves all hermitian functions of the two doublets up to dimension four and can be parameterized in the most general case as follows \cite{Wu1994,Branco:2011iw},
\begin{align}
V=&\,m^2_{11} \Phi^{\dagger}_1 \Phi_1+ m^2_{22} \Phi^\dagger_2 \Phi_2
-\left[m^2_{12} \Phi^\dagger_1 \Phi_2 + h.c.\right]\nonumber\\
&+ \frac{1}{2} \lambda_1 (\Phi^\dagger_1 \Phi_1)^2+\frac{1}{2} \lambda_2 (\Phi^\dagger_2 \Phi_2)^2+ \lambda_3 (\Phi^\dagger_1 \Phi_1)(\Phi^\dagger_2 \Phi_2)+ \lambda_4 (\Phi^\dagger_1 \Phi_2)(\Phi^\dagger_2 \Phi_1)\nonumber\\
&+\left[\frac{1}{2} \lambda_5(\Phi^\dagger_1 \Phi_2)^2
+(\lambda_6 \Phi^\dagger_1\Phi_1+\lambda_7 \Phi^\dagger_2 \Phi_2)\Phi^\dagger_1 \Phi_2+h.c.\right]\label{eq:THDMpot}.
\end{align}
The parameters $m_{11}^2,m_{22}^2,\lambda_1,\lambda_2,\lambda_3,\lambda_4$ are real, while the parameters $m_{12}^2,\lambda_5,\lambda_6,\lambda_7$ are complex, yielding a total number of 14 real degrees of freedom for the potential. However, the component fields of the two Higgs doublets $\Phi_1$ and $\Phi_2$ do not correspond to mass eigenstates, and these doublets can be redefined using an SU(2) transformation without changing the physics, so that only 11 physical degrees of freedom remain~\cite{Branco:2011iw}.
For each Higgs doublet we demand that the fields develop a 
vacuum expectation value (vev) in the neutral component,
\begin{align}
  \langle \Phi_1 \rangle&=\langle 0| \Phi_1|0\rangle =\begin{pmatrix}0 \\ \frac{v_1}{\sqrt{2} }  \end{pmatrix},&
  \langle \Phi_2 \rangle=\langle 0| \Phi_2|0\rangle =\begin{pmatrix} 0 \\ \frac{v_2}{\sqrt{2}}  \end{pmatrix}.
\end{align}
It is non-trivial that such a stable minimum of the potential exists, 
restricting the allowed parameter space already strongly~\cite{Ferreira:2004yd}. 
In general, the vevs are complex (with a significant relative phase).
The Higgs doublets can be decomposed as follows,
\begin{align}
\Phi_1=\begin{pmatrix} \phi_1^+ \\ \frac{1}{\sqrt{2}}(\eta_1+i \chi_1+v_1)\end{pmatrix}, && \Phi_2=\begin{pmatrix}\phi_2^+ \\ \frac{1}{\sqrt{2}}(\eta_2+i \chi_2+v_2)\end{pmatrix},
\label{eq:decom}
\end{align}
with the charged fields $\phi_1^+,\phi_2^+$, the neutral CP-even fields $\eta_1,\eta_2$, and the neutral CP-odd fields $\chi_1,\chi_2$.

\paragraph{Additional constraints:}
Since the 11-dimensional parameter space of the potential is too large for early experimental analyses, we restrict the model in our analysis by imposing two additional conditions, motivated by experimental results:
\begin{itemize}
\item absence of flavour-changing neutral currents at tree level,
\item CP conservation in the Higgs sector (even though this holds only approximately).
\end{itemize}
The former requirement can be ensured by adding a discrete $\mathbb{Z}_2$ symmetry $\Phi_1 \rightarrow - \Phi_1$ (see Sect.~\ref{sec:YukawaL}). This condition implies that the parameters $\lambda_6$ and $\lambda_7$ vanish.  Permitting operators of dimension two that
violate this  $\mathbb{Z}_2$ symmetry softly,  non-zero values of  $m_{12}$ are still allowed~\cite{HHGuide:1990, Gunion:2002zf}.
Concerning the second condition, the potential is CP-conserving if and only if a basis of the Higgs doublets exists in which all parameters and the vevs are real~\cite{Gunion:2005ja}. For our description we assume that a transformation to such a basis has been done already (if the parameters or vevs were initially complex), so that we only have to deal with real parameters. This renders $m_{12}$ and $\lambda_5$ real.
However, at higher orders in perturbation theory CP-breaking terms and complex phases in the Higgs sector are generated radiatively through loop contributions
involving the quark mixing matrix. For our NLO analysis, this does not present a problem as they appear only beyond NLO in the specific processes we consider.
In addition we assume that a basis of the doublets is chosen in which $v_1,v_2>0$ (which is always possible as a redefinition $\Phi_i \to - \Phi_i$ changes the sign of the vacuum expectation value).
The potential~\eqref{eq:THDMpot} has then the following form,
\begin{align}
V={}&m^2_{11} \Phi^{\dagger}_1 \Phi_1+ m^2_{22} \Phi^\dagger_2 \Phi_2
-m^2_{12} (\Phi^\dagger_1 \Phi_2 + \Phi^\dagger_2 \Phi_1)\nonumber\\ 
&+ \frac{1}{2} \lambda_1 (\Phi^\dagger_1 \Phi_1)^2+\frac{1}{2} \lambda_2 (\Phi^\dagger_2 \Phi_2)^2+ \lambda_3 (\Phi^\dagger_1 \Phi_1)(\Phi^\dagger_2 \Phi_2)+ \lambda_4 (\Phi^\dagger_1 \Phi_2)(\Phi^\dagger_2 \Phi_1)\nonumber\\ 
&+\frac{1}{2} \lambda_5\left[(\Phi^\dagger_1 \Phi_2)^2+(\Phi^\dagger_2 \Phi_1)^2\right]. 
\label{eq:lambdapara}
\end{align}
Expanding the potential using the decomposition~\eqref{eq:decom} and ordering terms with respect to powers of the fields, leads to the form
\begin{align}
V={}& -t_{\eta_1} \eta_1 - t_{\eta_2} \eta_2\nonumber\\
&+\frac{1}{2} \,(\eta_1,\eta_2) \mathbf{M}_\eta \begin{pmatrix}\eta_1 \\ \eta_2 \end{pmatrix}
+\frac{1}{2} (\chi_1,\chi_2)\,  \mathbf{M}_\chi  \begin{pmatrix}\chi_1 \\ \chi_2 \end{pmatrix}
+(\phi_1^+,\phi^+_2) \, \mathbf{M}_\phi \begin{pmatrix}\phi^-_1 \\ \phi^-_2 \end{pmatrix}+\ldots,\label{eq:potinmatrix}
\end{align}
with the tadpole terms proportional to the tadpole parameters $t_{\eta_1}$, $t_{\eta_2}$ and linear in the fields. The mass terms contain the mass matrices 
$\mathbf{M}_\eta$, $\mathbf{M}_\chi$, and $ \mathbf{M}_\phi$ and are quadratic in the CP-even, CP-odd, and charged Higgs-boson fields, respectively. Terms cubic or quartic in the fields are suppressed in the notation here. Only the neutral CP-even scalar fields can develop non-vanishing tadpole terms, since they carry the quantum numbers of the vacuum. Further, in the mass terms, only particles with the same quantum numbers can mix, so that the three different 
types of scalars (neutral CP-even, neutral CP-odd, and charged) do not mix with one another. Of course, through the cubic and quartic terms, which are not shown here, these particles interact with each other. The tadpole parameters are
\begin{subequations}
\begin{align}
t_{\eta_1}=- m^2_{11} v_1- \lambda_1v_1^3/2 + v_2 (m^2_{12}-\lambda_{345}v_1 v_2/2),\\
t_{\eta_2}=- m^2_{22} v_2- \lambda_2v_2^3/2 + v_1 (m^2_{12}-\lambda_{345} v_1 v_2/2),
\end{align}%
\label{eq:tadpoleterms}%
\end{subequations}%
where we introduced the abbreviations $\lambda_{ij\ldots}=\lambda_i+\lambda_j+\ldots$, and the mass matrices are given by
\begin{subequations}
\begin{align}
\mathbf{M}_\eta&=\begin{pmatrix}  m^2_{11}+ 3 \lambda_1 v_1^2/2+ \lambda_{345} v_2^2/2 & -m^2_{12}+  \lambda_{345} v_1 v_2\\ -m^2_{12}+  \lambda_{345} v_1 v_2&  m^2_{22}+ 3 \lambda_2 v_2^2/2+ \lambda_{345} v_1^2/2\end{pmatrix},
\\
\mathbf{M}_\chi&=\begin{pmatrix} m^2_{11}+ \lambda_1 v_1^2/2+ (\lambda_{34}-\lambda_{5}) v_2^2/2 & -m^2_{12}+ \lambda_{5} v_1 v_2\\ 
\hspace{-0pt}-m^2_{12}+ \lambda_{5} v_1 v_2& 
\hspace{-0pt}m^2_{22}+ \lambda_2 v_2^2/2+ (\lambda_{34}-\lambda_{5}) v_1^2/2\end{pmatrix},
\\
\mathbf{M}_\phi&=
\begin{pmatrix} m^2_{11}+ \lambda_1 v_1^2/2+ \lambda_{3} v_2^2/2 
& -m^2_{12}+ \lambda_{45} v_1 v_2/2 \\ 
-m^2_{12}+ \lambda_{45} v_1 v_2/2 & m^2_{22}+ \lambda_2 v_2^2/2+ \lambda_{3} v_1^2/2
\end{pmatrix}.
\label{eq:massmatricesfirst}
\end{align}
\end{subequations}
The fields can be transformed into their mass eigenstate basis via
\begin{subequations}
\label{eq:rotations}
\begin{align}
\label{eq:Hhrot}
\begin{pmatrix}\eta_1 \\ \eta_2\end{pmatrix} &
= \begin{pmatrix}\cos{\alpha} & -\sin{\alpha}\\ \sin{\alpha}& \cos{\alpha}\end{pmatrix} \begin{pmatrix}H \\ h\end{pmatrix},
\\
\begin{pmatrix}\chi_1 \\ \chi_2\end{pmatrix}&
= \begin{pmatrix}\cos{\beta_n} & -\sin{\beta_n}\\ \sin{\beta_n}& \cos{\beta_n}\end{pmatrix} \begin{pmatrix}G_0 \\ A_0\end{pmatrix},
\\
\begin{pmatrix}\phi^\pm_1 \\ \phi^\pm_2\end{pmatrix}&
= \begin{pmatrix}\cos{\beta_c} & -\sin{\beta_c}\\ \sin{\beta_c}& \cos{\beta_c}\end{pmatrix} \begin{pmatrix}G^\pm \\ H^\pm\end{pmatrix}.
\end{align}
\end{subequations}
where $h$, $H$ correspond to the CP-even, $A_0$ to the CP-odd, and $H^\pm$ to the charged mass eigenstates\footnote{In order to avoid a conflict in our notation, we define $\alpha_\mr{em}=e^2/(4\pi)$ as electromagnetic coupling constant and consistently keep the symbol $\alpha$ for the rotation angle.}. The fields $G^\pm$, 
$G_0$ correspond to the Goldstone bosons.
After a rotation of the fields, the potential has the following form,
\begin{align}
\label{eq:NLObarepot}
V=& -t_\PH H - t_\Ph h\nonumber\\*
&+\frac{1}{2} (H,h) \begin{pmatrix}\MH^2 & M^2_{\PH\Ph}\\ M^2_{\PH\Ph}& \Mh^2\end{pmatrix} \begin{pmatrix}H \\ h \end{pmatrix}
+\frac{1}{2} (G_0,A_0) \begin{pmatrix}M^2_\PGO & M^2_{\PGO\PAO}\\ M^2_{\PGO\PAO}& M^2_{\PAO}\end{pmatrix} \begin{pmatrix}G_0 \\ A_0 \end{pmatrix}\nonumber\\
&+ (G^+,H^+) \begin{pmatrix}M_\PGP^2 & M^2_{\PG\PHP}\\ M^2_{\PG\PHP}& \MHP^2\end{pmatrix} \begin{pmatrix}G^- \\ H^- \end{pmatrix}+ \textrm{interaction terms} 
\end{align}
with the 
tadpole parameters
\begin{align}
\label{eq:tadpoledefs}
t_\PH=\ca t_{\eta_1}+\sa t_{\eta_2}, \qquad
t_\Ph=-\sa t_{\eta_1}+\ca t_{\eta_2},
\end{align}
where general abbreviations for the trigonometric functions $s_x\equiv\sin x $, $c_x\equiv\cos x$, $t_x\equiv\tan x$ are introduced.
After the elimination of $m_{11}$, $m_{22}$ using the above equations, the entries of the mass matrices contain also the tadpole parameters $t_h$, $t_H$. 
Using 
\begin{align}
v^2=v_1^2+v_2^2, \qquad 
\tan{\beta} = \frac{v_2}{v_1},
\label{eq:tanbeta}
\end{align}
we obtain for the mass parameters of the CP-even Higgs bosons,
\begin{subequations}
\label{eq:neutralmassdef}
\begin{align}
\MH^2={}&\frac{2s_{\alpha-\beta}^2}{s_{2\beta}}  m^2_{12}
+ \frac{v^2}{2} \left( 2\lambda_1 c_\beta^2 c_\alpha^2 +2\lambda_2 s_\alpha^2 s_\beta^2 +s_{2 \alpha} s_{2\beta} \lambda_{345}\right)
-2t_\PH \frac{s_\alpha^3 c_\beta+c_\alpha^3 s_\beta }{ v s_{2 \beta }}
-t_\Ph\frac{s_{2 \alpha } s_{\alpha -\beta }}{v  s_{2 \beta }},
\\
\Mh^2={}&\frac{2c_{\alpha - \beta}^2}{s_{2 \beta}} m^2_{12}
+ \frac{v^2}{2} \left( 2\lambda_1 c_\beta^2 s_\alpha^2 +2\lambda_2 c_\alpha^2 s_\beta^2-s_{2 \alpha} s_{2\beta} \lambda_{345}\right)
-t_\PH\frac{s_{2 \alpha } c_{\alpha -\beta }}{v s_{2 \beta }}
-2t_\Ph\frac{c_\alpha^3 c_\beta-s_\alpha^3 s_\beta }{ v s_{2\beta }},
\\
M_{\PH\Ph}^2={}& \frac{s_{2(\alpha-\beta)}}{s_{2 \beta}}m^2_{12}
+ \frac{v^2}{2} \left[ s_{2\alpha} (-c_\beta^2 \lambda_1+ s_\beta^2 \lambda_2)+ s_{2 \beta} c_{2\alpha} \lambda_{345}\right]
-t_\PH\frac{s_{2 \alpha } s_{\alpha -\beta }}{v s_{2 \beta }}
-t_\Ph\frac{s_{2 \alpha } c_{\alpha -\beta }}{v s_{2 \beta }},
\label{eq:Hhmassmixing}
\end{align}
\end{subequations}
the ones of the mass matrix of the CP-odd Higgs fields result in
\begin{subequations}
\label{eq:cpmassdef}
\begin{align}
\MAO^2&=2c_{\beta-\beta_n}^2\Big(\frac{m^2_{12}}{s_{2\beta}}- \frac{\lambda_5 v^2}{2}\Big) 
- 2t_\PH \frac{c_{\beta_n}^2 c_\beta s_\alpha + s_{\beta_n}^2 s_\beta c_\alpha}{v s_{2\beta}}
- 2t_\Ph \frac{c_{\beta_n}^2 c_\beta c_\alpha - s_{\beta_n}^2 s_\beta s_\alpha}{v s_{2\beta}}, 
\label{eq:A0massdef}\\
M_{\PGO}^2&=2s_{\beta-\beta_n}^2\Big(\frac{m^2_{12}}{s_{2\beta}}- \frac{\lambda_5 v^2}{2}\Big)
- 2t_\PH \frac{s_{\beta_n}^2 c_\beta s_\alpha + c_{\beta_n}^2 s_\beta c_\alpha}{v s_{2\beta}}
- 2t_\Ph \frac{s_{\beta_n}^2 c_\beta c_\alpha - c_{\beta_n}^2 s_\beta s_\alpha}{v s_{2\beta}},\\
M_{\PGO\PAO}^2&=-s_{2(\beta-\beta_n)}\Big(\frac{m^2_{12}}{s_{2\beta}}- \frac{\lambda_5 v^2}{2}\Big)- t_\PH \frac{s_{2\beta_n} s_{\alpha-\beta}}{v s_{2\beta}}- t_\Ph \frac{s_{2\beta_n} c_{\alpha-\beta}}{v s_{2\beta}}, \label{eq:goldstonemassmixing}
\end{align}
\end{subequations}
and the ones of the mass matrix of the charged Higgs-boson fields are
\begin{subequations}
\label{eq:chargedmassdef}
\begin{align}
\MHP^2&=c_{\beta-\beta_c}^2\Big[\frac{2m^2_{12}}{s_{2\beta}}-\frac{v^2}{2}(\lambda_4+\lambda_5) \Big]
- 2t_\PH \frac{c_{\beta_c}^2 c_\beta s_\alpha + s_{\beta_c}^2 s_\beta c_\alpha}{v s_{2\beta}}
- 2t_\Ph \frac{c_{\beta_c}^2 c_\beta c_\alpha - s_{\beta_c}^2 s_\beta s_\alpha}{v s_{2\beta}},
\label{eq:Hpmmassdef}\\
M_\PGP^2&=s_{\beta-\beta_c}^2\Big[\frac{2m^2_{12}}{s_{2\beta}}-\frac{v^2}{2}(\lambda_4+\lambda_5) \Big]
- 2t_\PH \frac{s_{\beta_c}^2 c_\beta s_\alpha + c_{\beta_c}^2 s_\beta c_\alpha}{v s_{2\beta}}
- 2t_\Ph \frac{s_{\beta_c}^2 c_\beta c_\alpha - c_{\beta_c}^2 s_\beta s_\alpha}{v s_{2\beta}},
\\
M_{\PG\PHP}^2&=-\frac{s_{2(\beta-\beta_c)}}{2} \Big[\frac{2m^2_{12}}{s_{2\beta}}-\frac{v^2}{2}(\lambda_4+\lambda_5) \Big]
- t_\PH \frac{s_{2\beta_c} s_{\alpha-\beta}}{v s_{2 \beta}}
- t_\Ph \frac{s_{2\beta_c} c_{\alpha-\beta}}{v s_{2 \beta}}.
\label{eq:chargedgoldstonemassmixing}
\end{align}
\end{subequations}
At tree level we demand vanishing tadpole terms corresponding to $t_\PH=t_\Ph=0$ and diagonal propagators, so that the mixing terms proportional to $M_{\mr{Hh}}^2$, $M_{\PGO\PAO}^2$, $M_{\PG\PHP}^2$ vanish at LO. This yields $\beta_c=\beta_n=\beta$ and fixes the angle $\alpha$ as well. 
At NLO such a diagonalization is not possible, as the propagators receive also mixing contributions from the field renormalization and from 
(momentum-dependent) one-loop diagrams, so that there is no distinct condition to define the mixing angles. Therefore we keep the bare mass mixing parameters $M_{\mr{Hh}}^2$, $M_{\PGO\PAO}^2$, $M_{\PG\PHP}^2$ and the tadpole terms $t_\PH$, $t_\Ph$ in this section, and 
specify defining conditions for the bare parameters $\alpha$, $\beta_n$, $\beta_c$ and the tadpole terms later.
With the above equations, $m_{12}$, $\lambda_1$, $\lambda_2$, $\lambda_3$, $\lambda_4$ can be traded for the masses of the physical bosons $M_\PH$, $M_\Ph$, $M_\PAO$, $\MHP$, and the mixing angle $\alpha$. The parameter $\lambda_5$ cannot be replaced by a mass or a mixing angle as it appears only in cubic and quartic Higgs couplings and acts like an additional coupling constant. Explicit relations can be obtained by inverting Eqs.~\eqref{eq:neutralmassdef}, 
\eqref{eq:A0massdef}, and~\eqref{eq:Hpmmassdef}.
The other parameters are related to the masses by
\begin{subequations}
\label{eq:repara}
\begin{align}
\lambda_1={}&\frac{1}{\cb^2 v^2} \left[\ca^2 \MH^2+\sa^2 \Mh^2- M_{\PH\Ph}^2 s_{2\alpha}-\sb^2 \biggl(\frac{\MAO^2}{ c_{\beta-\beta_n}^2}+\lambda_5 v^2\biggr) \right]
\nonumber\\
&+t_\PH\frac{c_{{\beta_{n}}} \left(2 s_{\beta } s_{{\beta_{n}}} c_{\alpha }+c_{\alpha +\beta }c_{\beta_n}\right)}{v^3 c_{\beta }^2 c_{\beta -{\beta_{n}}}^2}
 -t_\Ph \frac{c_{\beta_{n}} \left(2 s_{\beta } s_{\beta_{n}} s_{\alpha }+s_{\alpha +\beta }c_{\beta_n}\right)}{v^3 c_{\beta }^2 c_{\beta -\beta_{n}}^2},
\\
\lambda_2={}&\frac{1}{\sb^2 v^2} \left[\ca^2 \Mh^2+\sa^2 \MH^2+ M_{\PH\Ph}^2 s_{2\alpha}-\cb^2 \biggl(\frac{\MAO^2}{c_{\beta-\beta_n}^2}+\lambda_5 v^2\biggr) \right]\nonumber\\
&+ t_\PH \frac{s_{\beta_n} (2 c_\beta c_{\beta_n} s_\alpha- c_{\alpha+\beta} s_{\beta_n})}{v^3 s_\beta^2 c_{\beta-\beta_n}^2}+ t_\Ph \frac{s_{\beta_n} (2 c_\beta c_{\beta_n} c_\alpha+ s_{\alpha+\beta} s_{\beta_n})}{v^3 s_\beta^2 c_{\beta-\beta_n}^2},
\\
\lambda_3={}&\frac{1}{v^2 s_{2 \beta}} \left[s_{2\alpha}(\MH^2-\Mh^2) + 2 c_{2\alpha}M_{\PH\Ph}^2\right]-\frac{\MAO^2}{v^2c_{\beta-\beta_n}^2}+\frac{2 \MHP^2}{v^2c_{\beta-\beta_c}^2}-\lambda_5\nonumber 
\\
&+\frac{2t_\PH}{v^3 s_{2\beta }} 
\left[s_{\beta } c_{\alpha } \bigg(\frac{2 s^2_{\beta_c}}{c_{\beta -\beta_c}^2}-\frac{s^2_{\beta_n}}{c_{\beta -\beta_n}^2}\bigg)+c_{\beta } s_{\alpha }
   \bigg(\frac{2 c^2_{\beta_c}}{c_{\beta -\beta_c}^2}-\frac{c^2_{\beta_n}}{c_{\beta -\beta_n}^2}\bigg)\right]
\nonumber\\
&+  \frac{2t_\Ph}{v^3 s_{2\beta }} 
\left[c_{\beta } c_{\alpha }\bigg(\frac{2 c^2_{\beta_c}}{c_{\beta -\beta_c}^2}-\frac{c^2_{\beta_n}}{c_{\beta -\beta_n}^2}\bigg)+s_{\beta } s_{\alpha }
   \bigg(\frac{s^2_{\beta_n}}{c_{\beta -\beta_n}^2}-\frac{2 s^2_{\beta_c}}{c_{\beta -\beta_c}^2}\bigg)\right]
\label{eq:Lambda3NLO}\\
\lambda_4={}&\lambda_5+\frac{2 \MAO^2}{v^2 c_{\beta -\beta _n}^2}-\frac{2 \MHP^2}{v^2 c_{\beta -\beta _c}^2}+\frac{2 t_\mr{H} s_{\beta _c-\beta _n} \left(s_{\alpha +\beta -\beta _c-\beta
   _n}-s_{\beta -\alpha } c_{\beta _c-\beta _n}\right)}{v^3 c_{\beta -\beta _c}^2 c_{\beta -\beta _n}^2}\nonumber\\
&+\frac{2 t_\mr{h} s_{\beta _c-\beta _n} \left(c_{\alpha +\beta -\beta _c-\beta
   _n}+c_{\beta -\alpha } c_{\beta _c-\beta _n}\right)}{v^3 c_{\beta -\beta _c}^2 c_{\beta -\beta _n}^2},\\
m^2_{12}={}&\frac{1}{2}\lambda_5 v^2 s_{2\beta }
+\frac{\MAO^2 s_{2\beta }}{2c_{\beta -\beta _n}^2}
+\frac{t_\mr{H} \left(s_{\beta } c_{\alpha } s_{\beta _n}^2+c_{\beta } s_{\alpha } c_{\beta_n}^2\right)}{v c_{\beta -\beta _n}^2}
+\frac{t_\mr{h} \left(c_{\beta } c_{\alpha } c_{\beta _n}^2-s_{\beta }
   s_{\alpha } s_{\beta _n}^2\right)}{v c_{\beta -\beta _n}^2}.\label{eq:m12NLO}
\end{align}
\end{subequations}
The tree-level relations are
easily obtained by setting $\beta_c=\beta_n=\beta$ and $t_\PH=t_\Ph=M_{\mr{Hh}}^2=M_{\PGO\PAO}^2=M_{\PG\PHP}^2=0$,
\begin{subequations}
\begin{align}
\lambda_1&={}\frac{1}{\cb^2 v^2} \left[\ca^2 \MH^2+\sa^2
\Mh^2-\sb^2 (\MAO^2+\lambda_5 v^2) \right],\\
\lambda_2&={}\frac{1}{\sb^2 v^2} \left[\sa^2 \MH^2+\ca^2
\Mh^2-\cb^2 (\MAO^2+\lambda_5 v^2) \right],\\
\lambda_3&={}\frac{s_{2\alpha}}{s_{2\beta} v^2}
(\MH^2-\Mh^2)-\frac{1}{v^2} (\MAO^2-2\MHP^2)-\lambda_5,
\label{eq:lambda3tree}\\
\lambda_4&={}\frac{2(\MAO^2-\MHP^2)}{v^2}+\lambda_5,\\
m^2_{12}&={}\frac{s_{2\beta}}{2} (\MAO^2+\lambda_5 v^2).
\end{align}
\end{subequations}

\paragraph{Parameters of the gauge sector:}
Mass terms of the gauge bosons arise through the interaction of the gauge bosons with the vevs, 
analogous to the SM. After a rotation into fields corresponding to mass eigenstates, one obtains relations similar to the SM ones:
\begin{align}
\MW=g_2 \frac{v}{2},&& \MZ=\frac{v}{2} \sqrt{g_1^2+g_2^2},&& e= \frac{g_1g_2}{\sqrt{g_1^2+g_2^2}},
\label{eq:EWmassesTHDM}
\end{align}
where the electric unit charge $e$ is identified with the coupling constant of the photon field $A_\mu$ in the covariant derivative.
Inverting these relations and introducing the weak mixing angle $\theta_W$ via $\cos \theta_W=g_2/\sqrt{g_1^2+g_2^2}$, one can replace $v$ and the gauge couplings $g_1$ and $g_2$ by
\begin{align}
v&= \frac{2 \MW \sw}{e},& g_1&=\frac{e}{\cw},& g_2&=\frac{e}{\sw}.
\label{eq:EWtrafo}
\end{align}

\paragraph{Mass parameterization:}
The relations~\eqref{eq:tadpoledefs}, 
\refeq{eq:tanbeta},
(\ref{eq:repara}a,b,d,e), and \eqref{eq:EWtrafo} between the masses, angles, and the basic parameters can be used to reparameterize the Higgs Lagrangian and change the parameters 
\begin{align}
\{p_{\mr{basic}}\}=\{\lambda_1,\ldots,\lambda_5,m^2_{11},m^2_{22},m^2_{12},v_1,v_2,g_1,g_2\},\label{eq:basicpara}
\end{align}
in favour of the bare mass parameters including one parameter from scalar self-interactions
which we take as $\lambda_3$, 
\begin{align}
\{p'_{\mr{mass}}\}=\{\MH,\Mh,\MAO,\MHP,M_\mathrm{W},\MZ,e,\lambda_5,\lambda_3,\beta, t_\mr{H} ,t_\mr{h}\}.
\label{eq:physparaNLOprime}
\end{align}
Additionally, one has to keep in mind that we keep the mixing parameters $\alpha$, $\beta_n$, and $\beta_c$ generic, and they have to be fixed by additional conditions (which  will be given later).
One can use Eq.~(\ref{eq:repara}c) to trade $\lambda_3$ for $\alpha$, in which case the mixing angle becomes a free parameter of the theory. Then, one obtains the parameter set
\begin{align}
\{p_{\mr{mass}}\}=\{\MH,\Mh,\MAO,\MHP,M_\mathrm{W},\MZ,e,\lambda_5,\alpha,\beta, t_\mr{H} ,t_\mr{h} \}.
\label{eq:physparaNLO}
\end{align}

\subsection{Yukawa couplings}
\label{sec:YukawaL}

The Higgs mechanism does not only give rise to the gauge-boson mass terms (which are determined by the vevs), 
but via Yukawa couplings, it introduces 
masses to chiral fermions. Since both Higgs doublets can couple to fermions, the general Yukawa couplings have the form
\begin{align}
\mathcal{L}_\mathrm{Yukawa}=&-\sum_{k=1,2}\sum_{i,j} \left(\bar{L}'^\rL_i \zeta^{l,k}_{ij} l'^\rR_j \Phi_k+\bar{Q}'^\rL_i \zeta^{u,k}_{ij} u'^\rR_j \tilde{\Phi}_k+\bar{Q}'^\rL_i \zeta^{d,k}_{ij} d'^\rR_j \Phi_k+h.c.\right),
\label{LYuk}
\end{align}
with the mixing matrices $\zeta^{f,k}$, $k=1,2$, in generation space for the 
gauge-invariant interactions with $\Phi_1$ and $\Phi_2$, respectively, and the generation indices $i, j = 1,2,3$.
The left-handed SU(2) doublets of quarks and leptons are denoted 
${Q}'^\rL=\left(u'^\rL,d'^\rL\right)^\rT$ and 
${L}'^\rL=\left(\nu'^\rL,l'^\rL\right)^\rT$, 
while the right-handed up-type quark, down-type quark, and lepton singlets are  $u'^\rR$, $d'^\rR$, and $l'^\rR$, respectively. 
The primes indicate that we deal with fields 
in the interaction basis here;
fields without primes correspond to mass eigenstates.
The field $\tilde{\Phi}_k$, $k = 1, 2$, is the charge-conjugated field of $\Phi_k$.
Since, in the general THDM, there are two mass mixing matrices for each type $f$ of fermions, flavour-changing neutral currents (FCNC) can occur at tree level, which, however, are experimentally known to be strongly suppressed.
According to the Paschos--Glashow--Weinberg theorem~\cite{Glashow1977,Paschos1977a}, 
FCNC are absent at tree level
if each type of fermion couples only to one of the Higgs doublets. 
 This can be achieved by imposing an additional discrete $\mathbb{Z}_2$ symmetry. It should be noted that the soft-$\mathbb{Z}_2$-breaking term  proportional to $m_{12}$ in the Higgs potential does not introduce FCNC. 
The Yukawa Lagrangian reduces then to
\begin{align}
\mathcal{L}_\mathrm{Yukawa}=-\sum_{i,j} \left(\bar{L}'^\rL_i \zeta^{l}_{ij} l'^\rR_j \Phi_{n_1}+\bar{Q}'^\rL_i \zeta^{u}_{ij} u'^\rR_j \tilde{\Phi}_{n_2}+\bar{Q}'^\rL_i \zeta^{d}_{ij} d'^\rR_j \Phi_{n_3}+h.c.\right),
\end{align}
with $n_i$ being either $1$ or $2$. 
Depending on the exact form of the symmetry,
one distinguishes four types of THDMs. In Type~I models, all fermions couple to one Higgs doublet (conventionally $\Phi_2$, but this is equivalent to $\Phi_1$ due to possible basis changes) which can be ensured by demanding a $\Phi_1 \rightarrow -\Phi_1$ symmetry. In Type~II models, down-type fermions couple to the other doublet, which can be enforced by the symmetry $\Phi_1 \rightarrow -\Phi_1$, $d'^\rR_j \to- d'^\rR_j$, $l'^\rR \to -l'^\rR$. The other two possibilities are called ``lepton-specific'' (Type~X) and ``flipped'' (Type~Y) models. An overview over the couplings and symmetries of the different models is given in Tab.~\ref{tab:yukcoupl}.
\begin{table}
  \centering
\begin{tabular}{|c|c|c|c|c|}\hline
&\hspace{20pt} $u_i$\hspace{20pt} &\hspace{20pt} $d_i$\hspace{20pt}  &\hspace{20pt}  $e_i$\hspace{20pt} & $\mathbb{Z}_2$ symmetry \\\hline
Type I & $\Phi_2$ & $\Phi_2$ & $\Phi_2$ & $\Phi_1 \rightarrow - \Phi_1$ \\
Type II& $\Phi_2$ & $\Phi_1$ & $\Phi_1$ &$(\Phi_1, d_i,e_i) \rightarrow -(\Phi_1, d_i,e_i)$\\
Lepton-specific & $\Phi_2$ & $\Phi_2$ & $\Phi_1$ &$(\Phi_1, e_i) \rightarrow -(\Phi_1,e_i)$ \\
Flipped & $\Phi_2$ & $\Phi_1$ & $\Phi_2$ & $(\Phi_1, d_i) \rightarrow -(\Phi_1, d_i)$\\\hline
\end{tabular}  
  \caption{Different types of the THDM having in common that only one of the Higgs doublet couples to each type of fermions. This can be achieved by imposing appropriate $\mathbb{Z}_2$ symmetry charges to the fields. }
\label{tab:yukcoupl}
\end{table}
For each of the fermion types, a redefinition of the fields can be performed in order to get diagonal mass matrices, analogously to the SM case. Similar to the SM, 
the coupling of fermions to the Z boson is flavour conserving, and a CKM matrix appears in the coupling to the charged gauge bosons. 
By specifying the model type, the Higgs--fermion interaction is determined, and one can write them, 
widely following the notation of 
Ref.~\cite{Aoki2009}, as
\begin{align}
 \mathcal{L}_\mathrm{Yukawa,int}=
&- \sum_i \sum_{f=u,d,l} \frac{m_{f,i}}{v}\left( \xi^f_\Ph\, \bar{f_i}f_i h+ \xi^f_\PH \, \bar{f}_if_iH
-2\im I^3_{\PW,f} \xi^f_\PAO \, \bar{f}_i\gamma_5 f_i A_0
- 2\im I^3_{\PW,f} \bar{f}_i\gamma_5 f_i G_0\right)\nonumber\\
&-\sum_{i,j}\biggl[\frac{\sqrt{2}V_{ij}}{v} \bar{u}_i (-m_{u,i} \xi^u_\PAO \omega_- +m_{d,j} \xi^d_\PAO \omega_+)\, d_j H^+ + h.c. \biggr]
\nonumber\\
&- \sum_i \biggl[ \frac{\sqrt{2}m_{l,i} \xi^l_\PAO}{v}\, \bar{\nu}_i^\rL l_i^\rR H^+ + h.c. \biggr]
\nonumber\\
&-\sum_{i,j}\biggl[ \frac{\sqrt{2}V_{ij}}{v} \bar{u}_i (-m_{u,i} \omega_- +m_{d,j} \omega_+)\, d_j G^+ + h.c. \biggr]
- \sum_i \biggl[ \frac{\sqrt{2}m_{l,i} }{v}\, \bar{\nu}_i^\rL l_i^\rR G^+ + h.c. \biggr],
\label{eq:ffH}
\end{align}
where $m_{l,i}$, $m_{u,i}$, and $m_{d,i}$ are the lepton, the up-type, and the down-type quark masses, respectively,
and $V_{ij}$ are the coefficients of the CKM matrix.
Left- and right-handed fermion fields, $f^\rL$ and $f^\rR$, are obtained from the 
corresponding Dirac spinor $f$ by applying the chirality projectors
$\omega_\pm= ( 1\pm \gamma_5)/2$, i.e.\ $f=(\omega_+ +\omega_-)f=f^\rL+f^\rR$.
The coupling coefficients $\xi^f_{\PH,\Ph,\PAO}$ are defined as the couplings relative to the canonical SM value of $m_f/v$ and are shown in Tab.~\ref{tab:yukint}. 
\begin{table}
  \centering
\begin{tabular}{|c|c|c|c|c|}\hline
&Type I & Type II &  Lepton-specific & Flipped\\\hline
$\xi^l_\PH$ & $\sin{\alpha}/\sin{\beta}$ & $\cos{\alpha}/\cos{\beta}$ & $\cos{\alpha}/\cos{\beta}$  &$\sin{\alpha}/\sin{\beta}$\\
$\xi^u_\PH$ & $\sin{\alpha}/\sin{\beta}$ & $\sin{\alpha}/\sin{\beta}$ & $\sin{\alpha}/\sin{\beta}$ &$\sin{\alpha}/\sin{\beta}$ \\
$\xi^d_\PH$& $\sin{\alpha}/\sin{\beta}$ & $\cos{\alpha}/\cos{\beta}$ & $\sin{\alpha}/\sin{\beta}$ & $\cos{\alpha}/\cos{\beta}$ \\\hline
$\xi^l_\Ph$ & $\cos{\alpha}/\sin{\beta}$ & $-\sin{\alpha}/\cos{\beta}$ & $-\sin{\alpha}/\cos{\beta}$& $\cos{\alpha}/\sin{\beta}$  \\
$\xi^u_\Ph$ & $\cos{\alpha}/\sin{\beta}$ & $\cos{\alpha}/\sin{\beta}$ & $\cos{\alpha}/\sin{\beta}$ &$\cos{\alpha}/\sin{\beta}$ \\
$\xi^d_\Ph$& $\cos{\alpha}/\sin{\beta}$ & $-\sin{\alpha}/\cos{\beta}$ & $\cos{\alpha}/\sin{\beta}$ &$-\sin{\alpha}/\cos{\beta}$  \\\hline
$\xi^l_\PAO$ & $\cot{\beta}$ & $-\tan{\beta}$ &$-\tan{\beta}$ & $\cot{\beta}$  \\
$\xi^u_\PAO$ & $\cot{\beta}$ & $\cot{\beta}$ & $\cot{\beta}$&$\cot{\beta}$ \\
$\xi^d_\PAO$& $\cot{\beta}$ & $-\tan{\beta}$ & $\cot{\beta}$ & $-\tan{\beta}$ \\\hline
\end{tabular}  
\caption{The coupling strengths $\xi^f$ of $\PH,\Ph,\PAO$ to the fermions relative to the SM value of $m_f/v$, see Eq.~\eqref{eq:ffH}. Note that the sign of $\xi^f_\PAO$ is defined relative to
the coupling of the Goldstone-boson field $G_0$ and the relation $\beta_n=\beta_c=\beta$
is used here.}
\label{tab:yukint}
\end{table}
Note that we have used $\beta_n=\beta_c=\beta$ in Eq.~\refeq{eq:ffH} and \refta{tab:yukint},
which is most relevant in applications; the generalization to independent
$\beta_n$, $\beta_c$, $\beta$ is simple.

\section{The counterterm Lagrangian}
\label{sec:CTLagrangian}

The next step in calculating higher-order corrections is the renormalization of the theory. In this section we focus on electroweak corrections of $\order{\alpha_{\mr{em}}}$. The QCD renormalization of the THDM is straightforward and completely analogous to the SM case, since all scalar degrees of freedom are colour singlets and do not interact strongly. 
In the formulation of the basic Lagrangian in the previous section, we dealt with {\it bare} 
parameters and fields. To distinguish those from {\it renormalized} quantities, in the
following we indicate {\it bare} quantities by subscripts ``0'' consistently.
We perform a multiplicative renormalization, i.e.\ we split {\it bare} quantities into
{\it renormalized} parts and corresponding counterterms,
use dimensional regularization,
and allow for matrix-valued field renormalization constants in the case that there are several fields with the same quantum numbers.
The counterterm Lagrangian $\delta \mathcal{L}$ containing the full dependence on the renormalization constants can be split into several parts analogous to Eq.~\eqref{eq:LTHDM},
\begin{align}
\label{eq:CTLagrangian}
\delta \mathcal{L}_\mathrm{THDM}
=\delta \mathcal{L}_{\mr{Gauge}}+\delta \mathcal{L}_{\mr{Fermion}}+\delta \mathcal{L}_{\mr{Higgs,kin}}-\delta V_{\mr{Higgs}}+\delta \mathcal{L}_{\mr{Yukawa}},
\end{align}
where the Higgs part of the Lagrangian $\delta \mathcal{L}_{\mr{Higgs}}$ is split up into the kinetic part $\delta \mathcal{L}_{\mr{Higgs,kin}}$ and the Higgs potential part $\delta V_{\mr{Higgs}}$.
Since the gauge fixing is applied after renormalization, no gauge-fixing counterterms occur, and since ghost fields
occur only in loop diagrams, a renormalization of the ghost sector is not necessary at NLO for the calculation of S-matrix elements. Though, for analyzing Slavnov--Taylor or Ward identities a complete renormalization procedure would be advisable. 
{Our renormalization procedure, thus, widely parallels the treatment described for the SM in \cite{Denner:1991kt};
an alternative variant that is based on the transformation of fields in the gauge eigenstate basis, as suggested
for the SM in \citere{Bohm:1986rj} and for the MSSM in \citere{Dabelstein:1994hb}, is described in
\citere{Altenkamp:Dis}.}

\subsection{Higgs potential}
\label{sec:CTpot}
According to Eq.~\eqref{eq:lambdapara}, the Higgs potential contains 
8 independent parameters which have to be renormalized. In addition, there are two vevs and two gauge couplings completing the set of input parameters of \eqref{eq:basicpara}. 
We have carried out different renormalization procedures and in this paper discuss the renormalization of the Lagrangian in the mass parameterization
\begin{enumerate}
\renewcommand{\theenumi}{(\alph{enumi}}
\renewcommand{\labelenumi}{\theenumi)}
\item with renormalization of the mixing angles,
\item alternatively,
taking the mixing angles as dependent parameters and applying the field transformation after renormalization.
\end{enumerate}
Additionally we have performed a renormalization of the basic parameters and a 
subsequent transformation to renormalization constants and parameters of the mass parameter set similar to 
Dabelstein in the MSSM~\cite{Dabelstein:1994hb}%
\footnote{Details about this method can be found in \citere{Altenkamp:Dis}.}. The latter has been used, after changing to our 
conventions for field and parameter renormalization, to check the counterterm Lagrangian. 
In the following, we give a detailed description of method (a), while method~(b) is 
briefly described in App.~\ref{App:renormassa}.

\subsubsection{Renormalization of the mixing angles}
\label{sec:renmixangles}

{In this section we show that the counterterms of mixing angles 
that are not used to replace another free parameter of the theory 
are redundant in the sense that they can be absorbed by 
field renormalization constants.}
We sketch the argument for generic scalar fields $\varphi_{1},\varphi_{2}$, which are transformed into fields $h_1, h_2$ 
corresponding to mass eigenstates by a rotation by the angle~$\theta$,
\begin{align}
\begin{pmatrix}\varphi_{1,0} \\ \varphi_{2,0}\end{pmatrix}
& = \mathbf{R}_\varphi(\theta_{0}) \begin{pmatrix}h_{1,0} \\ h_{2,0}\end{pmatrix}
=\begin{pmatrix}c_{\theta,0} & -s_{\theta,0}\\ s_{\theta,0}& c_{\theta,0}\end{pmatrix}
\begin{pmatrix}h_{1,0} \\ h_{2,0}\end{pmatrix},
\label{eq:mixanglesrot}
\end{align}
where we added subscripts ``0'' to indicate bare quantities.
The general argument can be applied to the neutral CP-even, the neutral 
CP-odd, and the charged Higgs fields of the THDM by replacing
$h_1$, $h_2$ by $H$, $h$, or $G$, $A_0$ or $G^\pm$, $H^\pm$, respectively, and by substituting the angle $\theta$ by 
$\alpha$, $\beta_n$, or $\beta_c$.
The fields corresponding to mass eigenstates are renormalized using matrix-valued renormalization constants, so that the renormalization transformation reads
\begin{align}
\begin{pmatrix}h_{1,0} \\ h_{2,0} \end{pmatrix}&=\begin{pmatrix}1+\frac{1}{2}\delta Z_{\Ph_1 \Ph_1} &\frac{1}{2} \delta Z_{\Ph_1 \Ph_2} \\\frac{1}{2} \delta Z_{\Ph_2 \Ph_1} &1+\frac{1}{2}\delta Z_{\Ph_2 \Ph_2} \end{pmatrix} \begin{pmatrix}h_1 \\ h_2\end{pmatrix},&&
  \theta_{0} = \theta + \delta \theta.\label{eq:alpherendef}
\end{align}
Applying this renormalization transformation to Eq.~\eqref{eq:mixanglesrot} leads to
\begin{align}
\begin{pmatrix} \varphi_{1,0} \\ \varphi_{2,0}\end{pmatrix}={}
& \bigg[ \begin{pmatrix}c_{\theta}  & -s_{\theta} \\ s_{\theta}& c_{\theta} \end{pmatrix}
\begin{pmatrix}1+\frac{1}{2}\delta Z_{\Ph_1\Ph_1}  & \frac{1}{2}\delta Z_{\Ph_1\Ph_2} \\ \frac{1}{2}\delta Z_{\Ph_2\Ph_1}& 1+\frac{1}{2}\delta Z_{\Ph_2\Ph_2} \end{pmatrix}
+\begin{pmatrix}-s_{\theta}  & -c_{\theta} \\ c_{\theta}& -s_{\theta} \end{pmatrix} \delta \theta \bigg]
\begin{pmatrix}h_1 \\ h_2\end{pmatrix}
\nonumber\\
{}={}& \begin{pmatrix}c_{\theta}  & -s_{\theta} \\ s_{\theta}& c_{\theta} \end{pmatrix}
\begin{pmatrix}1+\frac{1}{2}\delta Z_{\Ph_1\Ph_1}  &\frac{1}{2}(\delta Z_{\Ph_1\Ph_2}- 2 \delta \theta) 
\\ 
	\frac{1}{2}(\delta Z_{\Ph_2\Ph_1}+ 2 \delta \theta)& 1+\frac{1}{2}\delta Z_{\Ph_2\Ph_2} \end{pmatrix}
\begin{pmatrix}h_1 \\ h_2\end{pmatrix}.  
\label{eq:mixanglesrot2}
\end{align}
One can easily remove the dependence on the mixing angle 
with a redefinition of the mixing 
field renormalization constants by introducing
\begin{align}
\label{eq:redeffielren}
 \delta \tilde{Z}_{\Ph_2\Ph_1}= \delta Z_{\Ph_2\Ph_1}+ 2 \delta \theta,\qquad
 \delta \tilde{Z}_{\Ph_1\Ph_2}= \delta Z_{\Ph_1\Ph_2}- 2 \delta \theta.
 \end{align}
Then, the Eq.~\eqref{eq:mixanglesrot2} reads
\begin{align}
\begin{pmatrix} \varphi_{1,0} \\ \varphi_{2,0}\end{pmatrix}=
& \begin{pmatrix}c_{\theta}  & -s_{\theta} \\ s_{\theta}& c_{\theta} \end{pmatrix}
\begin{pmatrix}1+\frac{1}{2}\delta Z_{\Ph_1\Ph_1}  &\frac{1}{2}\delta \tilde{Z}_{\Ph_1\Ph_2} \\ 
	\frac{1}{2}\delta \tilde{Z}_{\Ph_2\Ph_1}& 1+\frac{1}{2}\delta Z_{\Ph_2\Ph_2} \end{pmatrix}
\begin{pmatrix}h_1 \\ h_2\end{pmatrix}.
\end{align}
Obviously,
the dependence on $\delta \theta$ can always be removed from the Lagrangian 
by a redefinition of the field renormalization constants. 
As a simple shift of the mixing field renormalization constant is performing the task, the renormalization of the mixing angle $\theta$ can be seen as an additional field renormalization 
(as it is done, e.g., in Ref.~\cite{Aoki2009}). 
This argument is general and holds for any renormalization condition on $\theta$.
Without loss of generality one can even assume that such a redefinition has already been performed and set $\delta \theta=0$ from the beginning, as done in method (b) in App.~\ref{App:renormassa}.
Of course, the bookkeeping of counterterms depends on the way $\delta \theta$ is treated.
This can be seen by considering the mass term of the potential. 
The general bare mass term can be written using the rotation matrix $\mathbf{R_\varphi}(\theta_{0})$ as
\begin{align}
V_{\Ph_1\Ph_2} {}={} 
&\frac{1}{2} \,(h_{1,0}, h_{2,0})\, \mathbf{R}_\varphi^\rT(\theta_{0}) \mathbf{M}_{\varphi,0} 
\mathbf{R}_\varphi(\theta_{0}) \begin{pmatrix} h_{1,0}\\ h_{2,0} \end{pmatrix}
\nonumber\\
{}={} &\frac{1}{2} \, (h_{1,0}, h_{2,0})\, 
\mathbf{R}_\varphi^\rT(\theta+\delta \theta) 
\left(\mathbf{M_\varphi}+\delta \mathbf{M_\varphi}\right) 
\mathbf{R_\varphi}(\theta+\delta \theta) 
\begin{pmatrix} h_{1,0}\\ h_{2,0} \end{pmatrix}.
\end{align}
This expression can be expanded in terms of renormalized and counterterm contributions, yielding
\begin{align}
V_{\Ph_1\Ph_2}=
\frac{1}{2} \, (h_{1,0}, h_{2,0})\, \begin{pmatrix} 
\hspace{-0pt}M_{\Ph_1}^2+\delta M_{\Ph_1}^2 & 
\hspace{-5pt}\delta \theta (M_{\Ph_2}^2-M_{\Ph_1}^2)+f_{\theta}(\{ \delta p\}) \\ \delta \theta (M_{\Ph_2}^2-M_{\Ph_1}^2)+f_{\theta}(\{ \delta p\}) & 
\hspace{-0pt}M_{\Ph_2}^2+\delta M_{\Ph_2}^2\end{pmatrix} \begin{pmatrix} h_{1,0}\\ h_{2,0} \end{pmatrix},
\end{align}
where we obtain off-diagonal terms from the counterterm $\delta\theta$
of the mixing angle and from the renormalization of the mass matrix. The latter contribution depends on the independent counterterms $\{ \delta p\}$ and is abbreviated by $f_{\theta}(\{ \delta p\})$.
At NLO, the mixing entry of the mass matrix reads
\begin{align}
\delta M_{\Ph_1\Ph_2}^2= \delta \theta (M_{\Ph_2}^2-M_{\Ph_1})+f_{\theta}(\{ \delta p\}), \label{eq:NLOmassmixHh}
\end{align}
{and since the renormalization of the {\it redundant} mixing angle can be chosen freely,} 
counterterm contributions in the Lagrangian can be shifted arbitrarily from mixing terms to mixing-angle counterterms.
{Note that $f_{\theta}(\{ \delta p\})$ does not change
by such redistributions, since it is fixed by the remaining renormalization 
constants.}

\subsubsection{Renormalization with a diagonal mass matrix -- version (a)}
\label{sec:renormassb}

In this prescription, we use the angle $\alpha$ as an independent parameter instead of $\lambda_3$. 
{To define $\alpha$ at NLO, we demand that 
the mass matrix of the CP-even Higgs bosons (in the Lagrangian), written in terms of bare fields,
is diagonal at all orders (i.e.\ in terms of bare or renormalized parameters).}
Equation~\eqref{eq:Lambda3NLO} with
\begin{align}\label{eq:MHhchoice2}
M_{\mr{Hh},0}^2 = 0 + \delta M_{\mr{Hh}}^2=0, \qquad M_{\mr{Hh}}^2 = 0
\end{align}
then defines the parameter $\alpha$. 
{Note that (momentum-dependent)
loop diagrams tend to destroy the diagonality 
of the matrix-valued two-point functions (inverse propagators)
in the effective action as well.
Below,
the field renormalization will be chosen to compensate those loop effects  
at the mass shells of the propagating particles.}
{It is relation \refeq{eq:MHhchoice2} that distinguishes 
$\alpha$ from the case of a redundant mixing angle, such as $\theta$ in the
previous section, which can be chosen freely or absorbed by field
renormalization constants.}
The relation between $\alpha$ and $\lambda_3$ of Eq.~\eqref{eq:Lambda3NLO} is the same for bare and renormalized quantities 
and can be used to eliminate $\lambda_3$ from the theory.
For each of the independent parameters of the mass parameter set \eqref{eq:physparaNLO} we apply the renormalization transformation
\begin{align}
M_{\PH,0}^2&=\MH^2+\delta \MH^2,&M_{\Ph,0}^2&= \Mh^2+\delta \Mh^2,&M_{\PAO,0}^2 &= \MAO^2+\delta \MAO^2,\nonumber\\
M_{\PHP,0}^2&= \MHP^2+\delta \MHP^2,&M_{\PW,0}^2&= \MW^2+\delta \MW^2,&M_{\PZ,0}^2&= \MZ^2+\delta \MZ^2, \nonumber\\
e_0&= e+\delta e,&\lambda_{5,0} &= \lambda_5+\delta \lambda_5&\alpha_0 &= \alpha+\delta \alpha,\nonumber\\
\beta_0&= \beta+\delta \beta,&t_{\PH,0}&= 0+\delta t_\PH,&t_{\Ph,0}&= 0+\delta t_\Ph, \label{eq:physpararen2}
\end{align}
so that the 12 parameter renormalization constants are
\begin{align}
\{\delta p_\mr{mass}\}=\{\delta \MH^2, \delta \Mh^2, \delta \MAO^2, \delta \MHP^2, \delta \MW^2, \delta \MZ^2, \delta e, \delta \lambda_5, \delta \alpha,\delta \beta,\delta t_\PH, \delta t_\Ph\},
\label{eq:renconstsR}
\end{align}
corresponding to $\{p_\mr{mass}\}$. 
In this part we describe the commonly used tadpole renormalization (which is gauge dependent) and describe 
a gauge-independent scheme in Sect.~\ref{sec:FJscheme}.

The higher-order corrections of the mixing angles $\beta_n$ and $\beta_c$ are irrelevant according to Sect.~\ref{sec:renmixangles} and we can choose
\begin{align}
\beta_{n,0}=\beta_{c,0}= \beta_0 = \beta+\delta \beta,
\label{eq:mixanglect}
\end{align}
which defines the mixing terms uniquely and ensures that the angles $\beta_n,\beta_c$, and $\beta$ do not have to be distinguished at any order.
From these conditions, we can compute the mass mixing terms from Eq.~\eqref{eq:goldstonemassmixing} and Eq.~\eqref{eq:chargedgoldstonemassmixing} to
\begin{align}
 M_{\PGO\PAO,0}^2 &= 0+\delta M_{\PGO\PAO}^2
=-e\frac{ 
\delta t_\PH s_{\alpha -\beta }+
\delta t_\Ph c_{\alpha -\beta }
}{2 \MW \sw},&\nonumber\\
 M_{\PG\PHP,0}^2 &= 0+\delta M_{\PG\PHP}^2
=-e\frac{ 
\delta t_\PH s_{\alpha -\beta }+
\delta t_\Ph c_{\alpha -\beta }
}{2 \MW \sw}. \label{eq:Mabchoice2}
\end{align}
The field re\-nor\-ma\-li\-zation is performed 
for each field corresponding to mass eigenstates,
\begin{align}
\begin{pmatrix}H_0 \\ h_0 \end{pmatrix}&=\begin{pmatrix}1+\frac{1}{2}\delta Z_\PH &\frac{1}{2} \delta Z_{\PH\Ph} \\\frac{1}{2} \delta Z_{\Ph\PH} &1+\frac{1}{2}\delta Z_{\Ph} \end{pmatrix} \begin{pmatrix}H \\ h\end{pmatrix},\nonumber\\
\begin{pmatrix}G_{0,0} \\ A_{0,0} \end{pmatrix}&=\begin{pmatrix}1+\frac{1}{2}\delta Z_{\PGO} &\frac{1}{2}\delta  Z_{\PGO\PAO} \\\frac{1}{2}\delta Z_{\PAO\PGO} &1+\frac{1}{2}\delta Z_\PAO \end{pmatrix} \begin{pmatrix}G_0 \\ A_0\end{pmatrix},\nonumber\\
\begin{pmatrix}G^\pm_0 \\ H^\pm_0 \end{pmatrix}&=\begin{pmatrix}1+\frac{1}{2}\delta Z_{\mr{G}^+} &\frac{1}{2}\delta  Z_{\mr{GH}^+} \\\frac{1}{2}\delta Z_{\mr{HG}^+} &1+\frac{1}{2}\delta Z_{\PH^+} \end{pmatrix} \begin{pmatrix}G^\pm \\ H^\pm\end{pmatrix},\label{eq:physfieldren}
\end{align}
with the field renormalization constants $\delta Z_\PH$, $\delta Z_{\PH\Ph}$, $\delta Z_{\Ph\PH}$, $\delta Z_{\Ph}$,  $\delta Z_{\PGO}$, 
$\delta  Z_{\PGO\PAO}$, $\delta Z_{\PAO\PGO}$, $\delta Z_\PAO$,  $\delta Z_{\mr{G}^+}$, $\delta  Z_{\mr{GH}^+}$, $\delta Z_{\mr{HG}^+}$, and $\delta Z_{\PH^+}$ for the CP-even, the CP-odd, and the charged Higgs fields.
We denote the complete set of parameter and field renormalization constants with $\{\delta \mathbf{R}_\mr{mass}\}$. All renormalization constants 
are of $\order{\alpha_{\mr{em}}}$, i.e.\ all contributions of $\order{\alpha_{\mr{em}}^2}$ or higher are omitted. 

\begin{sloppypar}
Applying the renormalization transformations~\refeq{eq:physpararen2}, \refeq{eq:physfieldren} 
to the bare potential of Eq.~\eqref{eq:NLObarepot} and linearizing in the renormalization constants results in 
$V(\{p_\mr{mass}\})+\delta V(\{p_\mr{mass}\},\{\delta \mathbf{R}_\mr{mass}\})$
with the LO potential as in Eq.~\eqref{eq:NLObarepot}, but with renormalized quantities and the counterterm potential of $\order{\alpha_\mr{em}}$,
\begin{align}
\delta V(\{p_\mr{mass}\},\{\delta \mathbf{R}_\mr{mass}\})=&- \delta t_\PH H -\delta t_\Ph h \nonumber\\
&+\frac{1}{2} \left(\delta \MH^2 +\delta Z_{\PH} \MH^2\right) H^2 +\frac{1}{2} \left(\delta \Mh^2 +\delta Z_{\Ph} \Mh^2\right) h^2 \nonumber\\
&+\frac{1}{2} \left(\delta \MAO^2 +\delta Z_{\PAO} \MAO^2\right) A_0^2 + \left(\delta \MHP^2+\delta Z_{\PHP} \MHP^2\right)H^+ H^-\nonumber\\
&+ \frac{e}{4 \MW \sw} \left(
-\delta t_\PH c_{\alpha-\beta}
+\delta t_\Ph s_{\alpha-\beta}
\right) (G_0^2 + 2 G^+ G^-)\nonumber\\
&+ \frac{1}{2}\left( \MH^2\delta Z_{\PH\Ph}+\Mh^2 \delta Z_{\Ph\PH}\right) H h\nonumber\\
&+\frac{1}{2} \left( \MAO^2 \delta Z_{\PAO\PGO} 
+ 2\delta M_{\PGO\PAO}^2 \right) G_0 A_0
\nonumber\\
&+ \frac{1}{2} \left( \MHP^2 \delta Z_{\PH\PGP} 
+2\delta M_{\PG\PH+}^2 \right)(H^+ G^-+G^+ H^-)
\nonumber \\
&+ \textrm{interaction terms}.\label{eq:dVH2p}
\end{align}
The interaction terms are derived in
the same way, but they are very lengthy and not shown here. 
\end{sloppypar}

The prescription for the field renormalization~\eqref{eq:physfieldren} is non-minimal in the sense that a renormalization of the doublets with two renormalization constants,
\begin{align}
\Phi_{1,0} = Z_{\PH_1}^{1/2} \Phi_1& \textstyle = \Phi_1 \left(1+\frac{1}{2} \delta Z_{\PH_1}\right),\nonumber\\
\Phi_{2,0} = Z_{\PH_2}^{1/2} \Phi_2& \textstyle = \Phi_2 \left(1+\frac{1}{2} \delta Z_{\PH_2}\right),\label{eq:inputfieldren}
\end{align}
actually would be
sufficient to cancel the UV divergences. However, the prescription with matrix-valued renormalization constants allows us to renormalize each field on-shell. The UV-divergent parts of 
the renormalization constants in Eq.~\refeq{eq:physfieldren}
cannot be independent, and relations between the UV-divergent parts of the two prescriptions exist. They can be obtained by applying the renormalization prescription \eqref{eq:inputfieldren} to the left-hand side and \eqref{eq:physfieldren} to the right-hand side of Eqs.~\eqref{eq:rotations}, transforming thereafter the interaction states on the left-hand side to mass eigenstates and comparing both sides.
This results in
\begin{align}
\delta Z_\Ph\big|_\mr{UV}&=s_{\alpha}^2\;\delta Z_{\mr{H}_1}\big|_\mr{UV}+c_{\alpha}^2\; \delta Z_{\mr{H}_2}\big|_\mr{UV},\nonumber\\
\delta Z_\PH\big|_\mr{UV}&=c_{\alpha}^2\; \delta Z_{\mr{H}_1}\big|_\mr{UV}+s_{\alpha}^2\; \delta Z_{\mr{H}_2}\big|_\mr{UV},\nonumber\\
\delta Z_{\PH\Ph}\big|_\mr{UV}&=s_{\alpha} c_{\alpha}\; \left(-\delta Z_{\mr{H}_1}\big|_\mr{UV}+\delta Z_{\mr{H}_2}\big|_\mr{UV}\right)+2 \delta \alpha\big|_\mr{UV},\nonumber\\
\delta Z_{\Ph\PH}\big|_\mr{UV}&=s_{\alpha} c_{\alpha}\; \left(-\delta Z_{\mr{H}_1}\big|_\mr{UV}+\delta Z_{\mr{H}_2}\big|_\mr{UV}\right)-2 \delta \alpha\big|_\mr{UV},\nonumber\\
\delta Z_\PAO\big|_\mr{UV}&=\delta Z_{\PH^+}\big|_\mr{UV}=s_{\beta}^2\;\delta Z_{\mr{H}_1}\big|_\mr{UV}+c_{\beta}^2\; \delta Z_{\mr{H}_2}\big|_\mr{UV},\nonumber\\
\delta Z_\PGO\big|_\mr{UV}&=\delta Z_{G^+}\big|_\mr{UV}=c_{\beta}^2\; \delta Z_{\mr{H}_1}\big|_\mr{UV}+s_{\beta}^2\; \delta Z_{\mr{H}_2}\big|_\mr{UV},\nonumber\\
\delta Z_{\PGO\PAO}\big|_\mr{UV}&=\delta Z_{\PG\PHP}\big|_\mr{UV}=s_{\beta} c_{\beta}\; \left(-\delta Z_{\mr{H}_1}\big|_\mr{UV}+\delta Z_{\mr{H}_2}\big|_\mr{UV}\right)+2 \delta \beta\big|_\mr{UV},\nonumber\\
\delta Z_{\PAO\PGO}\big|_\mr{UV}&=\delta Z_{\PH\PGP}\big|_\mr{UV}=s_{\beta} c_{\beta}\; \left(-\delta Z_{\mr{H}_1}\big|_\mr{UV}+\delta Z_{\mr{H}_2}\big|_\mr{UV}\right)-2 \delta \beta\big|_\mr{UV}.\label{eq:fieldrenconstrel}
\end{align}
We will use these relations to derive UV-divergent parts for specific renormalization constants in Sect.~\ref{sec:diffrenschemes}. In App.~\ref{App:renormassa} we discuss a different choice of the mixing angles which is suited for 
the renormalization with $\lambda_3$ as an independent parameter. 

\subsection{The Higgs kinetic part}
\label{sec:Hkinren}

After expressing the Higgs kinetic term
\begin{align}
 \mathcal{L}_\mr{H,kin}=(D_\mu\Phi_1)^\dagger (D^\mu \Phi_1)+(D_\mu\Phi_2)^\dagger (D^\mu \Phi_2)
\end{align}
in terms of
bare physical fields, mixing angles, and parameters, one can apply the renorma\-lization transformations \eqref{eq:physpararen2} and~\eqref{eq:physfieldren} to obtain the counterterm part of the kinetic Lagrangian which introduces 
scalar--vector mixing terms. The explicit terms are stated in App.~\ref{App:SV-terms}.

\subsection{Fermionic and gauge parts}
\label{sec:fermgauge}
Since the THDM extension of the SM does not affect the gauge and the fermion parts of the Lagrangian,  
the renormalization of these parts is identical to the SM case. 
{It is described in detail in Ref.~\cite{Denner:1991kt} in 
BHS convention, which is included in the standard implementation of the \textsc{FeynArts} package~\cite{Hahn2001}.
Other renormalization prescriptions
can, e.g., be found in \citeres{Denner:2004bm,Kniehl:2009kk}.}
Therefore, we here do not repeat the renormalization procedure of the CKM matrix, which does not change in the 
transition from the SM to the THDM, and spell out the renormalization of the fermionic parts only for the
case where the CKM matrix is set to the unit matrix, i.e.\ $V_{ij}=\delta_{ij}$.
The transformation of the left- and right-handed fermions and of the gauge-boson fields are
\begin{align}
 f^\sigma_{i,0} & = \left(1 + \textstyle\frac{1}{2}\delta Z^{f,\sigma}_i\right) f^\sigma_i,\qquad  
\rlap{$f=\nu,l,u,d, \quad \sigma=\rL,\rR, \quad i=1,2,3,$}\\
 \begin{pmatrix}Z_0 \\ A_\mr{bare}\end{pmatrix}&=\begin{pmatrix}1+\frac{1}{2}\delta Z_{\PZ\PZ} &\frac{1}{2} Z_{\PZ\PA} \\\frac{1}{2} \delta Z_{\PA\PZ} &1+\frac{1}{2}\delta Z_{\PA\PA} \end{pmatrix} \begin{pmatrix}Z \\ A\end{pmatrix}, &
W^\pm_0 &= \left(1 + \textstyle\frac{1}{2} \delta Z_\PW\right) W^\pm,
\end{align}
where the bare photon field is denoted $A_\mr{bare}$ to distinguish it from the 
neutral CP-odd field~$A_0$.
Mixing between left-handed up- and down-type fermions does not occur, owing to charge conservation. Inserting this into the Lagrangian directly delivers the renormalized and the counterterm Lagrangians.

\subsection{Yukawa part}
\label{sec:YukRen}

The renormalization of the Yukawa sector is straightforward in Type I, II, lepton-specific, and flipped models and can be done by taking the Lagrangian of Eq.~\eqref{eq:ffH}, 
replacing the vev $v$, and applying the renormalization transformations of 
Sect.~\ref{sec:renormassb}, as well as a renormalization of the fermion masses,
\begin{align}
  m_{f,i,0} &= m_{f,i} + \delta m_{f,i}.
  \end{align}
The corresponding  counterterm couplings are stated in App.~\ref{App:Yuk-CT}.

\section{Renormalization conditions}
\label{sec:renconditions}

The renormalization constants are fixed using on-shell conditions 
for all parameters that are accessible by experiments. 
However, not all parameters of the THDM correspond to measurable quantities, so that we renormalize three parameters of the Higgs sector in the $\overline{\mr{MS}}$ scheme, where the renormalization constants only contain the standard UV divergence
\begin{align}
  \Delta_\mr{UV}=\frac{2}{4-D}-\gamma_\mr{E}+\ln{4\pi}=\frac{1}{\epsilon}-\gamma_\mr{E}+\ln{4\pi}
\end{align}
in $D=4-2 \epsilon$ dimensions and with the Euler--Mascheroni constant $\gamma_\mr{E}$.
In Sect.~\ref{sec:diffrenschemes}, four different options, resulting in four different renormalization schemes, are presented. 
An overview over the renormalization constants introduced in the previous section is shown in Tab.~\ref{tab:renconsts}. In the following, we adapt the notation of Ref.~\cite{Denner:1991kt}, i.e.\ we use the same symbols for the renormalized and the corresponding unrenormalized 
Green function, self-energies, etc., but denoting the renormalized quantities with a caret.

\begin{table}
  \centering
\begin{tabular}{|lrl|}\hline
Parameters:&&\\\hline
&EW (3): & $\delta \MZ^2,\delta \MW^2,\delta e, (\delta \cw,\delta \sw)$\\
&fermion masses (9): & $\delta m_{f,i},\qquad f=l,u,d,\quad i=1,2,3$\\
&Higgs masses (4): & $\delta \MH^2,\delta \Mh^2,\delta \MAO^2,\delta \MHP^2$\\
&Higgs potential (3): & $\delta \lambda_3 \text{ or } \delta \alpha,\delta \lambda_5, \delta \beta$\\
&tadpoles (2): & $\delta t_\PH,\delta t_\Ph$\\\hline
Fields:&&\\\hline
&EW (5): & $\delta Z_\PW,\delta Z_{\PZ\PZ},\delta Z_{\PZ\PA},\delta Z_{\PA\PZ},\delta Z_{\PA\PA}$\\
&left-handed fermions (12): & $\delta Z^{f,\rL}_i,\qquad f=\nu,l,u,d,\quad i=1,2,3$ \\
&right-handed fermions (9): & $\delta Z^{f,\rR}_i,\qquad f=l,u,d,\quad i=1,2,3$ \\
&Higgs (12): & $\delta Z_{\PH},\delta Z_{\PH\Ph},\delta Z_{\Ph\PH},\delta Z_\Ph$\\
&& $\delta Z_{\PAO},\delta Z_{\PAO\PGO},\delta Z_{\PGO\PAO},\delta Z_\PGO$\\
&&$\delta Z_{\PHP},\delta Z_{\PH\PGP},\delta Z_{\PG\PHP},\delta Z_{\PGP}$\\\hline
\end{tabular}  
  \caption{The renormalization constants used to describe the THDM, separated into sectors of parameter and field renormalization. The renormalization constants in parentheses are not independent, but useful for a better bookkeeping. The numbers in parentheses are the numbers of independent renormalization constants. In total there are 38 field and 20 parameter renormalization constants to fix. }
\label{tab:renconsts}
\end{table}%

\subsection{On-shell renormalization conditions}
\label{sec:onshellrenconditions}

\subsubsection{Higgs sector}

\paragraph{Tadpoles:}
We start with the 
(irreducible) renormalized one-point vertex functions
\begin{align}
\hat\Gamma^{\PH,\Ph}=
\ri\hat{T}^{\PH,\Ph}=\,\parbox{2.5cm}{\includegraphics[scale=1]{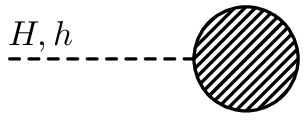}}
\hspace{20pt}.
\end{align}
At NLO the renormalized tadpole $\hat{T}$ consists of a counterterm contribution $\delta t$ and an unrenormalized one-loop irreducible one-point vertex function $T$
resulting from the diagrams shown in Fig.~\ref{fig:tadpoles}. 
\begin{figure}
\vspace{-30pt}
\input{diagrams/tadpoles}
\vspace{-40pt}
\caption{Generic tadpole diagrams. There is one diagram for each massive fermion $(f)$, 
scalar $(S)$, gauge-boson $(V)$, and ghost field $(u)$.}
\label{fig:tadpoles}
\end{figure}
In the conventional, but gauge-dependent tadpole treatment one demands that these two contributions cancel each other,
\begin{align}
\hat{T}_\PH&=\delta t_\PH+T_\PH=0,&\hat{T}_\Ph&=\delta t_\Ph+T_\Ph=0,
\label{eq:tadpoleren}
\end{align}
which means that explicit tadpole diagrams can be omitted from the 
set of one-loop diagrams for any process. 
However, as a remnant of the tadpole diagrams the tadpole counter\-terms 
appear also in various coupling 
counterterms and need to be calculated. 
It should be noted 
that the condition on the tadpoles does not affect physical observables 
as long as physically equivalent renormalization conditions are imposed on
the input parameters. This is, in particular, the case for on-shell renormalization,
where input parameters are tied to measurable quantities.
That means, changing the tadpole renormalization condition shifts contributions between 
Green functions and counterterms and merely changes the bookkeeping,
but the dependence of predicted observables on renormalized input parameters remains the same.
The situation changes if an \MSbar{} renormalization condition is used, where the counterterm
is not fixed by a measurable quantity, 
but by a divergence in a specific Green function, so that
the gauge-dependent tadpole terms can affect the relation between renormalized input parameters and
observables.
The gauge-independent treatment of tadpole contributions is based on a different renormalization condition and discussed in Sect.~\ref{sec:FJscheme}. 
\vspace{10pt}

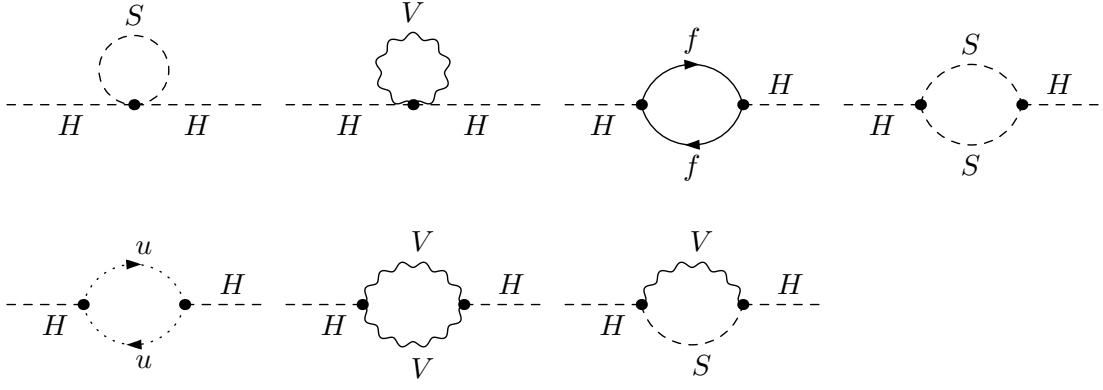
\begin{figure}
\vspace{-10pt}
\input{diagrams/selfenergies}
\vspace{-20pt}
\caption{Generic self-energy diagrams for the heavy, neutral CP-even Higgs self-energy, for other scalar self-energies the diagrams are analogous. Only massive particles contribute.}
\label{fig:selfenergies}
\end{figure}

\paragraph{Scalar self-energies:}
For scalars, the irreducible two-point functions  with momentum transfer $k$ are
\begin{align}
\hat{\Gamma}^{{ab}}(k)&=\quad \parbox{3.0cm}{
\includegraphics[scale=1]{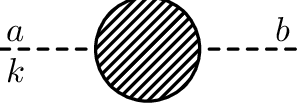}} \quad
=\im\delta_{{ab}} (k^2-M_{a}^2)+\im \hat{\Sigma}^{{ab}}(k),
\label{eq:two-def}
\end{align}
where both fields $a,b$ are incoming and $a,b=H,h,A_0,G_0,H^\pm,G^\pm$. The first term is the LO two-point vertex function, while the functions $\hat{\Sigma}^{{ab}}$ are the renormalized self-energies containing loop diagrams and counterterms. Generic diagrams contributing to the self-energies are shown in Fig.~\ref{fig:selfenergies}. 
Mixing occurs only between $H$ and $h$,
between $A_0$ and $G_0$, and between $H^\pm$ and $G^\pm$.
For the neutral CP-even fields we obtain
\begin{subequations}
\label{eq:neutralSE}
\begin{align}
\hat{\Sigma}^{\Ph\Ph}(k^2)&= \Sigma^{\Ph\Ph}(k^2) + \delta Z_\Ph(k^2-\Mh^2)  - \delta \Mh^2,\\
\hat{\Sigma}^{\PH\PH}(k^2)&= \Sigma^{\PH\PH}(k^2) + \delta Z_\PH(k^2-\MH^2)  - \delta \MH^2,\\
\hat{\Sigma}^{\PH\Ph}(k^2)&= \Sigma^{\PH\Ph}(k^2) + \frac{1}{2}\delta Z_{\PH\Ph}(k^2-\MH^2) +\frac{1}{2}\delta Z_{\Ph\PH}(k^2-\Mh^2) - \delta M_{\PH\Ph}^2, \label{eq:twopointHh}
\end{align}
\end{subequations}
and for the CP-odd fields
\begin{subequations}
\label{eq:CPoddSE}
\begin{align}
\hat{\Sigma}^{\PAO\PAO}(k^2)&= \Sigma^{\PAO\PAO}(k^2) + \delta Z_{\PAO}(k^2-\MAO^2)  - \delta \MAO^2,\\
\hat{\Sigma}^{\PGO\PGO}(k^2)&= \Sigma^{\PGO\PGO}(k^2) + \delta Z_\PGO k^2  - \delta M_\PGO^2,\\
\hat{\Sigma}^{\PGO\PAO}(k^2)&= \Sigma^{\PGO\PAO}(k^2) + \frac{1}{2}\delta Z_{\PAO\PGO}(k^2-\MAO^2) +\frac{1}{2}\delta Z_{\PGO\PAO} \;k^2 - \delta M_{\PGO\PAO}^2.
\end{align}
\end{subequations}
The charged sector involves the following self-energies,
\begin{subequations}
\label{eq:chargedSE}
\begin{align}
\hat{\Sigma}^{\mr{H}^+ \mr{H}^-}(k^2)&= \Sigma^{\mr{H}^+ \mr{H}^-}(k^2) + \delta Z_{\PHP}(k^2-\MHP^2)  - \delta \MHP^2,\\
\hat{\Sigma}^{\mr{G}^+\mr{G}^-}(k^2)&= \Sigma^{\mr{G}^+\mr{G}^-}(k^2) + \delta Z_{\mr{G}^+}\; k^2  - \delta M_\PGP^2,\\
\hat{\Sigma}^{\mr{G}^\pm\mr{H}^\mp}(k^2)&= \Sigma^{\mr{G}^\pm\mr{H}^\mp}(k^2) + \frac{1}{2}\delta Z_{\PH\PGP}(k^2-\MHP^2) +\frac{1}{2}\delta Z_{\PG\PHP}\; k^2 - \delta M_{\PG\PHP}^2.
\end{align}
\end{subequations}
The mass mixing constants are given in Eqs.~\eqref{eq:MHhchoice2}  and \eqref{eq:Mabchoice2}.
On these two-point functions we now impose our renormalization conditions. First, we  fix the renormalized mass parameters to the on-shell
values, so that the zeros of the real parts of the one-particle-irreducible two-point functions are located at the squares of the physical masses:
\begin{align}
\mr{Re}\,\hat{\Sigma}^{\PH\PH}(\MH^2)&=0,& \mr{Re}\,\hat{\Sigma}^{\Ph\Ph}(\Mh^2)&=0,\nonumber\\
\mr{Re}\,\hat{\Sigma}^{\PAO\PAO}(\MAO^2)&=0,&\mr{Re}\,\hat{\Sigma}^{\mr{H}^+\mr{H}^-}(\MHP^2)&=0.
\label{eq:massrenconddef}
\end{align}
Using Eqs.~\eqref{eq:neutralSE}, \eqref{eq:CPoddSE}, and \eqref{eq:chargedSE}, fixes  the mass renormalization constants to
\begin{align}
\delta \MH^2&= \mr{Re}\, \Sigma^{\PH\PH}(\MH^2),&\delta \Mh^2&= \mr{Re}\, \Sigma^{\Ph\Ph}(\Mh^2),\nonumber\\
\delta \MAO^2&= \mr{Re}\, \Sigma^{\PAO\PAO}(\MAO^2),&\delta M_{\PHP}^2&= \mr{Re}\, \Sigma^{\mr{H}^+\mr{H}^-}(M_{\PH^+}^2).
\end{align}
For the propagators of the fields, we demand that the residues of the particle poles are not changed by higher-order corrections.
This  determines the 
diagonal field renormalization constants by conditions on the one-particle-irreducible two-point functions,
\begin{align}
\lim_{k^2\to \MH^2}\mr{Re}\,\frac{\im\,\hat{\Gamma}^{\PH\PH}(k^2)}{k^2-\MH^2}&=-1,& 
\lim_{k^2\to \Mh^2}\mr{Re}\,\frac{\im\,\hat{\Gamma}^{\Ph\Ph}(k^2)}{k^2-\Mh^2}&=-1,\nonumber\\
\lim_{k^2\to \MAO^2}\mr{Re}\,\frac{\im\,\hat{\Gamma}^{\PAO\PAO}(k^2)}{k^2-\MAO^2}&=-1,&
\lim_{k^2\to \MHP^2}\mr{Re}\,\frac{\im\,\hat{\Gamma}^{\mr{H}^+\mr{H}^-}(k^2)}{k^2-\MHP^2}&=-1,
\end{align}
which implies
\begin{align}
\delta Z_\PH&= -\mr{Re}\, \Sigma'^{\PH\PH}(\MH^2),
&\delta Z_\Ph&= -\mr{Re}\, \Sigma'^{\Ph\Ph}(\Mh^2), \label{eq:dZHdef}\nonumber\\
\delta Z_{\PAO}&= -\mr{Re}\, \Sigma'^{\PAO\PAO}(\MAO^2),
&\delta Z_{\mr{H}^+}&= -\mr{Re}\, \Sigma'^{\mr{H}^+\mr{H}^-}(\MHP^2),
\end{align}
where we introduced 
${\Sigma}'(k^2)$ as the derivative w.r.t.\  the argument $k^2$.
To fix the mixing renormalization constants, we enforce the condition that on-mass-shell fields do not mix, i.e.
\begin{align}
\mr{Re}\,\hat{\Sigma}^{\PH\Ph}(\MH^2)&=0,& \mr{Re}\,\hat{\Sigma}^{\PH\Ph}(\Mh^2)&=0,
\nonumber\\
\mr{Re}\,\hat{\Sigma}^{\PGO\PAO}(\MAO^2)&=0,&\mr{Re}\,\hat{\Sigma}^{\PG^+\PH^-}(\MHP^2)&=0.
\label{eq:mixrencond}
\end{align}
After inserting the renormalized self-energies we obtain
\begin{align}
\delta Z_{\PH\Ph}&= 2 \, \frac{\delta M_{\PH\Ph}^2-\mr{Re}\,\Sigma^{\PH\Ph}(\Mh^2)}{\Mh^2-\MH^2},
&\delta Z_{\Ph\PH}&= 2 \, \frac{\delta M_{\PH\Ph}^2-\mr{Re}\,\Sigma^{\PH\Ph}(\MH^2)}{\MH^2-\Mh^2},
\nonumber\\
\delta Z_{\PGO\PAO}&= 2 \, \frac{\delta M_{\PGO\PAO}^2-\mr{Re}\,\Sigma^{\PGO\PAO}(\MAO^2)}{\MAO^2},
&\delta Z_{\PG\mr{H}^+}&= 2 \, \frac{\delta M_{\PG\PHP}^2-\mr{Re}\,\Sigma^{\mr{H}^+\mr{G}^-}(\MHP^2)}{\MHP^2}.\label{eq:mixfieldrendef}
\end{align}
Since Goldstone-boson fields 
do not correspond to physical states, we do not render Green functions with external 
Goldstone bosons finite, so that we need not fix the constants 
$\delta Z_{\PGO}$, $\delta Z_{\PAO\PGO}$, $\delta Z_{\PGP}$, $\delta Z_{\PH\PGP}$;
we could even set them to zero consistently.
The possible 
$\PZ\PAO$ and $\mr{W}^\pm \PH^\mp$ mixings
vanish for physical on-shell gauge bosons due to the Lorentz structure of the two-point function 
and the fact that polarization vectors $\varepsilon^\mu$ are orthogonal to the corresponding momentum.
Using the convention
\begin{align}
\hat\Gamma^{\PZ\PAO}_\mu(k) &= k_\mu \, \hat\Sigma^{\PZ\PAO}(k^2)  = 
k_\mu \left[ \Sigma^{\PZ\PAO}(k^2)-\textstyle\frac{1}{2}\MZ\delta Z_{\PGO\PAO}\right],
\nonumber\\
\hat\Gamma^{\PW^\pm\PH^\mp}_\mu(k) &= k_\mu \, \hat\Sigma^{\PW^\pm\PH^\mp}(k^2)  = 
k_\mu \left[ \Sigma^{\PW^\pm\PH^\mp}(k^2)\pm\textstyle\frac{\im}{2}\MW\delta Z_{\PG\PHP}\right],
\end{align}
where all fields are incoming and $k$ is the incoming momentum of the gauge-boson fields,
the vector--scalar mixing self-energies obey
\begin{align}
\left.\varepsilon^\mu_\PZ k_\mu \,\Re\,\hat{\Sigma}^{\PZ\PAO}(k^2)\right|_{k^2= \MZ^2}=0,
\qquad &
\left.\varepsilon^\mu_\PW k_\mu \,\Re\,\hat{\Sigma}^{\PW^\pm\PH^\mp}(k^2)\right|_{k^2= \MW^2}=0.
\end{align}
The mixing self-energies on the other on-shell points 
$k^2=\MAO^2$ and $k^2=\MHP^2$, respectively, 
are connected to the mixing of 
$\PAO$ or $\PH^\pm$ with the Goldstone-boson fields of the
$\PZ$ or the $\PW$ boson and can be calculated from a BRST symmetry \cite{Becchi:1975nq}. The BRST variation of the Green functions of one anti-ghost and a Higgs field
\begin{align}
 \delta_\mr{BRST} \langle 0 | T \bar{u}^\PZ(x) A_0(y)|0\rangle=0, &&
 \delta_\mr{BRST} \langle 0 | T \bar{u}^\pm(x) H^\pm(y)|0\rangle=0,
\end{align}
implies Slavnov--Taylor identities. While the variation of the anti-ghost fields yields the gauge-fixing term, the variation of the Higgs fields introduces ghost contributions which 
vanish for on-shell momentum resulting in%
\footnote{A particularly simple, alternative 
way to derive these identities is to exploit the gauge invariance
of the effective action in the background-field gauge, as 
done in \citere{Denner:1994xt} for the SM. The respective Ward identities for the background
fields differ from the Slavnov--Taylor identities only by off-shell terms, which vanish
on the particle poles. Generalizing the derivation of \citere{Denner:1994xt} to the THDM
and adapting the results to our conventions for self-energies,
the desired Ward identities for the unrenormalized background fields read
\begin{align}
0 &= k^2 {\Sigma}^{\hat\PZ\hat\PA_0}(k^2)+ \MZ {\Sigma}^{\hat\PA_0\hat\PG_0}(k^2)
+\frac{e}{2\cw\sw}\left(T^{\hat\PH}s_{\beta-\alpha}-T^{\hat\Ph}c_{\beta-\alpha}\right),
\\
0 &= 
k^2 {\Sigma}^{\hat\PW^\pm\hat\PH^\mp}(k^2)\mp \im \MW {\Sigma}^{\hat\PG^\pm\hat\PH^\mp}(k^2)
\mp\frac{\im e}{2\sw}\left(T^{\hat\PH}s_{\beta-\alpha}-T^{\hat\Ph}c_{\beta-\alpha}\right),
\end{align}
where the carets on the fields indicate background fields.
Setting $k^2$ to $\MAO^2$ or $\MHP^2$, respectively, and adding the 
relevant renormalization constants, directly leads to the identities \refeq{eq:ZA-WI}
and \refeq{eq:WH-WI}.
}
\begin{align}
\label{eq:ZA-WI}
 \left[k^2 \hat{\Sigma}^{\PZ\PAO}(k^2)+ \MZ \hat{\Sigma}^{\PGO\PAO}(k^2)\right]_{k^2= \MAO^2}=0,\\
 \left[k^2 \hat{\Sigma}^{\mr{W}^\pm\PH^\mp}(k^2)\mp \im \MW \hat{\Sigma}^{\mr{G}^\pm\mr{H}^\mp}(k^2)\right]_{k^2= \MHP^2}=0.
\label{eq:WH-WI}
\end{align}
We have verified these identities analytically and numerically.
Together with the renormalization condition of 
Eq.~\eqref{eq:mixrencond} we conclude that 
\begin{align}
\hat{\Sigma}^{\PZ\PAO}(\MAO^2)=0, \quad \hat{\Sigma}^{\mr{W}^\pm\PH^\mp}(\MHP^2)=0.
\end{align}
This set of renormalization conditions ensures that no on-shell two-point vertex function obtains any one-loop corrections, 
and the corresponding external
self-energy diagrams do not have to be taken into account in any calculation. 

\subsubsection{Electroweak sector}

The fixing of the renormalization constants of the electroweak sector is identical to the SM case. The mass renormalization constants are fixed in such a way that the squares of the masses correspond to the 
(real parts of the) locations of the poles of the gauge-boson propagators. The field renormalization constants are fixed by the conditions that residues of on-shell gauge-boson propagators  do not obtain higher-order corrections, and that on-shell gauge bosons do not mix. For a better bookkeeping we also keep the dependent renormalization constants $\delta \cw$ and $\delta \sw$ in our calculation. This results in~\cite{Denner:1991kt}
\begin{align}
\delta \MW^2 &=\mr{Re}\, \Sigma^W_\mr{T}(\MW^2),&\delta Z_{\PW}&=-\mr{Re}\, \Sigma'^{\PW}_\mr{T}(\MW^2),\nonumber\\
\delta \MZ^2&=\mr{Re}\, \Sigma^{\PZ\PZ}_\mr{T}(\MZ^2),\nonumber\\
\delta Z_{\PZ\PZ}&=-\mr{Re}\, \Sigma'^{\PZ\PZ}_\mr{T}(\MZ^2),&\delta Z_{\PA\PA}&=-\mr{Re}\, \Sigma'^{\PA\PA}_\mr{T}(0),\nonumber\\
\delta Z_{\PA\PZ}&=-2\mr{Re}\, \frac{\Sigma^{\PA\PZ}_\mr{T}(\MZ^2)}{\MZ^2},&
\delta Z_{\PZ\PA}&=2\mr{Re}\, \frac{\Sigma^{\PA\PZ}_\mr{T}(0)}{\MZ^2},\nonumber\\
\delta \cw&=\frac{\cw}{2}\left(\frac{\delta \MW^2}{\MW^2}-\frac{\delta \MZ^2}{\MZ^2}\right), &\delta \sw&= -\frac{\cw}{\sw}\delta \cw.
\end{align}
The electric charge $e$ is defined via the $\mr{ee} \gamma $ coupling in the Thomson limit of on-shell external electrons and zero momentum transfer to the photon,
which yields in BHS convention \cite{Denner:1991kt}
\begin{align}
\delta Z_e = -\frac{1}{2} \left(\delta Z_{\PA\PA}+ \frac{\sw}{\cw} \delta Z_{\PZ\PA}\right).
\label{eq:dZe}
\end{align}

\subsubsection{Fermions}

The renormalization conditions for the fermions are identical to the ones in the SM, described in detail in Ref.~\cite{Denner:1991kt}. We demand that the (real parts of the locations of the)
poles of the fermion propagators correspond to the squared fermion masses, 
 and that on-shell fermion propagators do not obtain loop corrections. 
Assuming the CKM matrix 
equal to the unit matrix, the results for the renormalization constants simplify to
\begin{align}
\delta m_{f,i}&=\frac{m_{f,i}}{2}\mr{Re}\,\big[\Sigma^{f,\rL}_i(m^2_{f,i})+\Sigma^{f,\rR}_i(m^2_{f,i})+2 \Sigma^{f,\rS}_i(m^2_{f,i})\big],
\nonumber\\
\delta Z^{f,\sigma}_i&=-\mr{Re}\, \Sigma^{f,\sigma}_i(m^2_{f,i})-m_{f,i}^2\frac{\partial}{\partial k^2} \mr{Re} \big[\Sigma_i^{f,\rL}(k^2)+\Sigma^{f,\rR}_i(k^2)+2\Sigma^{f,\rS}_i(k^2)\big]\Big|_{k^2=m_{f,i}^2}, \quad \sigma=\rL,\rR,
\end{align}
where we have 
used the usual 
decomposition of the fermion self-energies into a left-handed, a right-handed, and a scalar part, $\Sigma^{f,\rL}_i$, $\Sigma^{f,\rR}_i$, and $ \Sigma^{f,\rS}_i$, respectively.
The expressions for a non-trivial CKM matrix can be found in \citere{Denner:1991kt}.

\subsection{\boldmath{$\overline{\mr{MS}}$ renormalization conditions}}
\label{sec:diffrenschemes}

In the four renormalization schemes we 
are going to present, the imposed on-shell conditions are identical, and the differences only occur in the choice of different $\overline{\mr{MS}}$ conditions.
The parameters $\alpha$ or $\lambda_3$ governing the mixing of the CP-even Higgs bosons, and the parameters $\beta$ and $\lambda_5$ need to be fixed.
A formulation of an on-shell condition for these parameters is not obvious.  
One could relate the parameters to 
some physical processes, such as Higgs-boson decays, and demand that these processes do not 
receive higher-order corrections. However, so far, no sign of further Higgs bosons has been observed, hence, there is no distinguished process, and such a prescription does not only require more calculational effort, but could introduce artificially large corrections to the corresponding parameters, which would spread to many other observables, as discussed in Refs.~\cite{Freitas:2002um,Krause:2016oke}. Therefore, we choose to renormalize these parameters within the $\overline{\mr{MS}}$ scheme,
though different variables (such as $\alpha$ or $\lambda_3$) can be chosen to parameterize the model. Imposing an $\overline{\mr{MS}}$ condition on either of the parameters leads to differences in the calculation of observables. In addition, gauge-dependent definitions of $\overline{\mr{MS}}$-renormalized parameters spoil the gauge independence of the relations between input parameters and observables. However, gauge dependences might be even
acceptable if the renormalization scheme yields stable results and a good convergence of the perturbation series. The price to pay is that subsequent calculations should be done in the same gauge
or properly translated into another gauge. We will discuss different renormalization schemes based on different treatments of 
$\alpha$ or $\lambda_3$ parameterizing the CP-even Higgs-boson mixing, of the parameter $\beta$, and of the Higgs coupling constant 
 $\lambda_5$ 
in the following. We begin with the so-called $\overline{\mr{MS}}(\alpha)$ scheme.

\subsubsection{\boldmath{$\overline{\mr{MS}}(\alpha)$ scheme}}
\label{sec:alphaMSscheme}
In this scheme the independent parameter set is $\{p_{\mr{mass}}\}$ of Eq.~\eqref{eq:physparaNLO}, so that the parameters $\beta$, $\alpha$, and $\lambda_5$ are renormalized in $\overline{\mr{MS}}$. The corresponding counterterm Lagrangian was derived in Sect.~\ref{sec:renormassb}.

\paragraph{The renormalization constant \boldmath{$\delta \beta$}:}

The renormalization constant 
$\delta \tan{\beta}=\delta \beta/{c_\beta^2}$
of the mixing angle $\beta$ is
related to the renormalization constants of the vevs
by demanding the defining relation $\tan\beta=v_2/v_1$ for bare and renormalized 
quantities.
In \MSbar{}, $\delta \beta$ can be most easily calculated using the minimal field
renormalization \refeq{eq:inputfieldren} with the following renormalization transformation
of the vevs,
\begin{align}
v_{1,0} &=  Z_{\PH_1}^{1/2}(v_1+\delta \overline{v}_1),&v_{2,0}&=Z_{\PH_2}^{1/2}(v_2+\delta \overline{v}_2),
 \label{eq:vevren}
\end{align}
Using the well-known relation~\cite{Sperling:2013eva}
\begin{align}
\delta \overline{v}_1/v_1-\delta \overline{v}_2/v_2=\mr{finite},\label{eq:vacrencond}
\end{align}
the general form of $\delta\beta$ in the \MSbar{} scheme
\begin{align}
\delta \beta
&= \frac{s_{2\beta}}{2}\left(
\frac{\delta \overline{v}_2}{v_2} -\frac{\delta \overline{v}_1}{v_1}- \frac{1}{2}\delta Z_{\PH_1} + \frac{1}{2} \delta Z_{\PH_2} 
\right)\bigg|_\mr{UV}
\end{align}
simplifies to 
\begin{align}
\delta \beta
&=\frac{s_{2\beta}}{4}(-\delta Z_{\PH_1}+\delta Z_{\PH_2})\big|_\mr{UV}
=\frac{s_{2\beta}}{4c_{2\alpha}} \left(\delta Z_\Ph -\delta Z_\PH\right)\big|_\mr{UV}
=\frac{s_{2\beta}}{4s_{2\alpha}} \left(\delta Z_{\Ph\PH} +\delta Z_{\PH\Ph}\right)\big|_\mr{UV},
\label{eq:tanbetarencond}
\end{align}
where $\big|_\mr{UV}$ indicates that we take only the UV-divergent parts, which are proportional to 
the standard divergence $\Delta_\mr{UV}$.
The explicit calculation of the UV-divergent terms of $\delta Z_\Ph,$ $\delta Z_\PH$ according to Eqs.~\eqref{eq:dZHdef} in 
't~Hooft-Feynman gauge reveals that only diagrams with closed fermion loops contribute to the counterterm, 
\begin{align}\label{eq:deltabeta}
\delta \beta = -\Delta_\mr{UV}\, \frac{e^2}{64\pi^2\MW^2\sw^2} \sum_f c_f \xi^f_\PAO m_f^2,
\end{align}
with the colour factors $c_\mr{quark}=3$, $c_\mr{lepton}=1$ and the coupling coefficients $\xi^f_{\PAO}$ 
as defined in Tab.~\ref{tab:yukint}. In the class of $R_\xi$ gauges this result is gauge independent at one-loop order \cite{Krause:2016oke,Denner:2016etu}.

\paragraph{Neutral Higgs mixing:}
In the neutral Higgs sector, relations between field renormalization constants can also be used to determine another parameter in $\overline{\mr{MS}}$.
The first four equations of Eqs.~\eqref{eq:fieldrenconstrel} can be solved for $\delta \alpha$ in various
ways, e.g., yielding
\begin{align}
\label{eq:deltaalphaderiv}
\delta \alpha\big|_\mr{UV}&= \frac{1}{4} \left(\delta Z_{\PH\Ph}-\delta Z_{\Ph\PH}\right) \big|_\mr{UV}.
\end{align}
The field renormalization constants can be inserted according to Eqs.~\eqref{eq:dZHdef}, \eqref{eq:mixfieldrendef} using 
$\delta M_{\PH\Ph}^2 = 0$, thus 
\begin{align}
\label{eq:deltaalphaderiv2}
\delta \alpha&= \left. \mr{Re}\frac{\Sigma^{\PH\Ph}(\MH^2)+\Sigma^{\PH\Ph}(\Mh^2)}{2(\MH^2-\Mh^2)}\right|_\mr{UV}.
\end{align}
An explicit calculation of the counterterm 
yields for the fermionic contribution, 
\begin{align}\label{eq:delalphaferm}
\delta \alpha\big|_\mr{ferm} = \Delta_\mr{UV}\, \frac{e^2 s_{2\alpha}}{64\pi^2\MW^2\sw^2 s_{2\beta}(\MH^2-\Mh^2)} 
\sum_f c_f \xi^f_\PAO m_f^2 (\MH^2+\Mh^2-12m_f^2),
\end{align}
and for the bosonic contribution
\begin{align}\label{eq:delalphabos}
\delta \alpha\big|_\mr{bos} = &
-\Delta_\mr{UV} \frac{\lambda_5^2 \MW^2 \sw^2 }{8\pi^2 e^2 (\Mh^2 - \MH^2) s_{2\beta}^2}
   \Big[s_{2 (\alpha - 3 \beta)} + 10 s_{2 (\alpha - \beta)} + 13 s_{2 (\alpha + \beta)}\Big]
\nonumber\\
& + \Delta_\mr{UV} \frac{\lambda_5 }{ 128\pi^2 (\Mh^2 - \MH^2)  s_{2\beta}^2} \Big[
   -4 \MH^2 (13 c_{2 \alpha} + 2 c_{2 (\alpha - 2 \beta)} - 27 c_{2 \beta}) s_{2 \alpha} 
\nonumber\\
& + 4 \Mh^2 (13 c_{2 \alpha} + 2 c_{2 (\alpha - 2 \beta)} + 27 c_{2 \beta}) s_{2 \alpha} 
  + 2 \MHP^2 (s_{2 (\alpha - 3 \beta)} - 6 s_{2 (\alpha - \beta)} + 13 s_{2 (\alpha + \beta)}) 
\nonumber\\
& - \MAO^2 (7 s_{2 (\alpha - 3 \beta)} + 86 s_{2 (\alpha - \beta)} + 91 s_{2 (\alpha + \beta)})
  + 4 (2 \MW^2 + \MZ^2) s_{2 (\alpha - \beta)} s_{2 \beta}^2 
\Big]
\nonumber\\
& + \Delta_\mr{UV} \frac{e^2 }{ 1024\pi^2 (\Mh^2 - \MH^2) \MW^2  \sw^2 s_{2\beta}^2} \Big[
- 2 \MH^4 (-36 c_{2 \alpha} + 5 c_{4 \alpha - 2 \beta} + 31 c_{2 \beta}) s_{2 \alpha} 
\nonumber\\
& + 4 \Mh^2 \MH^2 (5 c_{4 \alpha - 2 \beta} - 29 c_{2 \beta}) s_{2 \alpha} 
- 2 \Mh^4 (36 c_{2 \alpha} + 5 c_{4 \alpha - 2 \beta} + 31 c_{2 \beta}) s_{2 \alpha} 
\nonumber\\
& + 32 \MHP^4 s_{2 (\alpha - \beta)} s_{2 \beta}^2 
+ 2 \MH^2 \MHP^2 (3 s_{4 \alpha} + 4 s_{2 (\alpha - \beta)} + s_{4 (\alpha - \beta)} 
	+ 9 s_{4 \beta} - 12 s_{2 (\alpha + \beta)}) 
\nonumber\\
& - 2 \Mh^2 \MHP^2 (3 s_{4 \alpha} -4 s_{2 (\alpha - \beta)} + s_{4 (\alpha - \beta)} 
	+ 9 s_{4 \beta} + 12 s_{2 (\alpha + \beta)}) 
\nonumber\\
& - 2 \MAO^4 (5 s_{2 (\alpha - 3 \beta)} + 42 s_{2 (\alpha - \beta)} + 41 s_{2 (\alpha + \beta)}) 
\nonumber\\
& + \MAO^2 \MH^2 (-49 s_{4 \alpha} + 112 s_{2 (\alpha - \beta)} - 7 s_{4 (\alpha - \beta)} 
	+ s_{4 \beta} + 96 s_{2 (\alpha + \beta)})
\nonumber\\
& + \MAO^2 \Mh^2 (49 s_{4 \alpha} + 112 s_{2 (\alpha - \beta)} + 7 s_{4 (\alpha - \beta)} 
	- s_{4 \beta} + 96 s_{2 (\alpha + \beta)}) 
\nonumber\\
& + 4 \MAO^2 \MHP^2 (s_{2 (\alpha - 3 \beta)} - 6 s_{2 (\alpha - \beta)} + 13 s_{2 (\alpha + \beta)}) 
\nonumber\\
& + 4 (2 \MW^2 + \MZ^2) s_{2 (\alpha - \beta)} s_{2 \beta} 
	((\Mh^2 - \MH^2) s_{2 \alpha} + 2 \MAO^2 s_{2 \beta}) 
\nonumber\\
& + 48 (2 \MW^4 + \MZ^4) s_{2 (\alpha - \beta)} s_{2 \beta}^2 
\Big].
\end{align}
This result, which is derived in 't~Hooft Feynman gauge, is gauge dependent \cite{Krause:2016oke,Denner:2016etu}.

\begin{figure}
\vspace{-15pt}
\input{diagrams/triangles}
\vspace{-20pt}
\caption{Generic diagrams contributing to the $HA_0A_0$ vertex correction used for the renormalization of $\lambda_5$.}
\label{fig:dL5diags}
\end{figure}
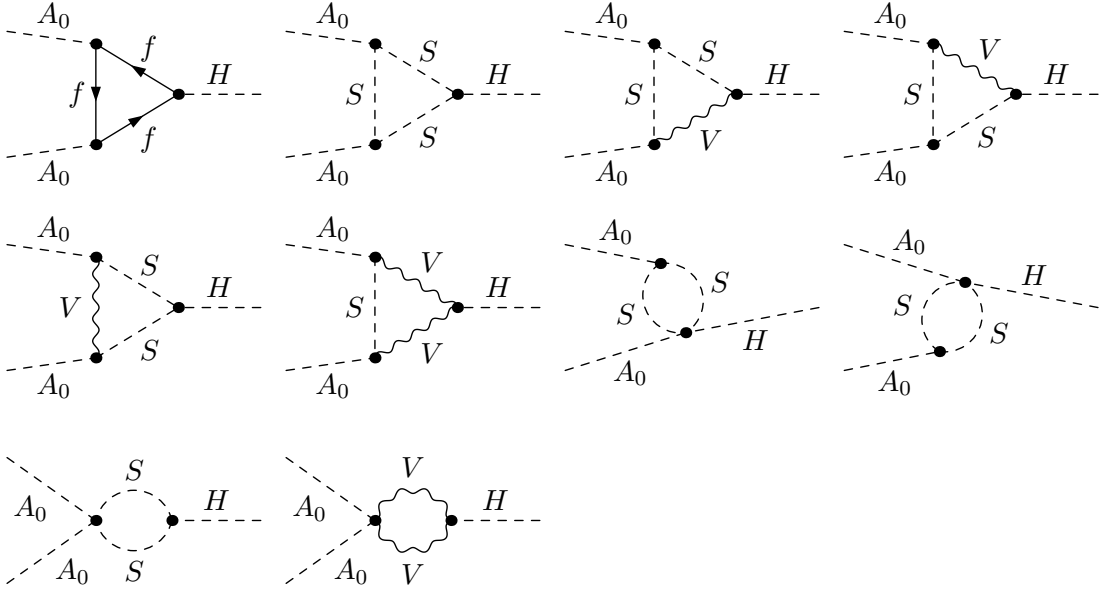

\paragraph{Higgs self-coupling:}
The Higgs self-coupling counterterm $\delta \lambda_5$ has to be fixed via a vertex correction. We define this renormalization constant  in $\overline{\mr{MS}}$ as well, as there is no distinguished process to fix it on-shell. Any 3- or 4-point vertex function with external Higgs bosons is suited to calculate the divergent terms.
Since the $\PH\PAO\PAO$ vertex correction involves fewest diagrams, it is our preferred choice. The condition is
\begin{align}
\hat{\Gamma}^{\PH\PAO\PAO}\big|_\mr{UV}=
\parbox{2.4cm}{\includegraphics[scale=1]{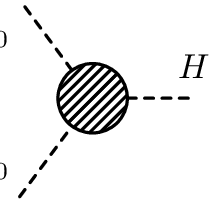}}\left.\rule{0cm}{1.1cm}\right|_\mr{UV}=0.
\end{align}
Solving this equation for $\delta \lambda_5$ fixes this renormalization constant. The generic one-loop diagrams appearing in this vertex correction are shown in Fig.~\ref{fig:dL5diags}, the contribution of the diagrams involving closed fermion loops is 
\begin{align}
\label{eq:L5ferm}
\delta \lambda_{5,\mr{ferm}}=\Delta_\mr{UV} 
\frac{e^2 \lambda_5}{16\pi^2 \MW^2\sw^2}\sum_f c_f \left(1+\frac{c_{2\beta}}{s_{2\beta}}\xi_{\PAO}^f\right) m_f^2.
\end{align}
The diagrams containing only bosons lead to 
\begin{align}
\label{eq:L5bos}
\delta \lambda_5\big|_\mr{bos}{}={}& {}
\Delta_\mr{UV} \frac{\lambda_5}{32 \pi^2}  
\left(2 \lambda_1 + 2\lambda_2+8\lambda_3 +12 \lambda_4- 9 g_2^2 - 3 g_1^2\right)
\nonumber\\
{}={}& - \Delta_\mr{UV}\frac{\lambda_5^2c^2_{2\beta}}{4 \pi^2 s^2_{2\beta}}
 +\Delta_\mr{UV}\frac{\lambda_5 e^2}{64 \pi^2\MW^2 \sw^2 s_{2\beta}^2} \Big[
  \MH^2 (2 + c_{2 (\alpha - \beta)} - 3 c_{2 (\alpha + \beta)}) 
\nonumber\\
& + \Mh^2 (2 - c_{2 (\alpha - \beta)} + 3 c_{2 (\alpha + \beta)})
+ \MAO^2 (1 - 5 c_{4 \beta})
- 4 \MHP^2 s_{2 \beta}^2
- 6(2 \MW^2+\MZ^2) s_{2 \beta}^2 
\Big].
\end{align}
Since $\lambda_5$ is a fundamental parameter of the Higgs potential, an $\overline{\mr{MS}}$  definition leads to a gauge-independent counterterm.

\subsubsection{\boldmath{$\overline{\mr{MS}}(\lambda_3)$ scheme}}
\label{sec:lambda3MSscheme}

In this scheme, the independent parameter set is $\{p'_{\mr{mass}}\}$ defined in Eq.~\eqref{eq:physparaNLOprime}.  The renormalization of $\beta$ and $\lambda_5$ is identical to the previous renormalization scheme and not stated again, but now the parameter $\lambda_3$ (instead of $\alpha$) is an independent parameter being renormalized in $\overline{\mr{MS}}$. This has the advantage that this parameter is gauge independent, as it is a defining parameter of the basic 
parameterization of the Higgs potential
and thus is safe against potentially gauge-dependent contributions appearing in relations between bare parameters. 
As stated above, the \MSbar{} renormalization of the 
parameter $\beta$ generally breaks gauge 
independence, but in $\mr{R}_\xi$ gauges the gauge dependence cancels at one loop \cite{Krause:2016oke,Denner:2016etu}, 
so that this scheme yields gauge-independent results at NLO. 
We take the counterterm potential of Sect.~\ref{sec:renormassb}, but treat $\delta \alpha$ as a dependent counterterm. 
As $\alpha$ is a pure mixing angle, we choose to apply the renormalization prescription of Sect.~\ref{sec:renormassb}, where the mixing angle diagonalizes the potential to all orders.
The relation between $\delta \alpha$ and the independent constants is given in Eq.~\eqref{eq:NLOmassmixHh} with $ \delta M_{\PH\Ph}^2=0$,
\begin{align}
\label{eq:alphadef32}
 \delta \alpha=\frac{f_\alpha(\{ \delta p'_\mr{mass}\})} {\MH^2-\Mh^2},
\end{align}
where $f_\alpha(\{ \delta p'_\mr{mass}\})$ can be obtained from Eq.~\eqref{eq:Hhmassmixing} by applying the renormalization transformation of Eq.~\eqref{eq:physpararen1} (which is identical to the renormalization transformation of Sect.~\ref{sec:renormassb}, but renormalizing $\lambda_3$ instead of $\alpha$). This yields
\begin{align}
f_\alpha(\{\delta p'_\mr{mass}\})={}&
\frac{1}{2} t_{2 \alpha } \left(\delta \Mh^2-\delta \MH^2\right)
+\frac{s_{2 \beta } \left(\delta \MAO^2-2 \delta \MHP^2\right)}{2 c_{2 \alpha }}
\nonumber\\
&+\frac{\delta \beta c_{2 \beta } \left(\MH^2- \Mh^2\right) t_{2 \alpha }}{s_{2 \beta }}
+\frac{2 \MW^2 s_{2\beta } (\delta \lambda_3+\delta \lambda_5) \sw^2}{e^2 c_{2 \alpha }}
\nonumber\\
&+\frac{s_{2 \beta } \left(\MAO^2-2 \MHP^2\right)+(\Mh^2-\MH^2)s_{2\alpha}}{c_{2\alpha}}
\left( \delta Z_e -\frac{\delta \sw}{\sw} -\frac{\delta \MW^2}{2\MW^2} \right)
\nonumber\\
&
-\frac{ e  \left[
\delta t_\PH \left(s_{\alpha -3 \beta }+3 s_{\alpha +\beta }\right)
+\delta t_\Ph \left(c_{\alpha -3 \beta }+3 c_{\alpha +\beta } \right) 
\right]}{8 \MW
   c_{2 \alpha } \sw}.\label{eq:MHhdependence}
\end{align}
The UV-divergent term of $\delta \alpha$ has been calculated in Eq.~\eqref{eq:deltaalphaderiv2}, and by renormalizing $\delta \lambda_3$ in $\overline{\mr{MS}}$ scheme, it is clear that the dependent $\delta \alpha$ must now have a finite part in addition. 
We choose this finite term in such a way that the finite part in $\delta \lambda_3$ 
(which results from $\delta \lambda_3$ by setting $\Delta_{\mr{UV}}$ to zero) vanishes and obtain
\begin{align}
\label{eq:alphainL3MS}
\delta \alpha\big|_{\overline{\mr{MS}}(\lambda_3)} =
\left. \mr{Re}\frac{\Sigma^{\PH\Ph}(\MH^2)+\Sigma^{\PH\Ph}(\Mh^2)}{2(\MH^2-\Mh^2)}\right|_\mr{UV}
+\frac{f_\alpha(\{ \delta p'_\mr{mass}\})} {\MH^2-\Mh^2}\bigg|_\mr{finite},
\end{align}
where $\delta \lambda_3$ drops out as it has no finite part. The divergent part of $\delta \lambda_3$ can be calculated by solving Eq.~\eqref{eq:alphadef32} and using the knowledge about the divergent parts of $\delta \alpha$ from Eqs.~\eqref{eq:delalphaferm} and \eqref{eq:delalphabos}. This results in 
\begin{align}
\delta \lambda_3= &\biggl[
\frac{e^2 c_{2\alpha}}{4 \MW^2 \sw^2 s_{2\beta}} 
\left(\mr{Re}\,\Sigma^{\PH\Ph}(\MH^2)+\mr{Re}\,\Sigma^{\PH\Ph}(\Mh^2)\right)
-\frac{\delta \beta e^2 s_{2 \alpha} c_{2 \beta} \left(\MH^2- \Mh^2\right) }{2 \MW^2 \sw^2 s_{2 \beta }^2}
\nonumber\\
& -\delta \lambda_5
-\frac{e^2 s_{2 \alpha }}{4 \MW^2 \sw^2 s_{2\beta}}  \left(\delta \Mh^2-\delta \MH^2\right)
-\frac{e^2 \left(\delta \MAO^2-2 \delta \MHP^2\right)}{4 \MW^2 \sw^2}
\nonumber\\
&-\frac{e^2\left(s_{2\beta}\left(\MAO^2-2 \MHP^2\right)+(\Mh^2-\MH^2) s_{2 \alpha }\right)}{2\MW^2 \sw^2 s_{2\beta}}
\left(\delta Z_e- \frac{\delta \sw}{\sw} -\frac{\delta \MW^2}{2 \MW^2} \right)
\nonumber\\
& +\frac{ e^3  \left[
\delta t_\PH \left(s_{\alpha -3 \beta }+3 s_{\alpha +\beta }\right)
+\delta t_\Ph \left(c_{\alpha -3 \beta }+3 c_{\alpha +\beta } \right) 
\right]}{16 \MW^3 \sw^3 s_{2\beta}}\biggr]_{\mr{UV}}.
\label{eq:deltaL3}
\end{align}
The fermionic contribution to $\delta \lambda_3$ is given by
\begin{align}
\nonumber
\delta \lambda_3\big|_\mr{UV,ferm}
& = -\delta \lambda_5\big|_\mr{UV,ferm}
-\Delta_\mr{UV} \frac{3e^4}{32\pi^2\MW^4\sw^4} \sum_i
\left(\xi_\PAO^u-\xi_\PAO^d\right)^2
m_{u,i}^2 m_{d,i}^2
\\ 
&\quad
- \Delta_\mr{UV} \frac{e^4}{64\pi^2\MW^4\sw^4}
\sum_f c_f \left(1+\frac{c_{2\beta}}{s_{2\beta}}\xi_\PAO^f\right) m_f^2
\left[\MAO^2-2\MHP^2+\frac{s_{2\alpha}}{s_{2\beta}}(\Mh^2-\MH^2)\right]
\end{align}
with the massive fermions $f = \Pe, \dots,\Pt$ and the generation index $i$. For the bosonic contribution we obtain
\begin{align}
\label{eq:L3bos}
\delta \lambda_3\big|_\mr{UV,bos}
= \,& \Delta_\mr{UV}\frac{1}{32 \pi^2} \big[(\lambda_1 + \lambda_2) (6 \lambda_3 + 2 \lambda_4) + 4 \lambda_3^2 + 2 \lambda_4^2 + 2 \lambda_5^2
-3 \lambda_3 (3 g_2^2 + g_1^2 )
\nonumber\\
& +\textstyle\frac{3}{4} (3 g_2^4 + g_1^4 - 2 g_2^2 g_1^2)
\big]
\nonumber\\
= &
\,\Delta_\mr{UV} \frac{\lambda_5^2}{2 \pi^2 s_{2\beta}^2}
+ \Delta_\mr{UV} \frac{e^2 \lambda_5}{128 \MW^2 \pi^2 \sw^2 s_{2\beta}^3}\Big[
12 (2 \MW^2 + \MZ^2) s_{2 \beta}^3 
+ \MAO^2 (27 s_{2 \beta} - s_{6 \beta}) 
\nonumber\\
&+ 2 \MHP^2 (-19 s_{2 \beta} + s_{6 \beta}) 
 + \MH^2 (-22 s_{2 \alpha} - 3 s_{2 (\alpha - 2 \beta)} - 8 s_{2 \beta} + s_{2 (\alpha + 2 \beta)}) 
\nonumber\\
& + \Mh^2 (22 s_{2 \alpha} + 3 s_{2 (\alpha - 2 \beta)} - 8 s_{2 \beta} - s_{2 (\alpha + 2 \beta)}) \Big]
\nonumber\\
& + \Delta_\mr{UV} \frac{e^4}{256 \MW^4 \pi^2 \sw^4 s_{2\beta}^3} \Big[
-2 \MH^4 (-3 + c_{2 (\alpha - \beta)} + 2 c_{2 (\alpha + \beta)}) s_{2 \alpha}  
\nonumber\\
& + 4 \Mh^2 \MH^2 (c_{2 (\alpha - \beta)} + 2 c_{2 (\alpha + \beta)}) s_{2 \alpha}
- 2 \Mh^4 (3 + c_{2 (\alpha - \beta)} + 2 c_{2 (\alpha + \beta)}) s_{2 \alpha}  
\nonumber\\
& - \MAO^4 (-7 s_{2 \beta} + s_{6 \beta}) 
- \MAO^2 \MH^2 (11 s_{2 \alpha} + s_{2 (\alpha - 2 \beta)} + 2 s_{2 \beta})
\nonumber\\
& + \MAO^2 \Mh^2 (11 s_{2 \alpha} + s_{2 (\alpha - 2 \beta)} - 2 s_{2 \beta})
+ 12 \MHP^4 s_{2 \beta}^3 
+ 16 \MH^2 \MHP^2 s_{2 \beta} s_{\alpha + \beta}^2
\nonumber\\
&+ 16 \Mh^2 \MHP^2 c_{\alpha + \beta}^2 s_{2 \beta}
+ 2 \MAO^2 \MHP^2 (-11 s_{2 \beta} + s_{6 \beta})
\nonumber\\
&+ 6 (2 \MW^2 + \MZ^2) s_{2 \beta}^2 ((\Mh^2 - \MH^2) s_{2 \alpha} + (\MAO^2 - 2 \MHP^2) s_{2 \beta})
\nonumber\\
&+ 6 (6 \MW^4 - 4 \MW^2 \MZ^2 + \MZ^4) s_{2 \beta}^3
\Big].
\end{align}

\subsubsection{The FJ tadpole scheme}
\label{sec:FJscheme}

Since tadpole loop contribution $T_S$ are gauge dependent \cite{Degrassi:1992ff}, the connection among bare parameters potentially becomes gauge dependent 
if $\delta t_S=-T_S$ enters the relations between bare parameters, as it is the case if renormalized tadpole parameters $t_S$ 
are forced to vanish. Note that these gauge dependences systematically cancel if on-shell renormalization conditions are employed, i.e.\ if predictions 
for observables are parameterized by directly measurable input parameters. If some input parameters are renormalized in the \MSbar{} scheme this cancellation of gauge dependences does not take place anymore in general, and the gauge dependence is ma\-ni\-fest 
in relations between predicted observables and input parameters at NLO.
In the $\overline{\mr{MS}}(\alpha)$ and the $\overline{\mr{MS}}(\lambda_3)$ renormalization schemes, the bare de\-finitions of $\alpha$ and $\beta$ contain tadpole terms 
leading to a gauge dependence (although the $\overline{\mr{MS}}(\lambda_3)$ scheme is gauge 
independent at NLO in $R_\xi$ gauges).
 
Fleischer and Jegerlehner \cite{Fleischer:125234} proposed a renormalization scheme for the SM, referred to as the FJ scheme in the following, that preserves gauge independence for all bare parameters, including the 
masses and mixing angles%
\footnote{A similar scheme, called $\beta_h$~scheme, was suggested in \citere{Actis:2006ra}.
A comparison of that approach to the conventional \MSbar{} and FJ schemes can be found in
\citere{Denner:2016etu}.}.
In this scheme, the parameters are defined in such a way that tadpole terms do not enter the definition of any bare parameter so that all relations among bare parameters remain gauge independent. 
This can be achieved by demanding that {\it bare} tadpole terms vanish,
$t_{S,0}=0$,
for all fields $S$ with the quantum numbers of the vacuum. Since tadpole conditions have no effect on physical observables and change only the bookkeeping, such a procedure is possible. The disadvantage is that now tadpole diagrams have to be taken explicitly into account 
in all higher-order calculations. In particular, the one-particle reducible tadpole contributions destroy the simple relation between propagators and 
two-point functions. In the SM, the FJ scheme does not affect observables if all parameters are 
renormalized using on-shell conditions---as usually done---except 
for the strong coupling constant $\alpha_\mr{s}$, which is, however, directly related to the strong gauge coupling, a model defining parameter.
A gauge-independent renormalization scheme for the THDM can be defined by applying the FJ prescription and imposing the \MSbar{} condition on  mixing angles \cite{Denner:2016etu, Krause:2016oke}.
The {\it bare} physical parameters defined in the FJ scheme 
differ by NLO tadpole contributions (including divergent and finite terms)
from the gauge-dependent definition of the bare parameters $\{p_{\mr{mass}}\}$ given in Eq.~\eqref{eq:physparaNLO}. 
Exceptions are $e$ and the parameter $\lambda_5$, which is a parameter of the basic potential and therefore gauge independent by construction. 
The renormalization of $\lambda_5$
in \MSbar{} is identical to the one in the previous schemes.

It should be noted that in Refs.~\cite{Denner:2016etu, Krause:2016oke} $m_{12}^2$ is chosen as independent
parameter in contrast to our choice of $\lambda_5$. The latter, however, is closer to common practice used in
the MSSM~\cite{Frank2007,Baro:2008bg}.
Moreover, in  Refs.~\cite{Krause:2016oke,Denner:2016etu}
tadpole counterterms are reintroduced by shifting the Higgs fields according to $\eta_i \to \eta_i + \Delta v_{\eta, i}$, $i = 1,2$, where 
the constants $\Delta v_{\eta, i}$ can be chosen arbitrarily, since physical observables do not depend on this shift,
which can be interpreted as an unobservable change of the integration variables in the path integral. In Refs.~\cite{Krause:2016oke,Denner:2016etu}, this freedom of choice is exploited, 
and $\Delta v_{\eta, i}$ 
are chosen in such a way that the fields $\eta_1$, $\eta_2$ do not develop vevs at all orders
. This affects the form of the counterterm Lagrangian and the definition of the renormalization constants with the consequence that the formulae given in Eq.~\eqref{eq:dVH2p} and Sect.\ref{sec:onshellrenconditions} cannot be applied.

We have implemented the FJ scheme following the strategy of Ref.~\cite{Denner:2016etu} by performing the shifts $\eta_i \to \eta_i+ \Delta v_{\eta, i}$ and in an alternative, simpler (but physically equivalent) way. 
In this simplified approach we keep the dependence of the Lagrangian in terms of gauge-dependent masses and couplings. In addition we keep the tadpole renormalization 
condition~\eqref{eq:tadpoleren}, so that the definitions of the renormalization constants of the on-shell parameters and the $Z$ factors according to Sect.~\ref{sec:onshellrenconditions} remain valid (otherwise we   needed to take into account actual tadpole diagrams everywhere).
In this simplified approach the counterterms for $\alpha$ and $\beta$ which reproduce the
results in the FJ scheme result from the previously derived $\delta\alpha$ and $\delta\beta$
by adding appropriate finite terms,
\begin{align}
\delta \alpha\big|_\mr{FJ}&= \delta \alpha+\mbox{finite terms},\nonumber\\
\delta \beta\big|_\mr{FJ} &= \delta \beta +\mbox{finite terms},
\end{align}
which depend on the (finite parts of the) tadpole contributions $T_\PH$ and $T_\Ph$.

Before performing the full calculations, we outline the strategy of the derivation of
those finite terms for $\beta$; for $\alpha$ everything works analogously.
We start by exploiting the fact that the form of the tadpole renormalization cannot
change physical results if all counterterms for independent parameters are determined 
by the same physical conditions. This means, as mentioned above, that we can simply define
the bare tadpoles to vanish, but this forces us to include all explicit tadpole
contributions to Green functions. 
We indicate quantities in this variant by a superscript ``$t$'' in the following. 
We get the same physical predictions in this ``$t$-variant'' if we use the
counterterm
\begin{align}
\delta \beta^t &= \delta \beta +\Delta\beta^t(T_\PH ,T_\Ph)
\label{eq:dbt_def}
\end{align}
instead of $\delta\beta$,
where $\delta\beta^t$ is calculated in the same way as $\delta \beta$, but with
tadpole counterterms omitted and 
explicit tadpole diagrams (including divergent and finite parts)
in the occurring Green functions taken into account.
Note that the \MSbar{} prescription to include only divergent terms, which is employed to define
$\delta \beta$, is not applied to the new tadpole contribution $\Delta\beta^t(T_\PH ,T_\Ph)$.
Otherwise the new $\Delta\beta^t$ terms could not be fully compensated by explicit 
tadpole contributions occuring elsewhere, so that
there would be differences in the renormalized amplitudes.
In fact, applying the \MSbar{} prescription to $\Delta\beta^t(T_\PH ,T_\Ph)$ as well
defines the FJ renormalization scheme,
\begin{align}
\delta \beta^t\big|_\mr{FJ} &= \delta \beta +\Delta\beta^t(T_\PH ,T_\Ph)\big|_\mr{UV}.
\end{align}
The quantity $\delta \beta^t\big|_\mr{FJ}$ is the gauge-independent counterterm for $\beta$
introduced in Ref.~\cite{Denner:2016etu} which is to be used in the $t$-variant,
where all explicit tadpole diagrams are included in Green functions 
(or equivalently are redistributed by the $\Delta v$ shift as described Ref.~\cite{Denner:2016etu}).
We can translate the FJ renormalization prescription back to our 
renormalization scheme (with vanishing renormalized tadpoles) by the counterpart
of Eq.~\refeq{eq:dbt_def}, but now formulated in the FJ scheme,
\begin{align}
\delta \beta^t\big|_\mr{FJ} &= \delta \beta\big|_\mr{FJ} +\Delta\beta^t(T_\PH ,T_\Ph),
\label{eq:dbtFJ_def}
\end{align}
i.e.\ $\delta \beta\big|_\mr{FJ}$ is the counterterm for $\beta$ to be used in our
counterterm Lagrangian in order to calculate renormalized amplitudes in the FJ scheme.
Combining the above formulas, we obtain the finite difference between
$\delta \beta$ in the (gauge-dependent) \MSbar{} scheme and 
$\delta \beta\big|_\mr{FJ}$ in the (gauge-independent) FJ scheme,
\begin{align}
\delta \beta\big|_\mr{FJ} &= \delta \beta^t\big|_\mr{FJ} -\Delta\beta^t(T_\PH ,T_\Ph)
\nonumber\\
&= \delta \beta +\Delta\beta^t(T_\PH ,T_\Ph)\big|_\mr{UV} -\Delta\beta^t(T_\PH ,T_\Ph)
\nonumber\\
&= \delta \beta -\Delta\beta^t(T_\PH ,T_\Ph)\big|_\mr{finite}.
\end{align}

\paragraph{The renormalization constant \boldmath{$\delta  \beta\big|_\mr{FJ}$}:}
We begin our calculation of $\delta\beta|_\mr{FJ}$ with an alternative 
computation of $\delta \beta$ in the \MSbar{}$(\alpha)$ scheme, because 
Eq.~\eqref{eq:vacrencond} cannot be applied in the FJ scheme.
To avoid the use of Eq.~\eqref{eq:vacrencond},
we calculate the counterterm in the \MSbar{}$(\alpha)$ scheme from the field 
renormalization constant by employing the last two equations of Eq.~\eqref{eq:fieldrenconstrel}. 
This results in
\begin{align}
 \delta \beta=\frac{1}{4} \left(\delta Z_{\PGO\PAO}-\delta Z_{\PAO\PGO}\right)\big|_\mr{UV}
=\left.\frac{
2\delta M_{\PGO\PAO}^2
-\mr{Re}\,\Sigma^{\PGO\PAO}(\MAO^2)-\mr{Re}\,\Sigma^{\PGO\PAO}(0)
}{2 \MAO^2}\right|_\mr{UV},
 \label{eq:betaMSaltern}
\end{align}
with $\delta M_{\PGO\PAO}^2$ as given in Eq.~\eqref{eq:Mabchoice2}. In the second step,  $\delta Z_{\PGO\PAO}$ from Eq.~\eqref{eq:mixfieldrendef} and
\begin{align}
\delta Z_{\PAO\PGO}=2 \, \frac{-\delta M_{\PGO\PAO}^2+\mr{Re}\,\Sigma^{\PGO\PAO}(0)}{\MAO^2}
\label{eq:dZAG}
\end{align}
have been used. Equation~\refeq{eq:dZAG} results from demanding finiteness of the
$G_0A_0$ mixing self-energy at zero-momentum transfer, $k^2=0$, but actually any other 
value of $k^2$ would be possible as well, since we only have to remove all UV-divergent terms
in the mixing. 
The non-vanishing tadpole counter\-terms in the \MSbar{}$(\alpha)$ scheme are $\delta t_S=-T_S$.
In the transition to the $t$-variant, $\delta \beta$ gets modified by two kind of terms:
First, there are no tadpole counterterms, i.e.\ the $\delta M_{\PGO\PAO}^2$ term is absent,
and second, there are explicit tadpole contributions to $\Sigma^{\PGO\PAO}$.
This implies
\begin{align}
\Delta\beta^t(T_\PH,T_\Ph) &= \delta\beta^t-\delta\beta
= -\frac{\delta M_{\PGO\PAO}^2}{\MAO^2} 
-\left.\frac{\mr{Re}\,\Sigma^{t,\PGO\PAO}(\MAO^2)+\mr{Re}\,\Sigma^{t,\PGO\PAO}(0)}{2 \MAO^2}\right|_\mr{T_\PH,T_\Ph},
\end{align}
where the superscript ``$t$'' indicates that one-particle-reducible tadpole diagrams are 
included in the self-energies.
The subscript ``$T_\PH,T_\Ph$'' means that only the those explicit tadpole contributions
are taken into account here.
Inserting $\delta M_{\PGO\PAO}^2$ from Eq.~\refeq{eq:Mabchoice2}
and evaluating the (momentum-independent) tadpole diagrams for the $G_0A_0$ mixing,
$\Delta\beta^t$ evaluates to
\begin{align}
\lefteqn{\Delta\beta^t(T_\PH,T_\Ph)} 
\nonumber\\*
&= -\frac{1}{\MAO^2} \left[
\delta M_{\PGO\PAO}^2
+\left.\mr{Re}\,\Sigma^{t,\PAO\PGO}(0)\right|_\mr{T_\PH,T_\Ph} \right]
\nonumber\\
&=-\frac{1}{ \MAO^2}\left[
\delta M_{\PGO\PAO}^2
+ \parbox[c][1cm]{2.8cm}{\includegraphics[scale=1]{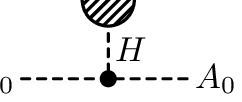}}+\parbox[c][1cm]{2.8cm}{\includegraphics[scale=1]{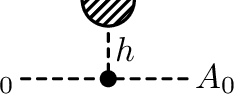}}\right]
\nonumber\\
&=-\frac{e}{2 \MW \sw\MAO^2} \left[
T_\PH s_{\alpha -\beta }
+T_\Ph c_{\alpha -\beta }
+ T_\PH \frac{\left(\MAO^2-\MH^2\right) s_{\alpha -\beta }}{\MH^2} 
+T_\Ph \frac{\left(\MAO^2-\Mh^2\right) c_{\alpha -\beta }}{\Mh^2}
\right]
\nonumber\\
&= -\frac{e}{2 \sw \MW} \left(
T_\PH \frac{ s_{\alpha -\beta }}{\MH^2}
+T_\Ph \frac{ c_{\alpha -\beta }}{\Mh^2} 
\right).
\end{align}
The counterterm $\delta\beta^t\big|_\mr{FJ}$ of the FJ scheme in the $t$-variant,
thus, reads
\begin{align}
\delta \beta^t\big|_\mr{FJ} &= \delta \beta +\Delta\beta^t(T_\PH ,T_\Ph)\big|_\mr{UV}
= \delta \beta 
-\left.\frac{e}{2 \sw \MW} \left(
T_\PH \frac{ s_{\alpha -\beta }}{\MH^2}
+T_\Ph \frac{ c_{\alpha -\beta }}{\Mh^2} 
\right)\right|_\mr{UV},
\label{eq:dbtFJ_expl}
\end{align}
which is in agreement with Ref.~\cite{Krause:2016oke,Denner:2016etu}. 
This translates to our treatment of tadpoles as
\begin{align}
\label{eq:dbFJ_expl}
\delta \beta\big|_\mr{FJ} &= \delta \beta 
-\Delta\beta^t(T_\PH ,T_\Ph)\big|_\mr{finite}
= \delta \beta
+ \left.\frac{e}{2 \sw \MW} \left(
T_\PH \frac{ s_{\alpha -\beta }}{\MH^2}
+T_\Ph \frac{ c_{\alpha -\beta }}{\Mh^2} 
\right)\right|_\mr{finite},
\end{align}
where again ``finite'' means that $\Delta_\mr{UV}$ is set to zero in the tadpole contribution.
Using this counterterm, it is possible to keep the form of the
counterterm Lagrangian derived 
in Sect.~\ref{sec:CTLagrangian}
to obtain results in the gauge-independent
FJ scheme, although the above counterterm Lagrangian employs a gauge-dependent
(but very convenient) tadpole renormalization.

Alternatively, $\delta\beta$ could be fixed by an analogous consideration
of the $\PZ\PAO$ mixing, leading to 
\begin{align}
\delta\beta|_\mr{UV}
=\left[ \frac{s_{2\beta}}{4c_{2\alpha}}\left(\delta Z_{\PH}-\delta Z_{\Ph}\right) 
+ \frac{\Sigma^{\PZ\PAO}(k^2)}{\MZ} \right]_\mr{UV},
\end{align}
which is independent of $k^2$ and does not use relation~\refeq{eq:vacrencond}.
The transition to the FJ scheme then simply amounts to replacing the
one-particle-irreducible self-energy $\Sigma^{\PZ\PAO}$ by
$\Sigma^{t,\PZ\PAO}$, which includes tadpole diagrams. 
The result $\delta \beta^t\big|_\mr{FJ}$ of this procedure
is again given by Eq.~\refeq{eq:dbtFJ_expl}, as it should be.

\paragraph{The renormalization constant \boldmath{$\delta \alpha\big|_\mr{FJ}$}:}
We apply the same method to the renormalization constant 
$\delta \alpha\big|_\mr{FJ}$, starting from Eq.~\refeq{eq:deltaalphaderiv2}.
The difference between $\delta \alpha$ and $\delta \alpha^t$ is entirely given
by the explicit tadpole diagrams that appear in the change from 
$\Sigma^{\PH\Ph}$ to $\Sigma^{t,\PH\Ph}$ in Eq.~\refeq{eq:deltaalphaderiv2},
\begin{align}
\Delta\alpha^t(T_\PH,T_\Ph) &= \delta\alpha^t-\delta\alpha
= \left. \mr{Re}\frac{\Sigma^{t,\PH\Ph}(\MH^2)+\Sigma^{t,\PH\Ph}(\Mh^2)}{2(\MH^2-\Mh^2)}
\right|_\mr{T_\PH,T_\Ph},
\end{align}
which evaluates to
\begin{align}
\Delta\alpha^t(T_\PH,T_\Ph) 
& = \left. \mr{Re}\frac{\Sigma^{t,\PH\Ph}(\Mh^2)}{\MH^2-\Mh^2} \right|_\mr{T_\PH,T_\Ph}
=\frac{1}{\MH^2-\Mh^2}\left[\parbox[c][1cm]{2.5cm}{\includegraphics[scale=1]{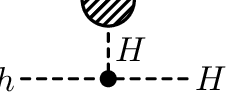}}+
\parbox[c][1cm]{2.5cm}{\includegraphics[scale=1]{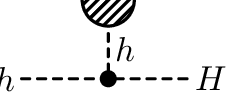}} \right]
\nonumber\\
&=\frac{e}{\MH^2-\Mh^2} \left( 
T_\PH \frac{C_{\Ph\PH\PH}}{\MH^2}
+T_\Ph \frac{C_{\Ph\Ph\PH}}{\Mh^2} 
\right),
\end{align}
with the coupling factors of the $\Ph\PH\PH$ and $\Ph\Ph\PH$ vertices
\begin{subequations}
\begin{align}
C_{\Ph\PH\PH} &=   \frac{e s_{\beta-\alpha} }{2\MW\sw s_{2\beta}}\biggl[ 
-(3s_{2\alpha}+s_{2\beta})
\left(\MAO^2 +4 \lambda_5 \frac{\MW^2 \sw^2}{e^2}\right)
+ s_{2 \alpha} \left(\Mh^2 + 2 \MH^2\right)  \biggr], 
\\
C_{\Ph\Ph\PH} &= \frac{ec_{\beta-\alpha}}{2\MW\sw s_{2\beta}}  \biggl[
(3s_{2\alpha}-s_{2\beta})
\left( \MAO^2 + 4 \lambda_5 \frac{\MW^2 \sw^2}{e^2}\right)
- s_{2\alpha}\left(2 \Mh^2  + \MH^2  \right) \biggr].
\end{align}
\end{subequations}
The counterterm $\delta\alpha^t\big|_\mr{FJ}$ of the FJ scheme in the $t$-variant,
thus, reads
\begin{align}
\delta \alpha^t\big|_\mr{FJ} &= \delta \alpha +\Delta\alpha^t(T_\PH ,T_\Ph)\big|_\mr{UV}
= \delta \alpha 
+\frac{e}{\MH^2-\Mh^2} \left( 
T_\PH \frac{C_{\Ph\PH\PH}}{\MH^2}+
T_\Ph \frac{C_{\Ph\Ph\PH}}{\Mh^2} 
\right)\bigg|_\mr{UV},
\label{eq:datFJ_expl}
\end{align}
which is again in agreement with Ref.~\cite{Krause:2016oke,Denner:2016etu}. 
This translates to our treatment of tadpoles as
\begin{align}
\delta \alpha\big|_\mr{FJ} &= \delta \alpha 
-\Delta\alpha^t(T_\PH ,T_\Ph)\big|_\mr{finite}
= \delta \alpha
+\frac{e}{\Mh^2-\MH^2} \left( 
T_\PH \frac{C_{\Ph\PH\PH}}{\MH^2}
+T_\Ph \frac{C_{\Ph\Ph\PH}}{\Mh^2} 
\right)\bigg|_\mr{finite}.
\end{align}
Concerning the use of $\delta \alpha\big|_\mr{FJ}$ in our counterterm Lagrangian
to obtain renormalized amplitudes in the gauge-independent FJ scheme,
the same comments made above for $\delta\beta\big|_\mr{FJ}$ apply.

\subsubsection{\boldmath{The FJ($\lambda_3$) scheme}}
\label{sec:FJL3scheme}

In the $\overline{\mr{MS}}(\lambda_3)$ scheme, the parameters $\lambda_{3}$ 
and $\lambda_{5}$ are defining parameters of the basic para\-meterization and gauge independent by construction. Therefore, the condition on $\delta \beta$ is the only renormalization condition potentially being gauge dependent. To provide a fully
gauge-independent renormalization scheme where $\lambda_3$ is an independent quantity, we apply the FJ scheme to the parameter $\beta$ and keep the renormalization of 
$\lambda_{3}$ and $\lambda_{5}$ 
as in the $\overline{\mr{MS}}(\lambda_3)$. We call the resulting scheme the FJ($\lambda_3$) scheme. The renormalization of the parameters reads:
\begin{align}
 \delta \beta\big|_\mr{FJ}\text{ as in Eq.~\eqref{eq:dbFJ_expl}},\nonumber\\
 \delta \lambda_3 \text{ as in Eqs.~\eqref{eq:deltaL3}}{-}\refeq{eq:L3bos}.\nonumber
\end{align}

\subsection{Conversion between different renormalization schemes}
\label{sec:inputconversion}

In the previous section, we have
presented four different renormalization schemes, which treat the mixing parameters differently. When observables calculated in different renormalization schemes are compared, particular care has to be taken that the input parameters are consistently translated from one scheme to the other. The bare values of identical independent
parameters are equal and independent of the renormalization scheme. 
Exemplarily, for a parameter $p$, the renormalized values $p^{(1)}$ and $p^{(2)}$ in two different renormalization schemes~1 and 2 are connected via the bare parameter $p_0$, 
\begin{align}
p_0=p^{(1)}+\delta p^{(1)} (p^{(1)})=p^{(2)} +\delta p^{(2)} (p^{(2)}),
\end{align}
within the considered order. 
If $p$ is a dependent parameter in one or both schemes, it must be calculated from the independent renormalized para\-meters and their counterterms from the relations between bare and renormalized quantities. For converting an input value from one scheme to another, one can solve for one renormalized quantity
\begin{align}
\label{eq:convertinputparam}
 p^{(1)}=p^{(2)} +\delta p^{(2)} (p^{(2)})-\delta p^{(1)} (p^{(1)}).
\end{align}
At NLO, this equation can be linearized by substituting the input value of $p^{(1)}$ by $p^{(2)}$ 
in the computation of the last counterterm. The differences to an exact solution are of higher order and beyond our desired NLO
accuracy. However, large counterterms or small tree-level values can spoil the approximation so that in this case a proper solution using numerical techniques could improve the results. 
Another benefit of a full solution of the implicit equation is the possibility that one can 
switch to another scheme and back in a self-consistent way, while start and end scenarios
in scheme~(1) do not exactly coincide when switching from scheme~(1) to (2) and back to (1)
using the linearized approximation.
The comparison of both methods allows for a consistency check of the computation and for an analysis of perturbative stability.
We have derived the Higgs mixing angles $\alpha$ and $\beta$ and their counterterm in all schemes. The finite parts of the gauge-dependent counterterms $\delta \alpha$, $\delta \beta$ are given here for the different renormalization schemes indicated by the respective index: 
\begin{subequations}
\label{eq:fintermsalpha}
\begin{align}
 \delta \alpha\big|_{\overline{\mr{MS}}(\alpha),\mr{finite}} &= 0,\\
 \delta \alpha\big|_{\overline{\mr{MS}}(\lambda_3),\mr{finite}} 
&= \delta \alpha\big|_{\mr{FJ}(\lambda_3),\mr{finite}} 
= \frac{f_\alpha\{\delta p'_\mr{mass}\}}{\MH^2-\Mh^2}\Big|_\mr{finite},\\
 \delta \alpha\big|_{\mr{FJ}(\alpha),\mr{finite}} &= -\Delta\alpha^t(T_\PH,T_\Ph)\big|_\mr{finite}
=\left.\frac{e}{\Mh^2-\MH^2} \left( 
T_\PH \frac{C_{\Ph\PH\PH}}{\MH^2}
+T_\Ph \frac{C_{\Ph\Ph\PH}}{\Mh^2} 
\right)\right|_\mr{finite}.
\end{align}
\end{subequations}
For the angle $\beta$ we obtain the following finite terms 
in the $\overline{\mr{MS}}$ and the FJ schemes,
\begin{subequations}
\label{eq:fintermsbeta}
\begin{align}
 \delta \beta\big|_{\overline{\mr{MS}(\alpha)},\mr{finite}} 
&= \delta \beta\big|_{\overline{\mr{MS}}(\lambda_3),\mr{finite}} 
= 0,\\
\delta \beta\big|_{\mr{FJ}(\alpha),\mr{finite}} 
&= \delta \beta\big|_{\mr{FJ(\lambda_3)},\mr{finite}} 
 = -\Delta\beta^t(T_\PH,T_\Ph)\big|_\mr{finite}=\left.\frac{e}{2 \sw \MW} \left(
T_\PH \frac{ s_{\alpha -\beta }}{\MH^2}
+T_\Ph \frac{ c_{\alpha -\beta }}{\Mh^2} 
\right)\right|_\mr{finite}.
\end{align}
\end{subequations}
%
%
With these formulae we can convert
the input variables for $\alpha$ and $\beta$
easily into each other. 
For instance, the conversion of the input values of $\alpha$ and $\beta$ 
defined in the $\overline{\mr{MS}}(\alpha)$ scheme into 
the other renormalization schemes reads
\begin{subequations}
\begin{align}
\alpha\big|_{\overline{\mr{MS}}(\lambda_3)} 
& = \alpha\big|_{\overline{\mr{MS}}(\alpha)}
+\left.\frac{f_\alpha\{\delta p'_\mr{mass}\}}{\Mh^2-\MH^2}\right|_\mr{finite}, &
\beta\big|_{\overline{\mr{MS}}(\lambda_3)} &=
\beta\big|_{\overline{\mr{MS}}(\alpha)},
\label{eq:convertinput1}%
\\[.3em]
\alpha\big|_{\mr{FJ}(\alpha)}
& = \alpha\big|_{\overline{\mr{MS}}(\alpha)}
+\Delta\alpha^t(T_\PH,T_\Ph)\big|_\mr{finite}, &
\beta\big|_{\mr{FJ}(\alpha)}
& = \beta\big|_{\overline{\mr{MS}}(\alpha)}
+\Delta\beta^t(T_\PH,T_\Ph)\big|_\mr{finite},
\label{eq:convertinput2}%
\\[.3em]
\alpha\big|_{\mr{FJ}(\lambda_3)}
& = \alpha\big|_{\overline{\mr{MS}}(\alpha)}
+\left.\frac{f_\alpha\{\delta p'_\mr{mass}\}}{\Mh^2-\MH^2}\right|_\mr{finite}, & 
\beta\big|_{\mr{FJ}(\lambda_3)}
& = \beta\big|_{\overline{\mr{MS}}(\alpha)}
+\Delta\beta^t(T_\PH,T_\Ph)\big|_\mr{finite}.
\label{eq:convertinput3}%
\end{align}
\end{subequations}%
Within a given scheme, $\lambda_3$ and $\alpha$ can be translated into each other
using the tree-level relation~\refeq{eq:lambda3tree}.
Note that, thus, the numerical values of $\alpha$, $\beta$, and $\lambda_3$ corresponding
to a given physical scenario of the THDM are different in different renormalization schemes.
In turn, fixing the input values in the four renormalization schemes to the same values
corresponds to different physical scenarios.
In particular, this means that the ``alignment limit'', in which $s_{\beta-\alpha}\to1$
so that $\Ph$ is SM like 
{(see, e.g., \citeres{Haber:1995be,Gunion:2002zf,Bernon:2015qea}),} is a notion that depends on the renormalization scheme
(actually even on the scale choice in a given scheme).%
\footnote{In this brief account of results, we do not consider the (phenomenologically
disfavoured, though not excluded) possibility that the heavier CP-even Higgs boson $\PH$ is SM-like,
which is discussed in detail in \citere{Bernon:2015wef}.}

\begin{figure}
\centering
\subfigure[]{
\label{fig:plot_Umrechnung-von-alpha-LM}
\includegraphics{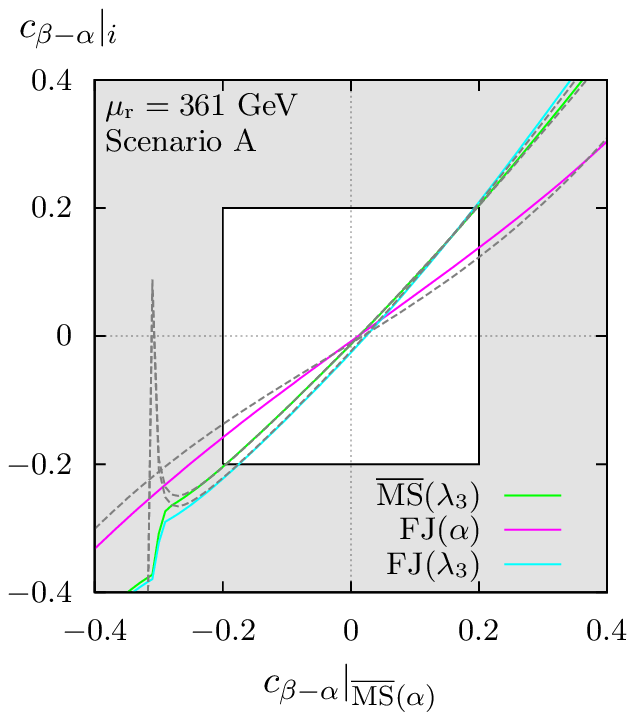}
}
\hspace{15pt}
\subfigure[]{
\label{fig:plot_Umrechnung-nach-alpha-LM}
\includegraphics{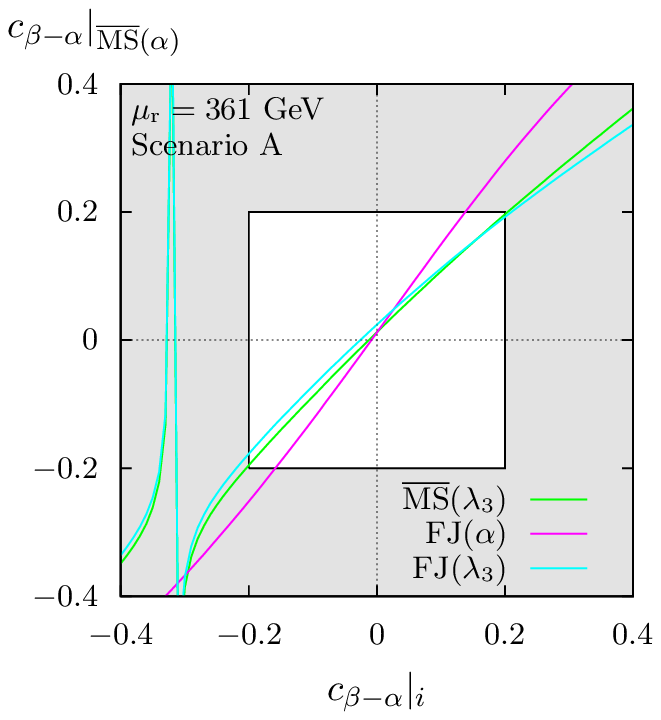}
}
\vspace*{-.5em}
\caption{(a) Conversion of the value of $c_{\beta-\alpha}$ from $\overline{\mr{MS}}(\alpha)$ 
to the $\overline{\mr{MS}}(\lambda_3)$ (green), FJ$(\alpha)$ (pink), and FJ($\lambda_3$) 
schemes (turquoise) for scenario~A. Panel~(b) shows the conversion to the $\overline{\mr{MS}}(\alpha)$ scheme using the same colour coding. 
The solid lines are obtained 
by solving the implicit equations \refeq{eq:convertinputparam} numerically, the dashed lines
correspond to the linearized approximation.
The phenomenologically relevant region is highlighted in the centre.}
\label{fig:plotconversionA}%
\end{figure}

Exemplarily, the 
conversions of $c_{\beta-\alpha}$  from the \MSbar{}$(\alpha)$ scheme into the $\overline{\mr{MS}}(\lambda_3)$ (green), FJ$(\alpha)$ (pink), and 
FJ$(\lambda_3)$ (turquoise) schemes are shown in Fig.~\ref{fig:plot_Umrechnung-von-alpha-LM}. The results of the transformations 
in the inverse directions are displayed in Fig.~\ref{fig:plot_Umrechnung-nach-alpha-LM}, and all other conversions can be seen as a combination of the presented ones. 
The input values (defined before the conversion) correspond to the low-mass scenario called ``A''
of a THDM of Type~I (based on a benchmark scenario of Ref.~\cite{Haber2015})
with  
\begin{align}
\Mh=125 \text{ GeV}, \quad
 \MH=300 \text{ GeV}, \quad \MAO=\MHP=460 \text{ GeV}, \quad \lambda_5=-1.9,\quad \tan \beta=2. 
\label{eq:scenario}
 \end{align}
Specifically, scenario~A is a scan in $c_{\beta-\alpha}$ in the mass parameterization, 
Aa and Ab are points of the scan region used to analyze the scale dependence:
\begin{subequations}
\begin{align}
\mbox{A:}  \quad \cos{(\beta-\alpha)} &= -0.2\ldots0.2, \\
\mbox{Aa:} \quad \cos{(\beta-\alpha)} &= +0.1, \\
\mbox{Ab:} \quad \cos{(\beta-\alpha)} &= -0.1. 
\end{align}
\label{eq:cba_A}
\end{subequations}
The \MSbar{} parameters are defined at the scale
\begin{align}
\label{eq:centralscale}
 \mu_0=\frac{1}{5}(\Mh+\MH+\MAO+2\MHP).
\end{align}
The motivation for this choice will become clear below.
The remaining input parameters for the SM part are given in 
App.~\ref{app:smparams}.
In both plots, we highlight the phenomenologically relevant region in the centre. 
The solid lines are the result obtained 
by solving the implicit equations \refeq{eq:convertinputparam} numerically,
the dashed lines correspond to a linearized conversion.
All curves show only minor conversion effects in the parameter values, i.e.\ the 
solution of the implicit equations
agrees well with the approximate linearized conversion, affirming that the contributions of the
higher-order functions $\Delta\alpha^t$, $\Delta\beta^t$, and $f_\alpha$ of 
Eqs.~\refeq{eq:convertinput1}--\refeq{eq:convertinput3}
are small, and perturbation theory is applicable. Since the values of the parameters change when going from one renormalization scheme to another, 
the alignment limit does not persist in these transformations, i.e.\
in this scenario the alignment limit sensitively depends on the definition of the parameters at NLO.

{For the schemes with $\lambda_3$ as input parameter, some singular behaviour in the parameter
conversion can be observed in the phenomenologically disfavoured region where $c_{\beta-\alpha}\lsim-0.3$.
This artifact in the conversion appears when $c_{2\alpha}\to0$ (see, e.g., Eq.~\refeq{eq:MHhdependence}),
indicating the breakdown of the $\overline{\mr{MS}}(\lambda_3)$ and FJ($\lambda_3$) schemes in such
parameter regions. Already this case-specific study shows that stability issues of different
renormalization schemes have to be carefully carried out for all interesting parameter regions
and that the applicability of a specific scheme in general does not cover the full THDM 
parameter space, a fact that was also pointed out in \citere{Krause:2016oke}
for the THDM and that is known from NLO calculations in the MSSM (see, e.g., \citeres{Heinemeyer:2010mm,Chatterjee:2011wc}).
Specifically, if the $\overline{\mr{MS}}(\lambda_3)$ and FJ($\lambda_3$) schemes are not
applicable in a region that might be favoured by future data analyses, it would
be desirable and straightforward to replace $\lambda_3$ by $\lambda_1$ or $\lambda_2$
as independent parameter, thereby defining analogous schemes like $\overline{\mr{MS}}(\lambda_1)$, etc..

To address this issue properly, was our basic motivation to introduce and compare different renormaliztion
schemes. We will continue this discussion in more detail in a forthcoming publication,
where further THDM scenarios are considered.}

\section{\boldmath{The running of the \MSbar{} parameters}}
\label{sec:running}

Parameters renormalized in the $\overline{\mr{MS}}$ scheme depend on an unphysical renormalization scale $\mu_\mr{r}$. The one-loop $\beta$-function of a parameter $p$  can be obtained from the UV-divergent parts of its 
counterterm $\delta p$, 
\begin{align}
\beta_p(\mu^2_\mr{r})=\frac{\partial}{\partial \ln{\mu_\mr{r}^2}} p(\mu_\mr{r}^2)
=\frac{\partial}{\partial \Delta_\mr{UV}} \delta p.
\label{eq:betafct}
\end{align}
Since the renormalization constants are computed in a perturbative manner, the $\beta$-functions have a perturbative expansion in the coupling parameters.
Note that the last equality in Eq.~\refeq{eq:betafct} holds in the FJ schemes
for $\alpha$ and $\beta$
only in the $t$-variant explained above, because the finite contributions
$\Delta\alpha^t$ and $\Delta\beta^t$ depend on the scale $\mu_\mr{r}$.

As discussed in the previous sections, 
the ratio of the vevs, $\tan \beta$, the Higgs mixing parameter $\alpha$ or $\lambda_3$, and the Higgs self-coupling $\lambda_5$ are renormalized in the $\overline{\mr{MS}}$ scheme. For each renormalization scheme described in Sect.~\ref{sec:diffrenschemes}, one obtains a set of coupled RGEs involving the $\beta$-functions of the independent parameters. Therefore, the scale dependence varies when different schemes are applied. In the perturbative expansion of the $\beta$-function we consider only the one-loop term, being second order in the coupling constants,
e.g., in the \MSbar$(\alpha)$ scheme
\begin{align}
\beta_p(\mu^2_\mr{r})= A_p \alpha_\mr{em}+ B_p \lambda_5+C_p \lambda_5^2/\alpha_\mr{em}.
\end{align}
The dependence on the strong coupling constant vanishes at one-loop order as the parameters renormalized in \MSbar{} appear only in couplings of particles that do not interact strongly. The coefficients $A_p, B_p, C_p$ of the respective renormalized parameter can be easily read 
from the divergent terms which have been derived in the previous section. We have checked them against the $\beta$-functions given for $\lambda_{3}$ and $\lambda_{5}$ in Ref.~\cite{Branco:2011iw} and for $\beta$ in Ref.~\cite{Sperling:2013eva} (supersymmetric contributions need to be omitted).

In general, RGEs, which are a set of coupled differential equations, cannot be solved analytically. 
Usually numerical techniques, such as a Runge--Kutta method, need to be employed to solve the RGEs and to compute the values of the parameters at a desired scale.
Moreover, we emphasize that the renormalization-group flow of a running parameter
depends on the renormalization scheme of the full set of independent parameters.
That means the fact that we use on-shell quantities, such as all the Higgs-boson masses,
to fix most of the scalar self-couplings has a significant impact on the running
of our \MSbar{} parameters.
The renormalization-group flow in other schemes was, e.g., investigated in
\citeres{Cvetic:1997zd,Branco:2011iw,Bijnens:2011gd,Chakrabarty:2014aya,Ferreira:2015rha}.

\begin{figure}
  \centering
  \subfigure[]{
\label{fig:running_cba0.1-LM}
\includegraphics{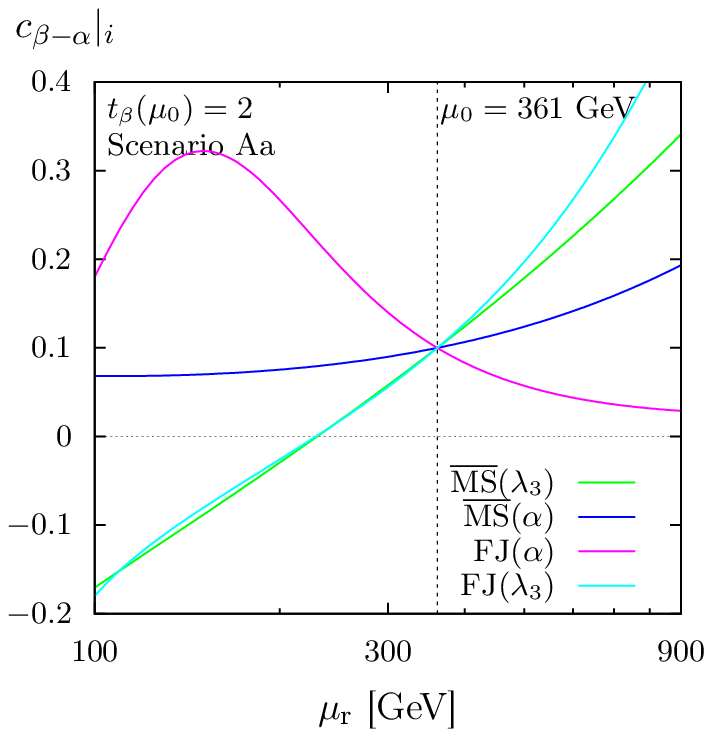}
}
\hspace{15pt}
\subfigure[]{
\label{fig:running_cba-0.1-LM}
\includegraphics{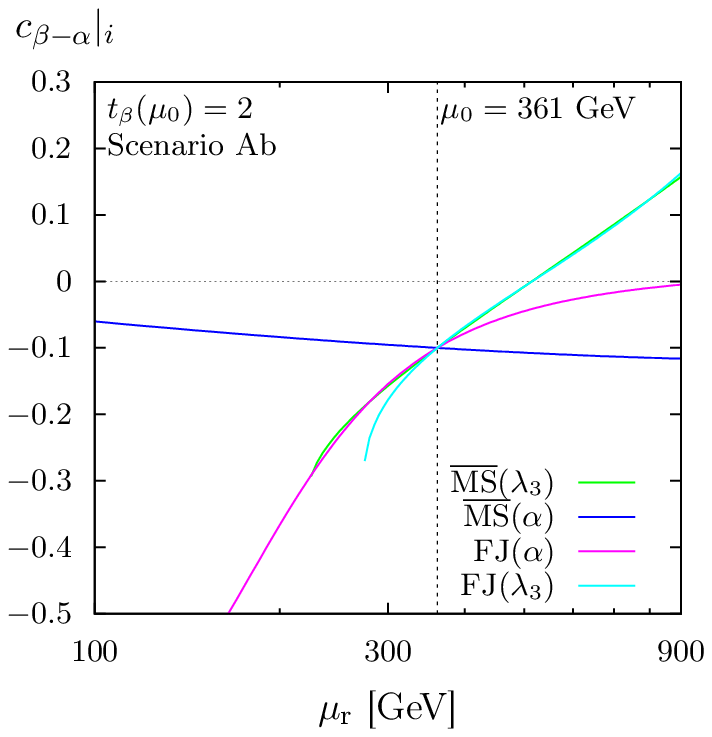}
}
\vspace*{-.5em}
  \caption{The running of $c_{\beta-\alpha}$ for the low-mass scenario~A with $c_{\beta-\alpha}=0.1$ (a) and $c_{\beta-\alpha}=-0.1$ (b) in the $\overline{\mr{MS}}(\alpha)$ (blue), $\overline{\mr{MS}}(\lambda_3)$ (green), FJ($\alpha)$ (pink), and FJ($\lambda_3$) (turquoise) schemes.}
\label{fig:plotrunningA}
\end{figure}

The scale dependence of $c_{\beta-\alpha}$ 
for $\mu = 100{-}900$~GeV is plotted in Fig.~\ref{fig:plotrunningA}, for the scenario defined in Eq.~\eqref{eq:scenario} with $c_{\beta-\alpha}=0.1$  (l.h.s) and $c_{\beta-\alpha}=-0.1$ (r.h.s) and input values given at the central scale $\mu_0$ stated in
Eq.~\refeq{eq:centralscale}. 
We observe that the choice of  the renormalization scheme has a large impact on the scale dependence. While the $\overline{\mr{MS}}(\alpha)$ scheme introduces only a mild running, the other schemes show a much stronger scale dependence, so that excluded and unphysical values of input parameters can be reached quickly. A similar observation has also been made in supersymmetric models for the parameter~$\tan\beta$~\cite{Freitas:2002um}. 
Gauge-dependent $\overline{\mr{MS}}$ schemes have a small scale dependence while replacing the parameters by gauge-independent ones like in the FJ schemes introduce additional terms in the $\beta$-functions which induce a stronger scale dependence.
In Fig~\ref{fig:running_cba-0.1-LM} one can also see that the curves for the $\overline{\mr{MS}}(\lambda_3)$ and the FJ($\lambda_3$) schemes terminate around 250 GeV. 
At this scale, the running of $\lambda_3$ yields unphysical values for which 
Eq.~\refeq{eq:lambda3tree}
with the given Higgs masses becomes 
overconstrained, and no solution with $|s_{2\alpha}|\le1$ exists. This is unique to the $\lambda_3$ running as only there an 
implicit equation needs to be solved to obtain the input parameter $\alpha$. For the other cases we prevent the angles from running out of their domain of definition by solving the running for the tangent function of the angles.

\section{Implementation into a \textsc{FeynArts} Model File}
\label{sec:FAmodelfile}
The \textsc{Mathematica} package \textsc{FeynRules} (FR)~\cite{Christensen2009} 
is a tool to generate Feynman rules from a given 
Lagrangian, providing the possibility to produce model files in various output formats
which can be employed by automated amplitude generators. 
We have inserted the Lagrangian into FR in its internal notation to obtain
the corresponding counter\-term Lagrangian after the renormalization transformations. 
Before this insertion, we have computed and simplified  
the Higgs potential and the corresponding counterterm potential~\eqref{eq:dVH2p} 
with inhouse \textsc{Mathematica} routines.
Using FR,
the tree-level and the counterterm Feynman rules as well as the renormalization conditions 
in the \MSbar{}$(\alpha)$ and \MSbar{}($\lambda_3$) schemes have been implemented into 
a model file for the amplitude generator \textsc{FeynArts} (FA)~\cite{Hahn2001}. 
The renormalization conditions of the FJ($\alpha)$ and the FJ($\lambda_3$) have not 
been included in the model file, because using 
Eqs.~\eqref{eq:fintermsalpha},~\eqref{eq:fintermsbeta} it is straightforward to 
implement the corresponding finite terms of $\delta \alpha$ and $\delta \beta$. 
With such a model file, NLO amplitudes for any process can be generated in an automated way. 

The FA NLO model file for the THDM, obtained with FR, has the following features:
\begin{itemize}
\item Type I, II, flipped, or lepton-specific THDM;
\item all tree-level and counterterm Feynman rules;
\item renormalization conditions according to the \MSbar{}$(\alpha)$ and \MSbar{}$(\lambda_3)$  schemes;
\item all renormalization constants are implemented additionally in $\overline{\mr{MS}}$ 
as well, which allows for fast checks of UV-finiteness;
\item BHS and HK conventions;
\item
CKM matrix set to the unit matrix (the generalization is straightforward).
\end{itemize}
This model file has been tested intensively, including checks of UV-finiteness for 
several processes, both numerically and analytically. 
This allows for the generation of amplitudes 
(and further processing with \textsc{FormCalc} \cite{Hahn1999}) for any process at the one-loop level, at any parameter point of the THDM.
The model file can be obtained from the authors upon request.

\section{\boldmath{Numerical results for \boldmath$ \Ph \to \PW\PW/\PZ\PZ \to 4f$}}
\label{se:numerics}

In this section we present first results from the computation of the decay of the light, neutral 
CP-even Higgs boson of the THDM into four fermions at NLO. The computer program 
\textsc{Prophecy4f}~\cite{Bredenstein:2006rh,Bredenstein:2006nk,Bredenstein:2006ha}%
\footnote{\tt http://prophecy4f.hepforge.org/index.html}
provides a ``\textbf{PROP}er description of the \textbf{H}iggs d\textbf{EC}a\textbf{Y} into \textbf{4 F}ermions'' and calculates 
observables for the decay process $\Ph {\to} \PW\PW/\PZ\PZ {\to} 4 f$ at NLO EW+QCD in the SM. 
We have extended this program to the calculation of the corresponding decay in the THDM in such a way that the usage of the program and its applicability as event
generator basically remains the same. 
{Owing to the fact that LO and real-emission amplitudes in the THDM receive only the
multiplicative factor $s_{\beta-\alpha}$ with respect to the SM, the bremsstrahlung corrections as well
as the treatment of infrared singularities could be taken over from the SM 
calculation~\cite{Bredenstein:2006rh,Bredenstein:2006ha}
via simple rescaling.}
The calculation in the THDM, the implementation in  \textsc{Prophecy4f}, as well as results of the application will be described in detail in an upcoming publication.
We just mention that we employ the complex-mass scheme~\cite{Denner:2005fg} to describe the 
W/Z~resonances, as already done in \citeres{Bredenstein:2006rh,Bredenstein:2006nk,Bredenstein:2006ha}
for the SM. Note that the W/Z-boson masses as well as the weak mixing angle are consistently taken as complex quantities in the complex-mass scheme to guarantee gauge invariance of all amplitudes
in resonant and non-resonant phase-space regions. 
Consequently all our renormalization constants of the THDM inherit imaginary parts from the complex
input values, but the impact of these spurious imaginary parts is beyond NLO and negligible
(as in the SM). Moreover, we mention that the modified version of \textsc{Prophecy4f}
makes use of the public \textsc{Collier} library~\cite{Denner:2016kdg} for the calculation 
of the one-loop integrals.
{Apart from performing two independent loop calculations, we have verified
our one-loop matrix elements by numerically comparing our results to
the ones obtained in \citere{Denner:2016etu} for the related
$\PW\Ph/\PZ\Ph$ production channels (including $\PW/\PZ$ decays)
using crossing symmetry.}

In this paper, we present first results in order to demonstrate the use and the 
self-consistency of our renormalization schemes, employing again the scenario
inspired by the first benchmark scenario of Ref.~\cite{Haber2015} where the additional 
Higgs bosons are not very heavy. 
The input values of the THDM parameters 
{for a Type~I THDM}
are given in Eqs.~\refeq{eq:scenario} and 
\refeq{eq:cba_A}.
Since $c_{\beta-\alpha}$ is the only parameter of the THDM appearing at LO, 
our process is most sensitive to this parameter. We vary $c_{\beta-\alpha}$ 
in the range $[-0.2,+0.2]$ in scenario~A for the computation of 
the partial decay width for $\Ph\to\PW\PW/\PZ\PZ\to4f$, $\Gamma^{\Ph\to4f}_\mr{THDM}$,
which is obtained by summing the partial widths of the h~boson
over all massless four-fermion final states $4f$.
The parameters of the SM part of the THDM are collected in App.~\ref{app:smparams}.
{Note that a non-trivial CKM matrix would not change our results,
since quark mass effects of the first two generations as well as mixing
with the third generation are completely negligible in the considered
decays.}

To perform scale variations we take two distinguished points named Aa and Ab with $c_{\beta-\alpha}=\pm0.1$. 
For the central renormalization scale we use the average mass 
$\mu_0$ defined in Eq.~\refeq{eq:centralscale}
of all scalar degrees of freedom.
The scale $\mu$ of $\alpha_\mr{s}$ is kept fixed at $\mu=\MZ$ which is the appropriate scale for the QCD corrections (which are dominated by the hadronic W/Z decays).

\subsection{Scale variation of the width}
\label{sec:LowMassscalevariation}

\begin{figure}
  \centering
  \subfigure[]{
\label{fig:plot_MUSCAN-Aa-L3MS}
\includegraphics{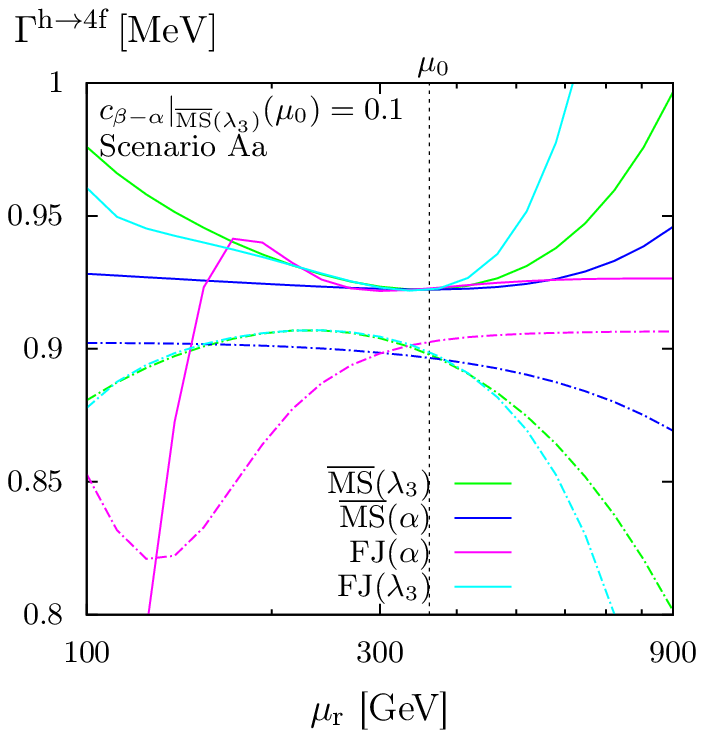}
}
\hspace{15pt}
  \subfigure[]{
\label{fig:plot_MUSCAN-Ab-alphaMS}
\includegraphics{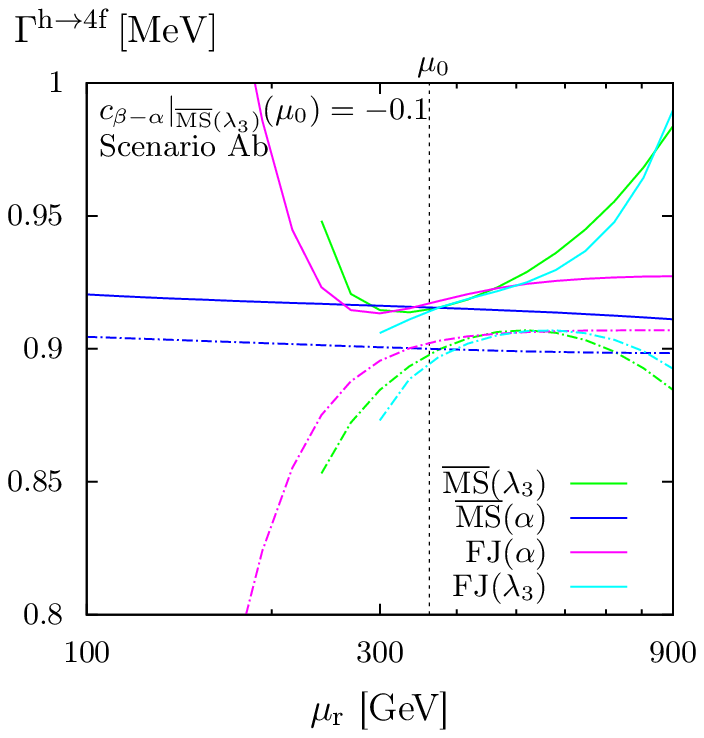}
}
\caption{The decay width for $\Ph \to 4f$ at LO 
(dashed) and NLO EW (solid) 
in dependence of the renormalization scale with $\beta$ and $\alpha$ defined in the $\overline{\mr{MS}}(\lambda_3)$ scheme. The result is computed in all four different renormalization schemes after converting the input at NLO (also for the LO curves)
and displayed for the benchmark points Aa~(a) and Ab~(b) using the colour code of Fig.~\ref{fig:plotrunningA}.  }
\label{fig:plotmuscan}
\end{figure}

{The running of the \MSbar-renormalized parameters $\alpha$ and $\beta$ is
induced by the Higgs-boson self-energies (and some scalar vertex for $\lambda_5$),
i.e.\ the relevant particles in the loops are all Higgs bosons, the W/Z bosons,
and the top quark. If all Higgs-boson masses are near the electroweak scale,
say $\sim 100{-}200\GeV$, 
where the W/Z-boson and top-quark masses are located, 
then the scale $\Mh$ turns out to be a reasonable scale, as expected.
{However, if some heavy Higgs-boson masses increase to some generic mass
scale $M_S$ and the mixing angle
$\beta-\alpha$ stays away from the alignment limit, there is no 
decoupling of heavy Higgs-boson effects, so that $M_S$ acts
as generic UV cutoff scale appearing in logarithms $\log(M_S/\mu_\mr{r})$.
The renormalization scale $\mu_\mr{r}$ has to go up with $M_S$ to
avoid that the logarithm drives the correction unphysically large.
The optimal choice of $\mu_\mr{r}$, though, is somewhat empirical.}
A good choice of the central scale $\mu_0$ should come close to the
stability point (plateau in the $\mu_{r}$ variation) in the major part of THDM
parameter space. Our choice \eqref{eq:centralscale} of $\mu_0$ 
effectively takes care of this and is eventually justified by the numerics.

To illustrate this and to estimate the theoretical uncertainties due to the
residual scale dependence, we compute the total width while the scale $\mu_\mr{r}$ is varied from $100{-}900\GeV$.}
{Results with central scale $\Mh$ are shown in App.~\ref{app:mu0mh},
proving that this would be not a good choice.}
The parameters $\alpha$ and $\beta$ are defined in the $\overline{\mr{MS}}(\lambda_3)$ scheme, and to compute results in other renormalization schemes their values are converted using
Eqs.~\refeq{eq:convertinput1}--\refeq{eq:convertinput3},
which are solved numerically without linearization.
Thereafter the scale is varied, the RGEs solved, and the width computed using the respective renormalization scheme. 
The results are shown 
in Fig.~\ref{fig:plotmuscan} at LO (dashed) and NLO EW (solid) for the benchmark points Aa and Ab.
The QCD corrections are not part of the EW scale variation and therefore omitted in these results. The benchmark point Aa shows almost textbook-like behaviour with the LO computation 
exhibiting a strong scale dependence for all renormalization schemes, resulting in sizable differences between the curves. However, each of the NLO curves shows a wide extremum with a large plateau, reducing the scale dependence drastically, as it is expected for NLO calculations. 
The central scale $\mu_\mr{r}=(\Mh+\MH+\MAO+2\MHP)/5$ lies perfectly in the middle of the plateau regions motivating this scale choice. In contrast, the naive scale choice $\mu_0=\Mh$ is not within the plateau region, leads to large, unphysical corrections, and should not be chosen. The breakdown of the FJ($\alpha)$ curve for small scales can be explained by the running which becomes unstable for these values (see Fig.~\ref{fig:running_cba0.1-LM}).
For all renormalization schemes, the plateaus coincide and the agreement between the renormalization schemes is improved at NLO w.r.t.~the LO results. This is expected, since results obtained with different  renormalization schemes should be equal up to higher-order terms, after the input parameters are properly converted. 
The relative renormalization scheme dependence at the central scale,
\begin{align}
\Delta_\mr{RS}=2 \,\frac{\Gamma^{\Ph \to 4f}_\mr{max}(\mu_0)-\Gamma^{\Ph \to 4f}_\mr{min}(\mu_0)}{\Gamma^{\Ph \to 4f}_\mr{max}(\mu_0)+\Gamma^{\Ph \to 4f}_\mr{min}(\mu_0)},
\label{eq:renschemedep}
\end{align}
 expresses the dependence of the result on the renormalization scheme. It can be computed from the difference of the smallest and largest width 
in the four renormalization schemes normalized to their average. 
In the calculation of $\Delta_\mr{RS}$, the full NLO EW+QCD corrections to the width
$\Gamma^{\Ph \to 4f}$ should be taken into account.
In Tab.~\ref{tab:schemevar}, $\Delta_\mr{RS}$ is given at LO and NLO and confirms the reduction of the scheme dependence in the NLO calculation. In addition, as already perceived when the running was analyzed, the $\overline{\mr{MS}}(\alpha)$ scheme shows the smallest dependence on the renormalization scale, which attests a good absorption of further corrections into the NLO prediction.\\
  \begin{table}
   \centering
   \renewcommand{\arraystretch}{1.2}
 \begin{tabular}{|c|cc|}\hline
  & Scenario Aa & Scenario Ab\\
    $\Delta^\mr{LO}_\mr{RS}$[\%]       & 0.67& 0.84  \\
    $\Delta^\mr{NLO}_\mr{RS}$ [\%]     & 0.08 & 0.34  \\\hline
  \end{tabular}  
   \caption{The variation $\Delta_\mr{RS}$
of the $\Ph {\to} 4f$ width using different renormalization schemes 
for input parameters defined in the $\overline{\mr{MS}}(\lambda_3)$ scheme.}
 \label{tab:schemevar}
 \end{table}%
The situation for the benchmark point Ab is more subtle. For negative values of $c_{\beta-\alpha}$ the truncation of the schemes involving $\lambda_3$ at 
$\mu_\mr{r}=250{-}300$ GeV as well as the breakdown of the running of the FJ($\alpha)$ scheme, which both were observed in the running in Fig.~\ref{fig:running_cba-0.1-LM}, are also ma\-ni\-fest in the computation of the $\Ph {\to} 4f$ width.
Therefore, the results vary much more, and the extrema with the plateau regions are not as distinct as for the benchmark point Aa. They are even missing for the truncated curves. Nevertheless, the situation improves at NLO. 
As for scenario~Aa, the central scale choice of $\mu_0$ 
is more appropriate in contrast than the choice of $\Mh$.
 
For both benchmark points, 
the estimate of the theoretical uncertainties by varying the scale by a factor of two from the central value for an arbitrary renormalization scheme is generally not appropriate. 
A proper strategy would be to identify the renormalization schemes which yield reliable results, and to use only those to quantify the theoretical uncertainties from the scale variation. In addition, the renormalization scheme dependence of those schemes should be investigated. 
This procedure should be performed for different parameter regions (and corresponding 
benchmark points) separately, which is beyond the scope of this work. 
 
\subsection{\boldmath{$c_{\beta-\alpha}$} dependence}
\label{sec:cbascanA}

\begin{figure}
\centering
\includegraphics{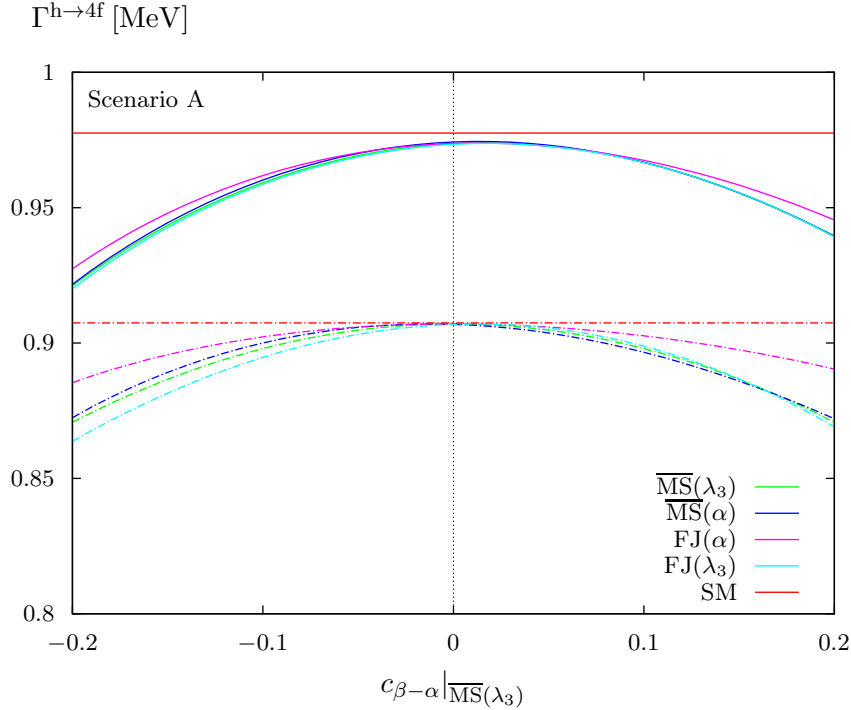}
\caption{The decay width for $\Ph \to 4f$ at LO 
(dashed) and full NLO EW+QCD (solid) for scenario A in dependence of $c_{\beta-\alpha}$. 
The input values are defined in the  $\overline{\mr{MS}}(\lambda_3)$ scheme and 
are converted to the other schemes at NLO (also for the LO curves). 
The results computed with different renormalization schemes are displayed with the colour code of Fig.~\ref{fig:plotrunningA}, and the SM 
(with SM Higgs-boson mass $\Mh$) is shown for comparison in red.}
\label{fig:plotscbascanA}
\end{figure}

\begin{figure}
  \centering
\includegraphics{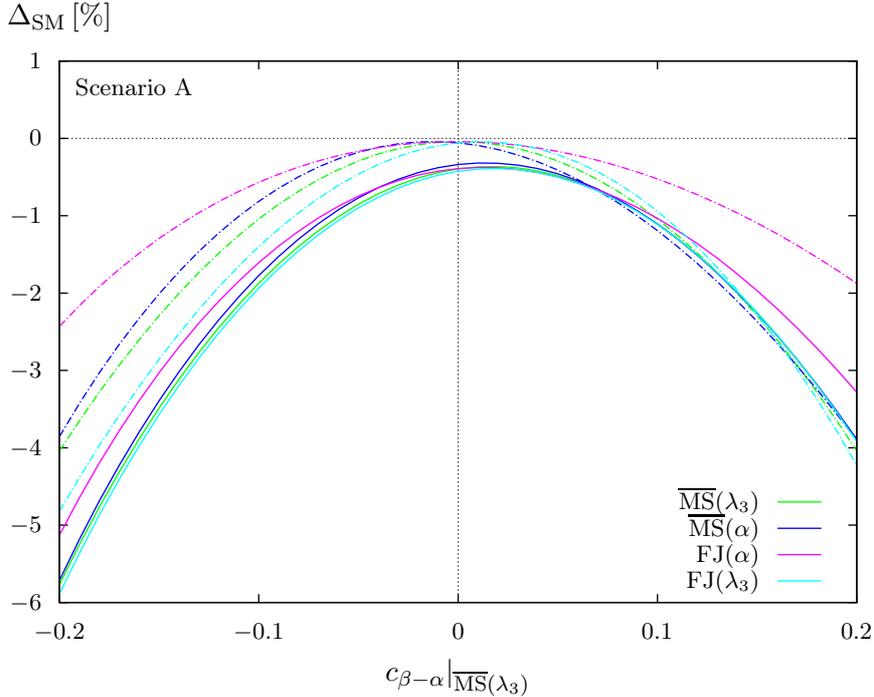}
  \caption{The relative difference of the decay width for $\Ph \to 4f$ in the THDM w.r.t.\ the SM prediction at LO (dashed) and NLO EW+QCD (solid). 
The input values are defined in the  $\overline{\mr{MS}}(\lambda_3)$ scheme and 
are converted to the other schemes at NLO (also for the LO curves). 
The results computed with different renormalization schemes are displayed with the colour code of Fig.~\ref{fig:plotrunningA}.}
\label{fig:plot_cbascanrelSM-A-diffschemes-L3MS}
\end{figure}

The decay width for $\Ph \to 4f$ in dependence of $c_{\beta-\alpha}$ in scenario~A is presented in Fig.~\ref{fig:plotscbascanA} for all renormalization schemes with 
the input values $\alpha$ and $\beta$ defined in the $\overline{\mr{MS}}(\lambda_3)$ scheme. 
The LO (dashed) and the full NLO EW+QCD total widths (solid) are computed in the different renormalization schemes after the NLO input conversion 
(without linearization) and using the constant default scale $\mu_0$ of Eq.~\eqref{eq:centralscale}. The SM values are illustrated in red. 
At tree level the widths show the suppression w.r.t. to the SM with the factor $s_{\beta-\alpha}^2$ originating from the 
$\PH\PW\PW$ and $\PH\PZ\PZ$ couplings. The differences between the renormalization schemes are due to the conversion of the input. As the conversion induces NLO differences in the LO results, a pure LO computation is identical for all renormalization schemes as the conversion vanishes at this order 
and is represented by the LO curve of the $\overline{\mr{MS}}(\lambda_3)$ scheme. The suppression w.r.t.\ the SM computation does not change at NLO, while the shape becomes slightly asymmetric,
and the NLO results show a significantly better agreement between the renormalization schemes.  
Deviations of the THDM results from the SM expectations can be investigated when the SM Higgs-boson mass is identified with the mass 
$\Mh$ of the light CP-even Higgs boson~h 
of the THDM. The relative deviation of the full width from the SM is then
\begin{align}
 \Delta_\mr{SM}= \frac{\Gamma_\mr{THDM}-\Gamma_\mr{SM}}{\Gamma_\mr{SM}},
\end{align}
which is shown in Fig.~\ref{fig:plot_cbascanrelSM-A-diffschemes-L3MS} at LO (dashed) and NLO (solid) in percent for parameters defined in the the $\overline{\mr{MS}}(\lambda_3)$ scheme.
The SM exceeds the THDM widths at LO and NLO. 
The LO  shape which is just given by 
$c_{\beta-\alpha}^2$ shows minor distortions due to the parameter conversions. At NLO, the shape is slightly distorted by an asymmetry of the EW corrections, and a small offset of $-0.5$\% is visible even in the alignment limit where the diagrams including heavy Higgs bosons still contribute. The NLO computations show larger negative deviations, and this could be used to improve current exclusion bounds or increase their significance. Nevertheless, in the whole scan region the deviation from the SM is within 6\% and for phenomenologically most interesting region with $|c_{\beta-\alpha}|<0.1$ even less than 2\%, which is challenging for experiments to measure.

\section{Conclusions}
\label{se:conclusion}

Confronting experimental results on Higgs precision observables with theory predictions
within extensions of the SM, provides an important alternative to search for
physics beyond the SM, in addition to the search for new particles.
The THDM comprises an extended scalar sector with regard to the SM Higgs sector and allows for a comprehensive study of the impact of new scalar degrees of freedom without introducing new fundamental symmetries or other new theoretical structures. 

In this article, we have considered the Type I, II, lepton-specific, and flipped versions of the 
THDM. We have introduced four different renormalization schemes which employ directly
measurable parameters such as masses as far as possible and make use of fields that 
directly correspond to mass eigenstates.
In all the schemes, the masses are defined via on-shell conditions, the electric charge is fixed via the Thomson limit, and the coupling $\lambda_5$ is defined with the $\overline{\mr{MS}}$ prescription. The fields are also defined on-shell which is most convenient in applications. The renormalization schemes differ in the treatment of the coupling $\lambda_3$ and the mixing angles $\alpha$ and $\beta$: In the $\overline{\mr{MS}}(\alpha)$ scheme, $\alpha$ and $\beta$ are renormalized using $\overline{\mr{MS}}$ conditions. In the $\overline{\mr{MS}}(\lambda_3)$ scheme instead $\lambda_3$ and $\beta$ are $\overline{\mr{MS}}$-renormalized parameters. In addition to the conventional treatment of tadpole contributions, we have
implemented an alternative prescription suggested by Fleischer and Jegerlehner where the mixing angles $\alpha$ and $\beta$ obtain extra terms of tadpole contributions, rendering these schemes gauge independent to all orders. 
It should, however, be noted that the $\overline{\mr{MS}}(\lambda_3)$ scheme is also gauge independent at NLO in the 
class of $R_\xi$ gauges. 
We have also discussed relations to renormalization procedures
suggested in the literature for the THDM. 

A comparison of these four different renormalization schemes allows for testing 
the perturbative consistency, and, for the parameter regions and renormalization schemes fulfilling this test,  estimating the theoretical uncertainty due to the truncation of the perturbation series.
To further investigate the latter,
we have investigated the scale dependence, solving the corresponding RGEs. 
One important observation is that it is crucial to be very careful and specific 
about the definitions of the parameters applied, i.e.\ the declaration of the renormalization
scheme of the parameters is vital if one aims at precision.
This is already relevant in the formulation of
benchmark scenarios, because a conversion to a different scheme might alter the physical properties of the scenario significantly. For example, the alignment limit 
may be reached with one specific set of parameters defined in a specific renormalization scheme, but converting these parameters consistently to parameters in a different renormalization scheme might shift the parameters away from the alignment limit.
 
The different renormalization schemes have been implemented into
a \textsc{FeynArts} model file and are thus ready for applications.%
\footnote{The model file is restricted to a unit CKM matrix, but can be generalized
to a non-trivial CKM matrix exactly as in the SM.}
As a first example, we have applied and tested the different schemes in the calculation of the decay width of a 
light CP-even Higgs boson decaying into four massless fermions.
We discuss the dependence of the total $\Ph{\to}4f$ decay width on the renormalization scale
and advocate a scale that is significantly higher than
the naive choice of $\mu_\mr{r} = \Mh$, taking care of the different mass scales in the
THDM Higgs sector.
In addition, results for various values of $\cos (\beta-\alpha)$, a parameter entering the prediction already at LO, are presented. The deviations of the SM are relatively small, in the phenomenologically interesting region they are about 
$2{-}6\%$---a challenge for future measurements. 
 
The detailed
description of the calculation of the decay width in the THDM 
and a survey of numerical results will be given in a forthcoming paper. 
This includes a deeper investigation in the renormalization scale dependence and the comparison of different renormalization schemes 
for more benchmark points
as well as differential distributions.

\subsection*{Acknowledgements}

We would like to thank Ansgar Denner, Howard Haber and Jean-Nicolas Lang for helpful discussions
{and especially Jean-Nicolas for an independent check of one-loop matrix elements
against the crossing-related amplitudes used in \citere{Denner:2016etu}.}
HR's work is partially funded by the Danish National Research Foundation, grant number DNRF90. HR acknowledges also support by the National Science Foundation under Grant No. NSFPHY11-25915. 
We thank the German Research Foundation~(DFG) and 
Research Training Group GRK~2044 for the funding and the support and
acknowledge support by the state of Baden--Württemberg through bwHPC and the 
DFG through grant no INST 39/963-1 FUGG.

\section*{Appendix}
\appendix

\section{Field rotation after renormalization -- version b}
\label{App:renormassa}

In this appendix, we present another 
technical variant of our renormalization procedure which is based on a renormalization of the bare potential~\eqref{eq:NLObarepot}. This prescription is similar to the one of Sect.~\ref{sec:renormassb}, however, the rotations of the fields are applied to the renormalized fields after the renormalization transformation. 
Therefore, $\alpha$, $\beta_n$, and $\beta_c$ are pure mixing angles, and $\lambda_3$ must be chosen to parameterize the potential (corresponding to the set $\{p'_\mr{mass}\}$). As no counterterms to the mixing angles exist, we can write their behaviour in the renormalization transformation schematically as 
 \begin{align}
 \alpha_0&= \alpha +0,&\beta_{c,0}&= \beta+0,& \beta_{n,0} &= \beta+0.
 \end{align}
This is analogous to the renormalization of the MSSM suggested in Ref.~\cite{Frank2007}, where the additional angle does not obtain any higher-order corrections.
Each parameter of Eq.~\eqref{eq:physparaNLOprime} has to be renormalized,
\begin{align}
M_{\PH,0}^2&=\MH^2+\delta \MH^2,&M_{\Ph,0}^2&= \Mh^2+\delta \Mh^2,&M_{\PAO,0}^2 &= \MAO^2+\delta \MAO^2,\nonumber\\
M_{\PHP,0}^2&= \MHP^2+\delta \MHP^2,&\beta_0&= \beta+\delta \beta,&\lambda_{3,0} &= \lambda_3+\delta \lambda_3, \nonumber\\
\lambda_{5,0} &= \lambda_5+\delta \lambda_5&M_{\PW,0}^2&= \MW^2+\delta \MW^2,&M_{\PZ,0}^2&= \MZ^2+\delta \MZ^2,,\nonumber\\
e_0&= e+\delta e, &t_{\PH,0}&= 0+\delta t_\PH,&t_{\Ph,0}&= 0+\delta t_\Ph, \label{eq:physpararen1}
\end{align}
so that the parameter renormalization constants are 
\begin{align}
 \{\delta p'_\mr{mass}\}=\{\delta \MH^2,\delta \Mh^2,\delta \MAO^2,\delta \MHP^2,\delta M_\mathrm{\PW}^2,\delta \MZ^2,\delta e,\delta \lambda_5\,\delta \lambda_3,\delta \beta,\delta t_\PH,\delta t_\Ph\}. \label{eq:renconstsRprime2}
\end{align}
In addition we renormalize each field according to Eq.~\eqref{eq:physfieldren}.
Applying the renormalization transformation of Eqs.\refeq{eq:physpararen1}, 
\refeq{eq:physfieldren} results in the potential
$V(\{p'_\mr{mass}\})+\delta V(\{p'_\mr{mass}\},\{\delta \mathbf{R}'_\mr{mass}\})$
with the already known LO potential and the counterterm potential up to quadratic terms
\begin{align}
\delta V(\{p'_\mr{mass}\},\{\delta \mathbf{R}'_\mr{mass}\})=&- \delta t_\PH H -\delta t_\Ph h \nonumber
\\
&+\frac{1}{2} (\delta \MH^2 +\delta Z_{\PH} \MH^2) H^2 +\frac{1}{2} (\delta \Mh^2 +\delta Z_{\Ph} \Mh^2) h^2 
\nonumber\\
&+\frac{1}{2} (\delta \MAO^2 +\delta Z_{\PAO} \MAO^2) A_0^2 + (\delta \MHP^2+\delta Z_{\PHP} \MHP^2)H^+ H^-
\nonumber\\
&+ \frac{e}{4 \MW \sw} (\delta t_\Ph s_{\alpha-\beta}-\delta t_\PH c_{\alpha-\beta}) (G_0^2 + 2 G^+ G^-)
\nonumber\\
&+  \frac{1}{2}\big(2\delta \overline{M}_{\PH\Ph}^2+ \MH^2\delta Z_{\PH\Ph}+ \Mh^2 \delta Z_{\Ph\PH}\big) H h
\nonumber\\
&+ \frac{1}{2}\big(2\delta \overline{M}_{\PGO\PAO}^2+ \MAO^2 \delta Z_{\PAO\PGO}\big) A_0 G_0
\nonumber\\
&+ \frac{1}{2}\big(2\delta \overline{M}^2_{\PG\PHP}+ \MHP^2 \delta Z_{\PH\PGP}\big) (H^+ G^-+G^+ H^-),
\label{eq:physCT}
\end{align}
with the $\PH\Ph$~mixing terms of Eq.~\refeq{eq:MHhdependence}
\begin{align}
\delta \overline{M}_{\PH\Ph}^2 = f_\alpha\{\delta p'_\mr{mass}\}.
\end{align}
and the mixing terms of the CP-odd and charged sectors
\begin{align}
\delta \overline{M}_{\PGO\PAO}^2={}& -\MAO^2  \delta  \beta-e\frac{ 
\delta t_\PH s_{\alpha -\beta }
+\delta t_\Ph c_{\alpha -\beta }
}{2 \MW \sw},\\
\delta \overline{M}_{\PG\PHP}^2={}& -\MHP^2  \delta  \beta-e\frac{ 
\delta t_\PH s_{\alpha -\beta }
+\delta t_\Ph c_{\alpha -\beta }
}{2 \MW \sw},
\end{align}
which are marked with a bar here to distinguish them from the corresponding
constants of our renormalization version~a.

\section{Supplemental results for counterterms}

In this appendix we supplement the derivation of the counterterm Lagrangian of
Sect.~\ref{sec:CTLagrangian} by some more details.

\subsection{Scalar--Vector mixing terms}
\label{App:SV-terms}

The scalar--vector mixing terms cancel at LO 
against terms in the gauge-fixing contribution. Since the gauge fixing is applied to
renormalized fields, NLO counterterms to the mixing contributions still survive\footnote{It is also possible to formulate the gauge fixing in terms of bare fields, however, one has to renormalize and fix all constants appearing in the gauge fixing separately, which has to be done carefully.}. The mixing of gauge-boson and scalar fields in terms of bare parameters and general mixing angles 
(without gauge fixing) is
\begin{align}
 \mathcal{L}_{SV}={}& \MZ c_{\beta-\beta_n} Z_\mu \partial^\mu G_0 -  \im \MW c_{\beta-\beta_c}(W^+_\mu \partial^\mu G^- - W^-_\mu \partial^\mu G^+) \nonumber \\
 &+\MZ s_{\beta-\beta_n} Z_\mu\partial^\mu A_0 -\im \MW s_{\beta-\beta_c} ( W^+_\mu \partial^\mu H^- - W^-_\mu \partial^\mu H^+).
\end{align}
Together with the renormalization transformation~\eqref{eq:physpararen2} and 
\eqref{eq:physfieldren}, one obtains the $SV$ mixing counterterms as
\begin{subequations}
 \begin{align}
  \delta \mathcal{L}_{\PZ\PGO}&= Z_\mu\partial^\mu G_0 (\MZ^2 \delta Z_{\PZ\PZ} + \delta \MZ^2)/(2\MZ),
\\
  \delta \mathcal{L}_{\PW\PGP}&= -\im ( W^+_\mu \partial^\mu G^- - W^-_\mu \partial^\mu G^+) (\MW^2\delta Z_{\PW} + \delta \MW^2)/(2 \MW), \\
 \delta \mathcal{L}_{\PZ\PAO}&= \MZ Z_\mu\partial^\mu A_0 (\delta Z_{\PGO\PAO}/2 +  \delta \beta-\delta \beta_n),\\
  \delta \mathcal{L}_{\PW\PHP}&= - \im \MW ( W^+_\mu \partial^\mu H^- - W^-_\mu \partial^\mu H^+) (\delta Z_{\PG\PHP}/2+ \delta \beta-\delta \beta_c),
 \end{align}
 \end{subequations}
where we have set the renormalization constants
$\delta Z_{\PGO}$, $\delta Z_{\PAO\PGO}$, $\delta Z_{\PGP}$, $\delta Z_{\PH\PGP}$
to zero.
When the counterterm definition~\eqref{eq:mixanglect} is inserted, the contributions from the mixing angle vanish.

\subsection{Counterterms to Yukawa couplings}
\label{App:Yuk-CT}
 The coupling counterterms of the neutral CP-even and pseudoscalar Higgs fields to the fermions factorize from the corresponding LO structure, while the couplings to the charged Higgs bosons obtain additional terms. 
Setting the CKM matrix to the unit matrix, the corresponding terms in the Lagrangian read
\begin{align}
 \delta \mathcal{L}_{\bar{f}f,\mr{mass}}={}& 
-m_f \bar{f} f \Big(\frac{1}{2} \delta Z^{f,\rR}+\frac{1}{2} \delta Z^{f,\rL}+\frac{\delta m_f}{m_f}\Big),
\nonumber\\
 \frac{\delta \mathcal{L}_{\bar{f}f\Ph}}{\mathcal{L}_{ff\Ph}}={}&\delta Z_e-\frac{\delta \MW^2}{2 \MW^2}- \frac{\delta \sw}{\sw}+ \frac{\delta m_f}{m_f}+\frac{1}{2} \delta Z^{f,\rR}+\frac{1}{2} \delta Z^{f,\rL}+ \frac{1}{2}\delta Z_\Ph+\frac{\delta \xi^f_\Ph}{\xi^f_\Ph} +\frac{\delta Z_{\PH\Ph} \xi^f_\PH}{2\xi^f_\Ph},
\nonumber \\
 \frac{\delta \mathcal{L}_{\bar{f}f\PH}}{\mathcal{L}_{ff\PH}}={}& \delta Z_e-\frac{\delta \MW^2}{2 \MW^2}- \frac{\delta \sw}{\sw}+ \frac{\delta m_f}{m_f}+\frac{1}{2} \delta Z^{f,\rR}+\frac{1}{2} \delta Z^{f,\rL}+ \frac{1}{2}\delta Z_\PH+\frac{\delta \xi^f_\PH}{\xi^f_\PH} +\frac{\delta Z_{\Ph\PH} \xi^f_\Ph}{2\xi^f_\PH},
\nonumber\\
  \frac{\delta \mathcal{L}_{\bar{f}f\PA_0}}{\mathcal{L}_{ff\PA_0}}={}&\delta Z_e-\frac{\delta \MW^2}{2 \MW^2}- \frac{\delta \sw}{\sw}+ \frac{\delta m_f}{m_f}+\frac{1}{2} \delta Z^{f,\rR}+\frac{1}{2} \delta Z^{f,\rL}+ \frac{1}{2}\delta Z_\PAO+\frac{\delta \xi^f_\PAO}{\xi^f_\PAO} +\frac{\delta Z_{\PGO\PAO}}{2\xi^f_\PAO},
\nonumber\\
    \frac{\delta \mathcal{L}_{\bar{f}f\PG_0}}{\mathcal{L}_{ff\PG_0}}={}&\delta Z_e-\frac{\delta \MW^2}{2 \MW^2}- \frac{\delta \sw}{\sw}+ \frac{\delta m_f}{m_f}+\frac{1}{2} \delta Z^{f,\rR}+\frac{1}{2} \delta Z^{f,\rL}+ \delta \xi^f_\PGO,
\nonumber\\
    \delta \mathcal{L}_{\bar{f}f\PH^+}={}&\Big(\delta Z_e-\frac{\delta \MW^2}{2 \MW^2}- \frac{\delta \sw}{\sw}+ \frac{1}{2}\delta Z_{\PHP} \Big) \mathcal{L}_{\bar{f}f\PH^+} + \frac{1}{2}\delta Z_{\PG\PHP}\left.\mathcal{L}_{\bar{f}f\PG^+}\right|_{G^+\to H^+} 
\nonumber\\
    &- \frac{e}{\sqrt{2} \MW\sw} H^+ \bar{u}\bigg[-m_u\xi_\PAO^u  \omega_-  \Big(\frac{\delta m_u}{m_u}+\frac{1}{2} \delta Z^{d,\rL}+\frac{1}{2} \delta Z^{u,\rR}  + \frac{\delta \xi_{\PH^+}^u}{\xi_\PAO^u}\Big) 
\nonumber\\
    &\qquad + m_d \xi_\PAO^d   \omega_+  \Big(\frac{\delta m_d}{m_d}+\frac{1}{2} \delta Z^{u,\rL}+\frac{1}{2} \delta Z^{d,\rR}+ \frac{\delta \xi_{\PH^+}^d}{\xi_\PAO^d}\Big)\bigg]d, 
\nonumber\\
    \delta \mathcal{L}_{\bar{f}f\PH^-}={}&\delta \mathcal{L}_{\bar{f}f\PH^+}^\dagger,
\nonumber\\
    \delta \mathcal{L}_{\bar{f}f\PG^+}={}&\Big(\delta Z_e-\frac{\delta \MW^2}{2 \MW^2}- \frac{\delta \sw}{\sw}+ \delta \xi_{\PG^+} \Big) \mathcal{L}_{\bar{f}f\PG^+} 
\nonumber\\
    &- \frac{e}{\sqrt{2} \MW\sw} G^+ \bar{u}\bigg[-m_u\, \omega_-  \Big(\frac{\delta m_u}{m_u}+\frac{1}{2} \delta Z^{d,\rL}+\frac{1}{2} \delta Z^{u,\rR}  \Big) 
\nonumber\\
    &\qquad + m_d \, \omega_+ \Big(\frac{\delta m_d}{m_d}+\frac{1}{2} \delta Z^{u,\rL}+\frac{1}{2} \delta Z^{d,\rR}\Big)\bigg]d,
\nonumber\\
    \delta \mathcal{L}_{\bar{f}f\PG^-}={}&\delta \mathcal{L}_{\bar{f}f\PG^+}^\dagger,
\end{align}
where the suffixes in the Lagrangian contributions $ \delta \mathcal{L}_{\dots}$
indicate the vertex which is represented.
The generation indices and the flavour summations are suppressed in the notation
and the renormalization constants
$\delta Z_{\PGO}$, $\delta Z_{\PAO\PGO}$, $\delta Z_{\PGP}$, $\delta Z_{\PH\PGP}$
are set to zero.
In contrast to the SM case, the counterterms in the Higgs--fermion interaction involve also the renormalization constants $\delta \beta$ (as vevs appear in the coupling constants), $\delta \beta_{n,c}$, and $\delta \alpha$ (through the general renormalization of the mixing angles) which are hidden in the $ \delta \xi$ factors. 
The values of the counterterms for the different types of THDM are summarized in Tab.~\ref{tab:deltayukint}. 
\begin{table}
   \renewcommand{\arraystretch}{1.3}
  \centering
\begin{tabular}{|c|c|c|c|c|}\hline
&Type I & Type II & Lepton-specific & Flipped\\\hline
$\delta \xi^l_\PH$ & $\frac{c_\alpha}{s_\beta}\delta \alpha-\frac{ s_\alpha c_\beta}{s_\beta^2} \delta \beta$ & $-\frac{s_\alpha} {c_\beta}\delta \alpha+ \frac{c_\alpha s_\beta}{c_\beta^2} \delta \beta$ & $-\frac{s_\alpha} {c_\beta}\delta \alpha+ \frac{c_\alpha s_\beta}{c_\beta^2} \delta \beta$  &$\frac{c_\alpha }{s_\beta}\delta \alpha -\frac{ s_\alpha c_\beta}{s_\beta^2} \delta \beta$
\\
$\delta \xi^u_\PH$ & $\frac{c_\alpha}{s_\beta}\delta \alpha-\frac{ s_\alpha c_\beta}{s_\beta^2} \delta \beta$ & $\frac{c_\alpha}{s_\beta}\delta \alpha-\frac{ s_\alpha c_\beta}{s_\beta^2} \delta \beta$ & $\frac{c_\alpha}{s_\beta}\delta \alpha-\frac{ s_\alpha c_\beta}{s_\beta^2} \delta \beta$ &$\frac{c_\alpha}{s_\beta}\delta \alpha-\frac{ s_\alpha c_\beta}{s_\beta^2} \delta \beta$ \\
$\delta \xi^d_\PH$& $\frac{c_\alpha}{s_\beta}\delta \alpha-\frac{ s_\alpha c_\beta}{s_\beta^2} \delta \beta$ & $-\frac{s_\alpha} {c_\beta}\delta \alpha+ \frac{c_\alpha s_\beta}{c_\beta^2} \delta \beta$ & $\frac{c_\alpha}{s_\beta}\delta \alpha-\frac{ s_\alpha c_\beta}{s_\beta^2} \delta \beta$ & $-\frac{s_\alpha} {c_\beta}\delta \alpha+ \frac{c_\alpha s_\beta}{c_\beta^2} \delta \beta$ \\\hline

$\delta \xi^l_\Ph$ & $-\frac{s_\alpha}{s_\beta}\delta \alpha- \frac{c_\alpha c_\beta}{s_\beta^2}\delta \beta$ & $-\frac{c_\alpha}{c_\beta} \delta \alpha- \frac{s_\alpha s_\beta}{c_\beta^2} \delta \beta$ & $-\frac{c_\alpha}{c_\beta} \delta \alpha- \frac{s_\alpha s_\beta}{c_\beta^2} \delta \beta$& $-\frac{s_\alpha}{s_\beta}\delta \alpha- \frac{c_\alpha c_\beta}{s_\beta^2}\delta \beta$  \\
$\delta \xi^u_\Ph$ & $-\frac{s_\alpha}{s_\beta}\delta \alpha- \frac{c_\alpha c_\beta}{s_\beta^2}\delta \beta$ & $-\frac{s_\alpha}{s_\beta}\delta \alpha- \frac{c_\alpha c_\beta}{s_\beta^2}\delta \beta$ & $-\frac{s_\alpha}{s_\beta}\delta \alpha- \frac{c_\alpha c_\beta}{s_\beta^2}\delta \beta$ &$-\frac{s_\alpha}{s_\beta}\delta \alpha- \frac{c_\alpha c_\beta}{s_\beta^2}\delta \beta$ \\
$\delta \xi^d_\Ph$& $-\frac{s_\alpha}{s_\beta}\delta \alpha- \frac{c_\alpha c_\beta}{s_\beta^2}\delta \beta$ & $-\frac{c_\alpha}{c_\beta} \delta \alpha- \frac{s_\alpha s_\beta}{c_\beta^2} \delta \beta$ & $-\frac{s_\alpha}{s_\beta}\delta \alpha- \frac{c_\alpha c_\beta}{s_\beta^2}\delta \beta$ &$-\frac{c_\alpha}{c_\beta} \delta \alpha- \frac{s_\alpha s_\beta}{c_\beta^2} \delta \beta$  
\\\hline

$\delta \xi^l_{\PAO,\PH^+}$ & $-\delta \beta_{n,c}-\frac{c_\beta^2}{s_\beta^2} \delta \beta$ & $-\delta \beta_{n,c}-\frac{s_\beta^2}{c_\beta^2} \delta \beta$ &$-\delta \beta_{n,c}-\frac{s_\beta^2}{c_\beta^2} \delta \beta$ & $-\delta \beta_{n,c}-\frac{c_\beta^2}{s_\beta^2} \delta \beta$  
\\
$\delta \xi^u_{\PAO,\PH^+}$ & $- \delta \beta_{n,c}-\frac{c_\beta^2}{s_\beta^2} \delta \beta$ & $- \delta \beta_{n,c}-\frac{c_\beta^2}{s_\beta^2} \delta \beta$ & $- \delta \beta_{n,c}-\frac{c_\beta^2}{s_\beta^2} \delta \beta$&$- \delta \beta_{n,c}-\frac{c_\beta^2}{s_\beta^2} \delta \beta$ 
\\
$\delta \xi^d_{\PAO,\PH^+}$& $-\delta \beta_{n,c}-\frac{c_\beta^2}{s_\beta^2} \delta \beta$ & $-\delta \beta_{n,c}-\frac{s_\beta^2}{c_\beta^2} \delta \beta$ & $-\delta \beta_{n,c}-\frac{c_\beta^2}{s_\beta^2} \delta \beta$ & $-\delta \beta_{n,c}-\frac{s_\beta^2}{c_\beta^2} \delta \beta$ 
\\\hline

$\delta \xi^l_{\PGO,\PG^+}$ & $\frac{c_\beta}{s_\beta} (\delta \beta_{n,c}-\delta \beta)$ & $-\frac{s_\beta}{c_\beta} (\delta \beta_{n,c}-\delta \beta)$ &$-\frac{ s_\beta}{c_\beta} (\delta \beta_{n,c}-\delta \beta)$ & $\frac{c_\beta}{s_\beta} (\delta \beta_{n,c}-\delta \beta)$  
\\
$\delta \xi^u_{\PGO,\PG^+}$ & $\frac{c_\beta}{s_\beta} (\delta \beta_{n,c}-\delta \beta)$ & $\frac{c_\beta}{s_\beta} (\delta \beta_{n,c}-\delta \beta)$ & $\frac{c_\beta}{s_\beta} (\delta \beta_{n,c}-\delta \beta)$&$\frac{c_\beta}{s_\beta} (\delta \beta_{n,c}-\delta \beta)$ 
\\
$\delta \xi^d_{\PGO,\PG^+}$& $\frac{c_\beta}{s_\beta} (\delta \beta_{n,c}-\delta \beta)$ & $-\frac{s_\beta}{c_\beta} (\delta \beta_{n,c}-\delta \beta)$ & $\frac{c_\beta}{s_\beta} (\delta \beta_{n,c}-\delta \beta)$ & $-\frac{ s_\beta}{c_\beta} (\delta \beta_{n,c}-\delta \beta)$ 
\\\hline
\end{tabular}  
  \caption{The dependence of the angular counterterms $\delta \xi$ for the different types of models.}
\label{tab:deltayukint}
\end{table}

\section{SM parameters}
\label{app:smparams}

In this appendix we collect the remaining input parameters used in the numerics,
which are necessary to define the SM part of the THDM. 
As recommended by the LHC Higgs Cross Section Working Group~\cite{deFlorian:2016spz}, 
we use the parameter values 
\begin{align}
 G_\mu={}& 0.11663787\cdot 10^{-4} \text{ GeV}^{-2},& \alpha_\mr{s}={} & 0.118,
\nonumber\\
  \MZ={} &91.1876 \text{ GeV}, & \MW={} &80.385 \text{ GeV},
\nonumber\\
 \Gamma_\mr{Z}={}&2.4952 \text{ GeV}, &\Gamma_\mr{\PW}={}&2.085 \text{ GeV}, \nonumber\\
  \Me={}&510.998928 \text{ keV},& m_\mu={}&105.6583715 \text{ MeV}, & m_\tau={}&1.77682 \text{ GeV},\nonumber\\
  \Mu={} &100 \text{ MeV},& \Mc={} &1.51 \text{ GeV}, & \Mt={} &172.5 \text{ GeV},\nonumber\\
 \Md={} &100 \text{ MeV},  &  \Ms={} &100  \text{ MeV},  & \Mb={} &4.92 \text{ GeV}, 
\end{align}
where $G_\mu$ is the Fermi constant, $\alpha_\mr{s}$ the strong coupling constant at the Z~pole, 
$\Gamma_\mr{Z}$ and $\Gamma_\mr{\PW}$ the total decay widths of the Z and W boson, respectively, 
and $\Me, \dots, \Mb$ the fermion masses. 
{The W/Z masses are ``on-shell masses'', which are combined with the W/Z~decay widths 
to complex pole masses; all Higgs-boson and fermion masses are (real) pole masses.}
The electromagnetic coupling is fixed in the $\GF$~scheme, i.e.\ calculated from the
muon decay constant according to
\begin{align}
\alpha_\mr{em} =
\frac{\sqrt{2}}{\pi}\GF\MW^2\left(1-\frac{\MW^2}{\MZ^2}\right),
\end{align}
since this choice is appropriate in the NLO calculation for
$\Ph\to4f$. 
In the $\GF$~scheme, the charge renormalization constant $\delta Z_e$ of Eq.~\refeq{eq:dZe}
receives an additional contribution $\Delta r$, which quantifies the NLO corrections 
to muon decay (see, e.g., \citere{Dittmaier:2001ay}). The correction $\Delta r$
was calculated in the THDM, for instance, in \citere{LopezVal:2012zb}.
For the conversion of THDM parameters between the different renormalization schemes
and the calculation of the \MSbar{} parameter running choosing the 
$\GF$~scheme plays only a minor role.

\begin{figure}
  \centering
  \subfigure[]{
\label{fig:plot_MUSCAN-Aa-L3MS_muzero125}
\includegraphics{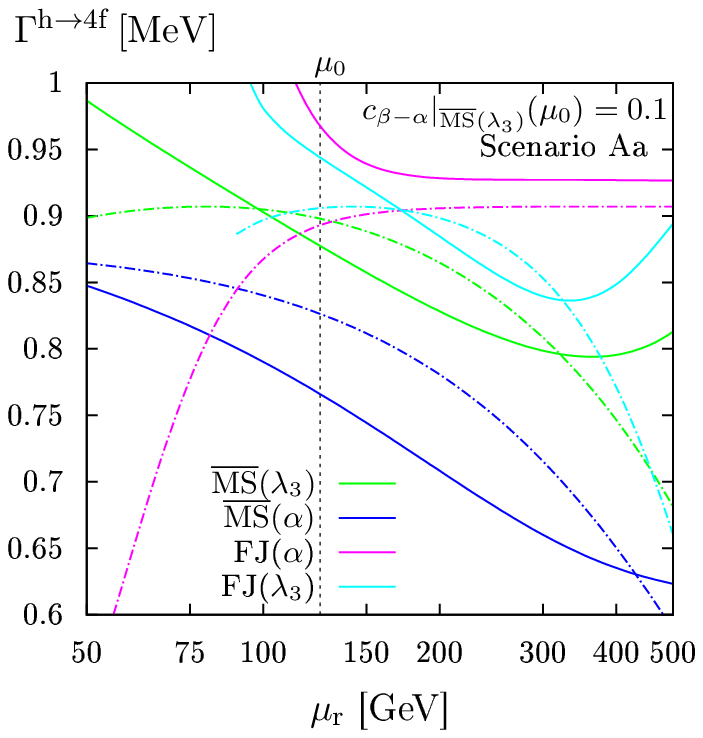}
}
\hspace{15pt}
  \subfigure[]{
\label{fig:plot_MUSCAN-Aa-L3MS_muzero125_woCon}
\includegraphics{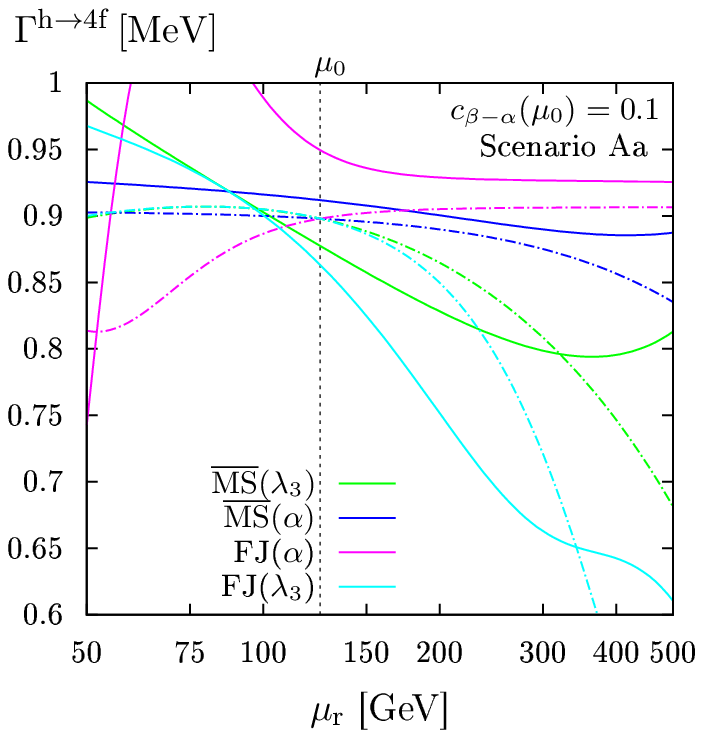}
}
\caption{Renormalization scale dependence of
the decay width for $\Ph \to 4f$ 
in LO (dashed) and NLO EW (solid)
for the benchmark point~Aa
using a central renormalization scale of $\mu_0=\Mh$ 
(in contrast to \reffi{fig:plotmuscan}).
In (a) the input for $\beta$ and $\alpha$ is
defined in the $\overline{\mr{MS}}(\lambda_3)$ scheme
and converted to the other schemes at NLO (also for the LO curves).
In (b) the input for $\beta$ and $\alpha$ is taken without
conversion between the schemes, so that $c_{\beta-\alpha}(\mu_0)=0.1$ 
in all schemes.}
\end{figure}

{\section{\boldmath{Results for the $\Ph\to4f$ decay width with central
renormalization scale $\Mh$}}
\label{app:mu0mh}

Figure~\ref{fig:plot_MUSCAN-Aa-L3MS_muzero125}
shows the renormalization scale variation
of the decay width for $\Ph\to4f$ in scenario~Aa ($\cos{(\beta-\alpha)} = 0.1$)
for the central scale $\mu_0=\Mh$,
in parallel to the results shown in \reffi{fig:plot_MUSCAN-Aa-L3MS}
for our default choice $\mu_0=\frac{1}{5}(\Mh+\MH+\MAO+2\MHP)$.
In contrast to \reffi{fig:plot_MUSCAN-Aa-L3MS},
we observe big discrepancies between the results in the different
renormalization schemes (with proper scheme conversion)
at LO and NLO, with no tendency of
improvement in the transition from LO to NLO.
The large differences in the LO predictions at the 
central scale already signal huge scheme conversion effects
due to unnaturally large corrections that cannot be made up by
NLO effects. 

Figure~\ref{fig:plot_MUSCAN-Aa-L3MS_muzero125_woCon}
shows the respective results without
parameter scheme conversion, so that the LO predictions coincide
at the central scale and reflect the $\mu_{r}$ dependence of
$s^2_{\beta-\alpha}$. 
Lacking the parameter conversion, no reduction of scheme
dependence can be expected here. We rather include this
figure to check whether and where the different schemes
show some reduction of the $\mu_{r}$ dependence in the
transition from LO to NLO.
Such stabilizations are observed at scales about
$300{-}400\GeV$, but not near $\Mh=125\GeV$.

Choosing $\mu_0=\frac{1}{5}(\Mh+\MH+\MAO+2\MHP)=361\GeV$,
the conversion effects and the NLO corrections, however, are
nicely under perturbative control, as discussed in \refse{se:numerics}.
Note that the results at $\mu_{r}=361\GeV$ 
neither in \reffi{fig:plot_MUSCAN-Aa-L3MS_muzero125_woCon},
nor in \reffi{fig:plot_MUSCAN-Aa-L3MS_muzero125}
correspond to $\mu_0=361\GeV$ in \reffi{fig:plot_MUSCAN-Aa-L3MS},
since the input parameters $\alpha$, $\beta$, and $\lambda_5$ are
defined at different renormalizations scales $\mu_0$.
}

\clearpage

\bibliographystyle{JHEPmod}            
\bibliography{bibliography}

\end{document}

%% file: diagrams/tadpoles.tex
\unitlength1bp{%

\begin{feynartspicture}(100,115)(1,1)
\FADiagram{}
\FAProp(0.,10.)(7.5,10.)(0.,){ScalarDash}{0}
\FALabel(3.75,9.18)[t]{$H,h$}
\FAProp(7.5,10.)(7.5,10.)(14.,10.){Straight}{1}
\FALabel(14.82,10.)[l]{$f$}
\FAVert(7.5,10.){0}
\end{feynartspicture}
\begin{feynartspicture}(100,115)(1,1)
\FADiagram{}
\FAProp(0.,10.)(7.5,10.)(0.,){ScalarDash}{0}
\FALabel(3.75,9.18)[t]{$H,h$}
\FAProp(7.5,10.)(7.5,10.)(14.,10.){ScalarDash}{0}
\FALabel(14.82,10.)[l]{$S$}
\FAVert(7.5,10.){0}
\end{feynartspicture}
\begin{feynartspicture}(100,115)(1,1)
\FADiagram{}
\FAProp(0.,10.)(7.5,10.)(0.,){ScalarDash}{0}
\FALabel(3.75,9.18)[t]{$H,h$}
\FAProp(7.5,10.)(7.5,10.)(14.,10.){GhostDash}{1}
\FALabel(14.82,10.)[l]{$u$}
\FAVert(7.5,10.){0}
\end{feynartspicture}
\begin{feynartspicture}(100,115)(1,1)
\FADiagram{}
\FAProp(0.,10.)(7.5,10.)(0.,){ScalarDash}{0}
\FALabel(3.75,9.18)[t]{$H,h$}
\FAProp(7.5,10.)(7.5,10.)(14.,10.){Sine}{0}
\FALabel(15.07,10.)[l]{$V$}
\FAVert(7.5,10.){0}

\end{feynartspicture}}

%% file: diagrams/selfenergies.tex
\unitlength=1bp%
\begin{feynartspicture}(432,104)(4,1)

\FADiagram{}
\FAProp(0.,10.)(10.,10.)(0.,){ScalarDash}{0}
\FALabel(5.,9.18)[t]{$H$}
\FAProp(20.,10.)(10.,10.)(0.,){ScalarDash}{0}
\FALabel(15.,9.18)[t]{$H$}
\FAProp(10.,10.)(10.,10.)(10.,15.5){ScalarDash}{0}
\FALabel(10.,16.32)[b]{$S$}
\FAVert(10.,10.){0}

\FADiagram{}
\FAProp(0.,10.)(10.,10.)(0.,){ScalarDash}{0}
\FALabel(5.,9.18)[t]{$H$}
\FAProp(20.,10.)(10.,10.)(0.,){ScalarDash}{0}
\FALabel(15.,9.18)[t]{$H$}
\FAProp(10.,10.)(10.,10.)(10.,15.5){Sine}{0}
\FALabel(10.,16.57)[b]{$V$}
\FAVert(10.,10.){0}

\FADiagram{}
\FAProp(0.,10.)(6.,10.)(0.,){ScalarDash}{0}
\FALabel(3.,9.18)[t]{$H$}
\FAProp(20.,10.)(14.,10.)(0.,){ScalarDash}{0}
\FALabel(17.,10.82)[b]{$H$}
\FAProp(6.,10.)(14.,10.)(0.8,){Straight}{-1}
\FALabel(10.,5.98)[t]{$f$}
\FAProp(6.,10.)(14.,10.)(-0.8,){Straight}{1}
\FALabel(10.,14.02)[b]{$f$}
\FAVert(6.,10.){0}
\FAVert(14.,10.){0}

\FADiagram{}
\FAProp(0.,10.)(6.,10.)(0.,){ScalarDash}{0}
\FALabel(3.,9.18)[t]{$H$}
\FAProp(20.,10.)(14.,10.)(0.,){ScalarDash}{0}
\FALabel(17.,10.82)[b]{$H$}
\FAProp(6.,10.)(14.,10.)(0.8,){ScalarDash}{0}
\FALabel(10.,5.98)[t]{$S$}
\FAProp(6.,10.)(14.,10.)(-0.8,){ScalarDash}{0}
\FALabel(10.,14.02)[b]{$S$}
\FAVert(6.,10.){0}
\FAVert(14.,10.){0}
\end{feynartspicture}
\vspace{-30pt}\\
\begin{feynartspicture}(432,104)(4,1)

\FADiagram{}
\vspace{-80pt}
\FAProp(0.,10.)(6.,10.)(0.,){ScalarDash}{0}
\FALabel(3.,9.18)[t]{$H$}
\FAProp(20.,10.)(14.,10.)(0.,){ScalarDash}{0}
\FALabel(17.,10.82)[b]{$H$}
\FAProp(6.,10.)(14.,10.)(0.8,){GhostDash}{-1}
\FALabel(10.,5.98)[t]{$u$}
\FAProp(6.,10.)(14.,10.)(-0.8,){GhostDash}{1}
\FALabel(10.,14.02)[b]{$u$}
\FAVert(6.,10.){0}
\FAVert(14.,10.){0}

\FADiagram{}
\FAProp(0.,10.)(6.,10.)(0.,){ScalarDash}{0}
\FALabel(3.,9.18)[t]{$H$}
\FAProp(20.,10.)(14.,10.)(0.,){ScalarDash}{0}
\FALabel(17.,10.82)[b]{$H$}
\FAProp(6.,10.)(14.,10.)(0.8,){Sine}{0}
\FALabel(10.,5.73)[t]{$V$}
\FAProp(6.,10.)(14.,10.)(-0.8,){Sine}{0}
\FALabel(10.,14.27)[b]{$V$}
\FAVert(6.,10.){0}
\FAVert(14.,10.){0}

\FADiagram{}
\FAProp(0.,10.)(6.,10.)(0.,){ScalarDash}{0}
\FALabel(3.,9.18)[t]{$H$}
\FAProp(20.,10.)(14.,10.)(0.,){ScalarDash}{0}
\FALabel(17.,10.82)[b]{$H$}
\FAProp(6.,10.)(14.,10.)(0.8,){ScalarDash}{0}
\FALabel(10.,5.98)[t]{$S$}
\FAProp(6.,10.)(14.,10.)(-0.8,){Sine}{0}
\FALabel(10.,14.27)[b]{$V$}
\FAVert(6.,10.){0}
\FAVert(14.,10.){0}

\end{feynartspicture}

%% file: diagrams/triangles.tex
\unitlength=1bp%

\begin{feynartspicture}(432,104)(4,1)

\FADiagram{}
\FAProp(0.,15.)(7.,14.)(0.,){ScalarDash}{0}
\FALabel(3.68385,15.3069)[b]{$A_0$}
\FAProp(0.,5.)(7.,6.)(0.,){ScalarDash}{0}
\FALabel(3.68385,4.69307)[t]{$A_0$}
\FAProp(20.,10.)(13.5,10.)(0.,){ScalarDash}{0}
\FALabel(16.75,10.82)[b]{$H$}
\FAProp(7.,14.)(7.,6.)(0.,){Straight}{1}
\FALabel(6.18,10.)[r]{$f$}
\FAProp(7.,14.)(13.5,10.)(0.,){Straight}{-1}
\FALabel(10.4513,12.6272)[bl]{$f$}
\FAProp(7.,6.)(13.5,10.)(0.,){Straight}{1}
\FALabel(10.4513,7.37284)[tl]{$f$}
\FAVert(7.,14.){0}
\FAVert(7.,6.){0}
\FAVert(13.5,10.){0}

\FADiagram{}
\FAProp(0.,15.)(7.,14.)(0.,){ScalarDash}{0}
\FALabel(3.68385,15.3069)[b]{$A_0$}
\FAProp(0.,5.)(7.,6.)(0.,){ScalarDash}{0}
\FALabel(3.68385,4.69307)[t]{$A_0$}
\FAProp(20.,10.)(13.5,10.)(0.,){ScalarDash}{0}
\FALabel(16.75,10.82)[b]{$H$}
\FAProp(7.,14.)(7.,6.)(0.,){ScalarDash}{0}
\FALabel(6.18,10.)[r]{$S$}
\FAProp(7.,14.)(13.5,10.)(0.,){ScalarDash}{0}
\FALabel(10.4513,12.6272)[bl]{$S$}
\FAProp(7.,6.)(13.5,10.)(0.,){ScalarDash}{0}
\FALabel(10.4513,7.37284)[tl]{$S$}
\FAVert(7.,14.){0}
\FAVert(7.,6.){0}
\FAVert(13.5,10.){0}

\FADiagram{}
\FAProp(0.,15.)(7.,14.)(0.,){ScalarDash}{0}
\FALabel(3.68385,15.3069)[b]{$A_0$}
\FAProp(0.,5.)(7.,6.)(0.,){ScalarDash}{0}
\FALabel(3.68385,4.69307)[t]{$A_0$}
\FAProp(20.,10.)(13.5,10.)(0.,){ScalarDash}{0}
\FALabel(16.75,10.82)[b]{$H$}
\FAProp(7.,14.)(7.,6.)(0.,){ScalarDash}{0}
\FALabel(6.18,10.)[r]{$S$}
\FAProp(7.,14.)(13.5,10.)(0.,){ScalarDash}{0}
\FALabel(10.4513,12.6272)[bl]{$S$}
\FAProp(7.,6.)(13.5,10.)(0.,){Sine}{0}
\FALabel(10.5824,7.15993)[tl]{$V$}
\FAVert(7.,14.){0}
\FAVert(7.,6.){0}
\FAVert(13.5,10.){0}

\FADiagram{}
\FAProp(0.,15.)(7.,14.)(0.,){ScalarDash}{0}
\FALabel(3.68385,15.3069)[b]{$A_0$}
\FAProp(0.,5.)(7.,6.)(0.,){ScalarDash}{0}
\FALabel(3.68385,4.69307)[t]{$A_0$}
\FAProp(20.,10.)(13.5,10.)(0.,){ScalarDash}{0}
\FALabel(16.75,10.82)[b]{$H$}
\FAProp(7.,14.)(7.,6.)(0.,){ScalarDash}{0}
\FALabel(6.18,10.)[r]{$S$}
\FAProp(7.,14.)(13.5,10.)(0.,){Sine}{0}
\FALabel(10.5824,12.8401)[bl]{$V$}
\FAProp(7.,6.)(13.5,10.)(0.,){ScalarDash}{0}
\FALabel(10.4513,7.37284)[tl]{$S$}
\FAVert(7.,14.){0}
\FAVert(7.,6.){0}
\FAVert(13.5,10.){0}
\end{feynartspicture}
\vspace{-25pt}\\
\begin{feynartspicture}(432,104)(4,1)
\FADiagram{}
\FAProp(0.,15.)(7.,14.)(0.,){ScalarDash}{0}
\FALabel(3.68385,15.3069)[b]{$A_0$}
\FAProp(0.,5.)(7.,6.)(0.,){ScalarDash}{0}
\FALabel(3.68385,4.69307)[t]{$A_0$}
\FAProp(20.,10.)(13.5,10.)(0.,){ScalarDash}{0}
\FALabel(16.75,10.82)[b]{$H$}
\FAProp(7.,14.)(7.,6.)(0.,){Sine}{0}
\FALabel(5.93,10.)[r]{$V$}
\FAProp(7.,14.)(13.5,10.)(0.,){ScalarDash}{0}
\FALabel(10.4513,12.6272)[bl]{$S$}
\FAProp(7.,6.)(13.5,10.)(0.,){ScalarDash}{0}
\FALabel(10.4513,7.37284)[tl]{$S$}
\FAVert(7.,14.){0}
\FAVert(7.,6.){0}
\FAVert(13.5,10.){0}

\FADiagram{}
\FAProp(0.,15.)(7.,14.)(0.,){ScalarDash}{0}
\FALabel(3.68385,15.3069)[b]{$A_0$}
\FAProp(0.,5.)(7.,6.)(0.,){ScalarDash}{0}
\FALabel(3.68385,4.69307)[t]{$A_0$}
\FAProp(20.,10.)(13.5,10.)(0.,){ScalarDash}{0}
\FALabel(16.75,10.82)[b]{$H$}
\FAProp(7.,14.)(7.,6.)(0.,){ScalarDash}{0}
\FALabel(6.18,10.)[r]{$S$}
\FAProp(7.,14.)(13.5,10.)(0.,){Sine}{0}
\FALabel(10.5824,12.8401)[bl]{$V$}
\FAProp(7.,6.)(13.5,10.)(0.,){Sine}{0}
\FALabel(10.5824,7.15993)[tl]{$V$}
\FAVert(7.,14.){0}
\FAVert(7.,6.){0}
\FAVert(13.5,10.){0}

\FADiagram{}
\FAProp(0.,15.)(7.5,13.5)(0.,){ScalarDash}{0}
\FALabel(4.00495,15.0448)[b]{$A_0$}
\FAProp(0.,5.)(9.5,8.)(0.,){ScalarDash}{0}
\FALabel(5.14147,5.74034)[t]{$A_0$}
\FAProp(20.,10.)(9.5,8.)(0.,){ScalarDash}{0}
\FALabel(14.9932,8.20296)[t]{$H$}
\FAProp(7.5,13.5)(9.5,8.)(0.8,){ScalarDash}{0}
\FALabel(5.55827,9.50573)[r]{$S$}
\FAProp(7.5,13.5)(9.5,8.)(-0.8,){ScalarDash}{0}
\FALabel(11.4417,11.9943)[l]{$S$}
\FAVert(7.5,13.5){0}
\FAVert(9.5,8.){0}

\FADiagram{}
\FAProp(0.,15.)(9.5,12.)(0.,){ScalarDash}{0}
\FALabel(5.14147,14.2597)[b]{$A_0$}
\FAProp(0.,5.)(7.5,6.5)(0.,){ScalarDash}{0}
\FALabel(4.00495,4.95525)[t]{$A_0$}
\FAProp(20.,10.)(9.5,12.)(0.,){ScalarDash}{0}
\FALabel(14.9932,11.797)[b]{$H$}
\FAProp(7.5,6.5)(9.5,12.)(0.8,){ScalarDash}{0}
\FALabel(11.4417,8.00573)[l]{$S$}
\FAProp(7.5,6.5)(9.5,12.)(-0.8,){ScalarDash}{0}
\FALabel(5.38,10.15)[r]{$S$}
\FAVert(7.5,6.5){0}
\FAVert(9.5,12.){0}
\end{feynartspicture}
\vspace{-25pt}\\
\begin{feynartspicture}(432,104)(4,1)
\FADiagram{}
\FAProp(0.,15.)(7.,10.)(0.,){ScalarDash}{0}
\FALabel(3.22439,11.9221)[tr]{$A_0$}
\FAProp(0.,5.)(7.,10.)(0.,){ScalarDash}{0}
\FALabel(3.77561,6.92215)[tl]{$A_0$}
\FAProp(20.,10.)(13.,10.)(0.,){ScalarDash}{0}
\FALabel(16.5,10.82)[b]{$H$}
\FAProp(13.,10.)(7.,10.)(0.8,){ScalarDash}{0}
\FALabel(10.,13.22)[b]{$S$}
\FAProp(13.,10.)(7.,10.)(-0.8,){ScalarDash}{0}
\FALabel(10.,6.78)[t]{$S$}
\FAVert(13.,10.){0}
\FAVert(7.,10.){0}

\FADiagram{}
\FAProp(0.,15.)(7.,10.)(0.,){ScalarDash}{0}
\FALabel(3.22439,11.9221)[tr]{$A_0$}
\FAProp(0.,5.)(7.,10.)(0.,){ScalarDash}{0}
\FALabel(3.77561,6.92215)[tl]{$A_0$}
\FAProp(20.,10.)(13.,10.)(0.,){ScalarDash}{0}
\FALabel(16.5,10.82)[b]{$H$}
\FAProp(13.,10.)(7.,10.)(0.8,){Sine}{0}
\FALabel(10.,13.47)[b]{$V$}
\FAProp(13.,10.)(7.,10.)(-0.8,){Sine}{0}
\FALabel(10.,6.53)[t]{$V$}
\FAVert(13.,10.){0}
\FAVert(7.,10.){0}

\FADiagram{}

\FADiagram{}
\end{feynartspicture}